\PassOptionsToPackage{unicode}{hyperref}
\PassOptionsToPackage{hyphens}{url}
\PassOptionsToPackage{dvipsnames,svgnames,x11names}{xcolor}
\documentclass[
  12pt]{article}

\usepackage{amsmath,amssymb}
\usepackage{iftex}
\usepackage{booktabs}
\usepackage{setspace}
\usepackage{subfigure}
\usepackage{xr}
\usepackage{comment}
\newcommand{\btheta}{ \mbox{\boldmath $\theta$}}

\newcommand{\balpha}{ \mbox{\boldmath $\alpha$}}
\newcommand{\bbeta}{ \mbox{\boldmath $\beta$}}

\newcommand{\bgamma}{ \mbox{\boldmath $\gamma$}}

\newcommand{\bepsilon}{ \mbox{\boldmath $\epsilon$}}

\newcommand{\bSigma}{ \mbox{\boldmath $\Sigma$}}

\newcommand{\bGamma}{ \mbox{\boldmath $\Gamma$}}

\newcommand{\bA}{ \mbox{\bf A}}

\newcommand{\bx}{ \mbox{\bf x}}
\newcommand{\bX}{ \mbox{\bf X}}
\newcommand{\bB}{ \mbox{\bf B}}
\newcommand{\bZ}{ \mbox{\bf Z}}

\newcommand{\bY}{ \mbox{\bf Y}}

\newcommand{\bV}{ \mbox{\bf V}}
\newcommand{\bG}{ \mbox{\bf G}}
\newcommand{\bC}{ \mbox{\bf C}}

\newcommand{\bI}{ \mbox{\bf I}}
\newcommand{\bD}{ \mbox{\bf D}}
\newcommand{\bM}{ \mbox{\bf M}}
\newcommand{\bW}{ \mbox{\bf W}}

\newcommand{\bT}{ \mbox{\bf T}}
\newcommand{\bQ}{ \mbox{\bf Q}}

\newcommand{\bU}{ \mbox{\bf U}}
\newcommand{\bS}{ \mbox{\bf S}}

\newcommand{\iid}{\stackrel{iid}{\sim}}
\newcommand{\indep}{\stackrel{indep}{\sim}}

\newcommand{\beq}{ \begin{equation}}
\newcommand{\eeq}{ \end{equation}}
\newcommand{\beqn}{ \begin{eqnarray}}
\newcommand{\eeqn}{ \end{eqnarray}}

\usepackage{caption}
\renewcommand{\arraystretch}{1.5}

\ifPDFTeX
  \usepackage[T1]{fontenc}
  \usepackage[utf8]{inputenc}
  \usepackage{textcomp} 
\else 
  \usepackage{unicode-math}
  \defaultfontfeatures{Scale=MatchLowercase}
  \defaultfontfeatures[\rmfamily]{Ligatures=TeX,Scale=1}
\fi
\usepackage{lmodern}
\ifPDFTeX\else  
\fi
\IfFileExists{upquote.sty}{\usepackage{upquote}}{}
\IfFileExists{microtype.sty}{
  \usepackage[]{microtype}
  \UseMicrotypeSet[protrusion]{basicmath} 
}{}
\makeatletter
\@ifundefined{KOMAClassName}{
  \IfFileExists{parskip.sty}{%
    \usepackage{parskip}
  }{
    \setlength{\parindent}{0pt}
    \setlength{\parskip}{6pt plus 2pt minus 1pt}}
}{
  \KOMAoptions{parskip=half}}
\makeatother
\usepackage{xcolor}
\makeatletter
\ifx\paragraph\undefined\else
  \let\oldparagraph\paragraph
  \renewcommand{\paragraph}{
    \@ifstar
      \xxxParagraphStar
      \xxxParagraphNoStar
  }
  \newcommand{\xxxParagraphStar}[1]{\oldparagraph*{#1}\mbox{}}
  \newcommand{\xxxParagraphNoStar}[1]{\oldparagraph{#1}\mbox{}}
\fi
\ifx\subparagraph\undefined\else
  \let\oldsubparagraph\subparagraph
  \renewcommand{\subparagraph}{
    \@ifstar
      \xxxSubParagraphStar
      \xxxSubParagraphNoStar
  }
  \newcommand{\xxxSubParagraphStar}[1]{\oldsubparagraph*{#1}\mbox{}}
  \newcommand{\xxxSubParagraphNoStar}[1]{\oldsubparagraph{#1}\mbox{}}
\fi
\makeatother

\usepackage{longtable,booktabs,array}
\usepackage{calc} 
\usepackage{etoolbox}
\makeatletter
\patchcmd\longtable{\par}{\if@noskipsec\mbox{}\fi\par}{}{}
\makeatother
\IfFileExists{footnotehyper.sty}{\usepackage{footnotehyper}}{\usepackage{footnote}}
\makesavenoteenv{longtable}
\usepackage{graphicx}
\makeatletter
\def\maxwidth{\ifdim\Gin@nat@width>\linewidth\linewidth\else\Gin@nat@width\fi}
\def\maxheight{\ifdim\Gin@nat@height>\textheight\textheight\else\Gin@nat@height\fi}
\makeatother
\setkeys{Gin}{width=\maxwidth,height=\maxheight,keepaspectratio}
\makeatletter
\def\fps@figure{htbp}
\makeatother

\makeatletter
\@ifpackageloaded{caption}{}{\usepackage{caption}}
\AtBeginDocument{%
\ifdefined\contentsname
  \renewcommand*\contentsname{Table of contents}
\else
  \newcommand\contentsname{Table of contents}
\fi
\ifdefined\listfigurename
  \renewcommand*\listfigurename{List of Figures}
\else
  \newcommand\listfigurename{List of Figures}
\fi
\ifdefined\listtablename
  \renewcommand*\listtablename{List of Tables}
\else
  \newcommand\listtablename{List of Tables}
\fi
\ifdefined\figurename
  \renewcommand*\figurename{Figure}
\else
  \newcommand\figurename{Figure}
\fi
\ifdefined\tablename
  \renewcommand*\tablename{Table}
\else
  \newcommand\tablename{Table}
\fi
}
\@ifpackageloaded{float}{}{\usepackage{float}}
\floatstyle{ruled}
\@ifundefined{c@chapter}{\newfloat{codelisting}{h}{lop}}{\newfloat{codelisting}{h}{lop}[chapter]}
\floatname{codelisting}{Listing}

\makeatother
\makeatletter
\@ifpackageloaded{caption}{}{\usepackage{caption}}
\@ifpackageloaded{subcaption}{}{\usepackage{subcaption}}
\makeatother

\ifLuaTeX
  \usepackage{selnolig}  
\fi
\usepackage[]{natbib}
\usepackage{bookmark}

\IfFileExists{xurl.sty}{\usepackage{xurl}}{} 
\hypersetup{
  pdftitle={Title},
  pdfauthor={Author 1; Author 2},
  pdfkeywords={3 to 6 keywords, that do not appear in the title},
  colorlinks=true,
  linkcolor={blue},
  filecolor={Maroon},
  citecolor={Blue},
  urlcolor={Blue},
  pdfcreator={LaTeX via pandoc}}

\newcommand{\anon}{1}

\begin{document}

\def\spacingset#1{\renewcommand{\baselinestretch}%
{#1}\small\normalsize} \spacingset{1}



\if1\anon
{
  \title{\bf A Spectral Confounder Adjustment for Spatial Regression with Multiple Exposures and Outcomes}
  \author{Shih-Ni Prim\thanks{Department of Statistics, North Carolina State University},
  Yawen Guan\thanks{Department of Statistics, Colorado State University}, 
  Shu Yang$^*$,
  Ana G Rappold\thanks{US EPA, Center for Public Health and Environmental Assessment},
  K. Lloyd Hill\thanks{Oak Ridge Associated Universities at US EPA, Center for Public Health and Environmental Assessment}, \\
  Corinna Keeler$^\ddag$,
  and Brian J Reich$^*$ \\
    }
  \maketitle
} \fi

\if0\anon
{
  \bigskip
  \bigskip
  \bigskip
  \begin{center}
    {\LARGE\bf A Spectral Confounder Adjustment for Spatial Regression with Multiple Exposures and Outcomes}
\end{center}
  \medskip
} \fi

\bigskip
\begin{abstract}

    Characterizing social vulnerability is fundamental to disaster response planning. Numerous vulnerability indicators have been developed, but they are typically not validated for their predictive power over the outcomes of interest. As with many environmental health studies where interventions are impractical or unethical, validation of social vulnerability against public health outcomes is observational and relies on spatially-dependent data.  Observational studies are susceptible to bias induced by unmeasured confounders.  This problem is exacerbated in spatial studies with multiple health outcomes and environmental exposure variables, as the source and magnitude of confounding bias may differ by spatial scale and across exposure/outcome pairs.  We propose to mitigate confounding effects in multivariate spatial studies using a tensor-regression model that allows exposure effects to vary by exposure, outcome, and spatial scale.  By extracting exposure effects that correspond to local spatial scales, we replace the strict causal assumption of no unmeasured confounders with a more realistic assumption of local unconfoundedness i.e., differences between nearby regions are unconfounded, allowing for causal interpretation. Our study of the Southern United States shows that economic resilience (Theme 1) demonstrates the strongest positive effect on diabetes and negative effect on hyperlipidemia. Our analysis reveals a concise representation of the effects of social vulnerability indices on an array of chronic health outcomes, offering interpretable epidemiological insights that adjust for multivariate spatial confounding.  

\end{abstract}

\noindent%
\vfill

\newpage
\spacingset{1.8} 

\section{Introduction}\label{s:intro}

Large electronic health databases empower researchers to characterize how the environment and social context give rise to adverse health outcomes and to map community resilience. This task requires untangling many-to-many causal relationships that are deeply intertwined geographically.  For example, poverty influences multiple diseases through pathways such as education, physical and physiological safety, and healthcare access. Consequently, poverty can drive heart failure and diabetes simultaneously, though with varying levels of severity. Because health conditions and poverty are intertwined with other spatially-varied factors, such as local environmental pollution and income, they become spatially confounded. In this paper, we offer a statistical method to study the many-to-many relationship when the assumption of no unmeasured spatial confounding is violated.

All observational studies are susceptible to missing-confounder bias, and in environmental applications, the missing confounders often have a spatial structure.  \cite{reichEffectsResidualSmoothing2006} demonstrated a profound impact of spatial confounding on the estimated regression coefficients in the models that include or exclude spatial random effects.  There are many approaches to remedying this problem in the univariate setting \citep{paciorekImportanceScaleSpatialconfounding2010,hughes2012dimension, pageEstimationPredictionPresence2017,Dupont2022SpatialPlus, khan2022RSR, marques2022mitigating,guan2023spatial,adin2023, dupont2023demystifying,thaden2018structural,wiecha2024twostage,wang2019blessings,reichReviewSpatialCausal2021,miao2023UC,schnell2020mitigating,gilbert2024causal, wiecha2024twostage}. Here, we build on the spectral method in \cite{guan2023spatial} to relax the no unmeasured confounder assumption. \cite{guan2023spatial} allow the exposure effect to vary by spatial scale, and take the effect estimate for the most local scale as the causal estimate.  This relaxes the stringent no unmeasured confounders assumption to the more realistic local unconfoundedness assumption.    

In addition to the presence of unmeasured confounders, mixtures as exposures (i.e., many simultaneous exposures) present unique problems such as high dimensionality, collinearity, and interactions between exposures \citep{joubert2022powering}. A variety of approaches has been proposed  \citep{ carrico2015characterization, bobb2015bayesian, reich2020integrative, keil2020gcomputation, wei2020sparse, ferrari2020identifying, kowal2021bayesian, gibson2021bayesian, roy2021perturbed, chen2022statistical, devick2022bayesian, gennings2022using}, but they do not address spatial confounding, leaving a critical gap. 
As many mixture studies are concerned with epidemiological and environmental problems, the causal inference framework has also been utilized \citep{zigler2021invited, gennings2022using, devick2022bayesian}. 

Among the methods for spatial confounding, few studies include many-to-many relationships with multiple treatments and multiple responses. Note that the terms ``treatment'' and ``exposure'' are used interchangeably. 
For non-spatial approaches, \cite{zheng2023sensitivity} work with one treatment and \cite{kang2023sensitivity} deal with multiple treatments; both rely on factor models and use sensitivity analysis to bound the causal effects of unmeasured confounder for multiple outcomes. \cite{miao2023UC} use auxiliary variables---which are not causally associated with the response---and null treatments to find causal effects from multiple treatments when unmeasured confounders exist. For spatial data, \cite{marques2022mitigating} investigate a single response and multiple covariates in their use of Gaussian random field to reduce spatial confounding.  \cite{Urdangarin2024simplified} extend the work of \cite{Dupont2022SpatialPlus}, who regress the covariates on the thin plate model, and the resulting residuals are used as the covariates while a penalty term for smoothness is included. \cite{Urdangarin2024simplified} find the spectral decomposition of the spatial precision matrix and express the covariate as linear combinations of eigenvectors from frequencies assumed confounded. 

We develop a novel method for modeling causal effects for spatial data with multiple exposures and outcomes in the presence of unmeasured, global spatial confounders. Motivated by \cite{guan2023spatial}, we project spatial data into the spectral domain and use the local-unconfoundedness assumption to identify causal effects.  The extension from univariate to multivariate requires a three-way tensor to quantify the effects by exposure, outcome and spatial scale. The proposed tensor structure captures complicated patterns with constraints on diminishing correlations at local scales to adhere to causal assumptions, and provides interpretable latent factors that lay bare the key relationships that drive the results. 
Compared to related methods, the proposed method allows for multiple exposures and outcomes unlike the univariate methods of \cite{guan2023spatial} and \cite{marques2022mitigating}.  Compared to \cite{Urdangarin2024simplified}, the proposed method provides a comprehensive assessment of how estimates vary by spatial scale and a low-rank and interpretable estimate of the  outcome/exposure structure, and avoids hard selection of the number of frequency terms that are assumed to be confounded.

The remainder of the paper proceeds as follows.  The motivating data are described in Section \ref{s:data}.  Section \ref{s:method} introduces the statistical method. The method is applied to the motivating data in Section \ref{s:data_analysis}. Section \ref{s:discussion} provides concluding remarks and discusses limitations. An extensive simulation study, computation, derivations, and additional results are provided in the Supplementary Materials.

\section{Research Objectives}\label{s:data}

We demonstrate the utility of the proposed method to establish a causal estimate of the strength of the effect between community resilience and adaptive capacity—--characterized by the CDC’s Social Vulnerability Index for Disaster Management (SVIDM)--—and the incidence of prevalent chronic diseases. Although the SVIDM is a composite index designed to identify communities at risk during disasters and public health emergencies, its  predictive validity regarding long-term public health burdens remains unknown. Furthermore, evaluating this relationship is heavily complicated by spatial confounding from unmeasured factors that vary geographically in a similar way to public health metrics. Because understanding this predictive power is necessary before the index can be deployed as an effective public health intervention tool, identifying the causal link between health burden and community resilience (Flanagan et al. 2011) is a critical step in operationalizing the index as a public health intervention tool. A strong predictive relationship enables public health officials to prioritize targeted interventions preemptively and allocate resources both during and in the wake of emergencies. The deployment of resources to vulnerable areas during disasters and emergencies serves as a salient motivating example to demonstrate the value of our proposed methodology.

Composite indices are common tools for mapping community-level vulnerability and managing environmental threats. These indices synthesize multiple variables into a single metric to summarize localized contexts, such as environmental quality \citep{messer2014eqi}, disaster readiness \citep{flanagan2011svi}, and neighborhood deprivation \citep{messer2006ndi}. Methodologically, researchers typically build these composite measures using factor models or principal component analysis \citep{reckien2018index, greco2018indices}. However, the most common drawback is that their predictive value is not validated. Moreover, the relationship between composite indices and public health burden is subject to confounding by unmeasured factors at multiple spatial scales, which can impact their predictive performance and lead to biased results. Therefore, we explore the relationship between health burden and a composite index of community resilience at various spatial scales
to demonstrate the value of the proposed methodology and to evaluate evidence of spatial confounding.

We proceed with the following objectives:
\begin{itemize}
     \item Test whether the relationship between SVIDM and chronic conditions varies by spatial scale, and thus caution against using results from commonly used methods (e.g. linear regression) that do not adjust for confounding.
    \item Estimate the causal effect of each component of SVIDM on the incidence of chronic conditions to help decision makers allocate resources on relevant SVIDM themes for a chronic outcome of interest. 
   \item Distill the many-to-many causal inference problem down to interpretable visualizations to motivate future studies on public health policies to improve disaster preparedness.
\end{itemize}
All three aims are policy relevant: the first calls into question results from naive analyses of the data; the second frames the problem causally which is needed for decision making; and the third focuses on the dimension reduction to help policymakers extract meaningful interpretations from the analysis.

\subsection{Data}

The outcome data consisted of incidence rates of five prevalent chronic conditions (hypertension, chronic kidney disease, hyperlipidemia, congestive heart failure, and diabetes) by ZIP code tabulated area of residence. We obtained individual level data from the Centers for Medicare and Medicaid Services' Chronic Conditions Warehouse, which included chronic health conditions and year specific ZIP-code of residence. Each incident health condition for each unique beneficiary contributed one count to the residential ZIP-code. The counts were aggregated to their corresponding ZIP code tabulated areas (ZCTAs), and the incident rate was calculated as the total counts of incident health conditions divided by the total beneficiaries in the ZCTA each year \citep{usdhhs}. We used the mean of the incident rates from 2017, 2018, and 2019.  

\begin{table}[h]
    \small
    \centering
    \small
    \caption{{\bf Summary statistics of response variables}: The mean and range are from incident rates---the proportion of beneficiaries with the condition in a ZCTA.}
    \begin{tabular}{cccccccc}
    \hline
    & & & \multicolumn{5}{c}{Correlation}  \\
    \hline
    & Mean & Range & Hypertension & CKD & Hyperlipidemia & CHF & Diabetes  \\
    \hline
    Hypertension & 0.130 & 0.530 & 1.000 & 0.718 & 0.765 & 0.621 & 0.685 \\ 
    CKD & 0.051 & 0.131 & 0.718 & 1.000 & 0.584 & 0.625 & 0.648 \\
    Hyperlipidemia & 0.110 & 0.394 & 0.765 & 0.584 & 1.000 & 0.425 & 0.527 \\
    CHF & 0.028 & 0.087 & 0.621 & 0.625 & 0.425 & 1.000 & 0.498 \\
    Diabetes & 0.032 & 0.137 & 0.685 & 0.648 & 0.527 & 0.498 & 1.000 \\
    \hline
    \end{tabular}
    \label{t:summary_response}
\end{table}

The Social Vulnerability Index for Disaster Management was created by the Centers for Disease Control \citep{svi_data} with the intent to understand adaptive capacity for natural disasters through the lens of public health. We retrieved the 2018 SVIDM index covering the period of 2014-2018 at the ZCTA level using the {\tt R} package {\tt findSVI} \citep{findSVI}. 
The index scores range between 0 and 1, representing the proportion of ZCTAs that are lower than or equal to the ZCTA of interest in terms of social vulnerability. Thus, higher scores signal worse abilities to respond to disasters.
The index scores are structured into four themes based on 15 factors: Theme 1 describes lack of economic resilience based on measures of poverty, unemployment, income, and lack of high school diploma; Theme 2 describes household-level adaptive capacity relying on measures of number of people aged 65 or older, aged 17 or younger, number of civilian population with a disability, and number of single-parent households; Theme 3 describes fraction of minority and non-English speaking population; Theme 4 describes the housing type and transportation (multi-unit structures, mobile homes, crowding, no vehicle, group quarters).


\begin{table}[h]
    \small
    \centering
    \caption{Cross correlations of SVIDM}
    \begin{tabular}{c|cccc}
      & Theme 1 & Theme 2 & Theme 3 & Theme 4  \\
       \hline
       Theme 1  & 1.000 & 0.538 & 0.233 & 0.404 \\
       Theme 2 & 0.538 & 1.000 & 0.019 & 0.249 \\
       Theme 3 & 0.233 & 0.019 & 1.000 & 0.588 \\
       Theme 4 & 0.404 & 0.249 & 0.588 & 1.000 
    \end{tabular}
    \label{t:summary_svidm}
\end{table}

\begin{figure}[]
    \centering
    \caption{{\bf Maps for exposures and outcomes}: Social Vulnerability Index for Disaster Management themes 1-4 in southern states are in (a)-(d). SVIDM ranges between $0$ and $1$, with the value indicating the proportion of ZCTAs with lower or equal social vulnerability. Two of the five chronic health outcomes in southern states are in (e) and (f). The maps show incidence rates---the proportion of beneficiaries with the condition in a ZCTA.}
    \subfigure[Theme 1]{\includegraphics[trim={.5cm, 1.5cm, 0cm, 1cm}, clip,width=0.49\linewidth]{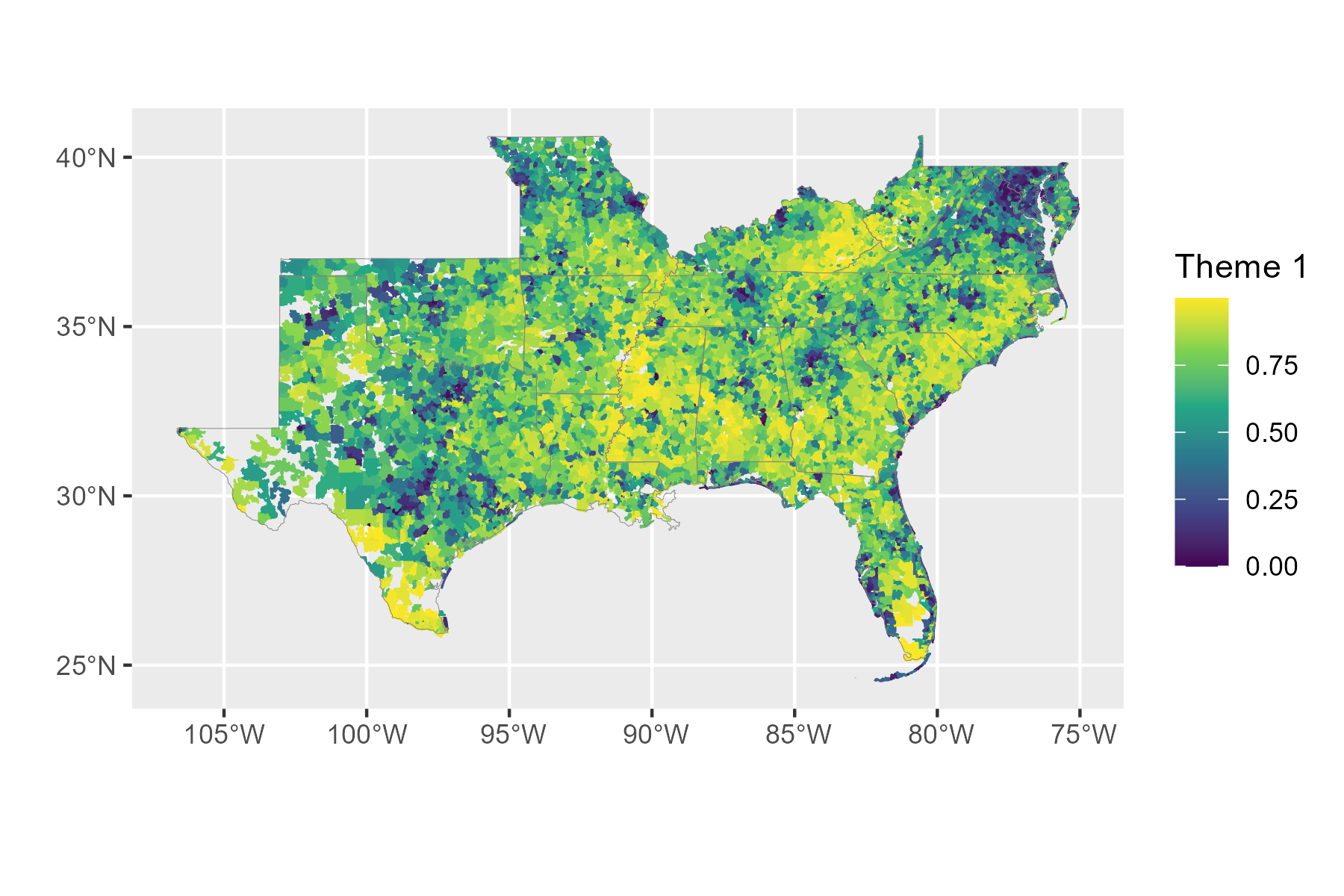}}
    \subfigure[Theme 2]{\includegraphics[trim={.5cm, 1.5cm, 0cm, 1cm}, clip,width=0.49\linewidth]{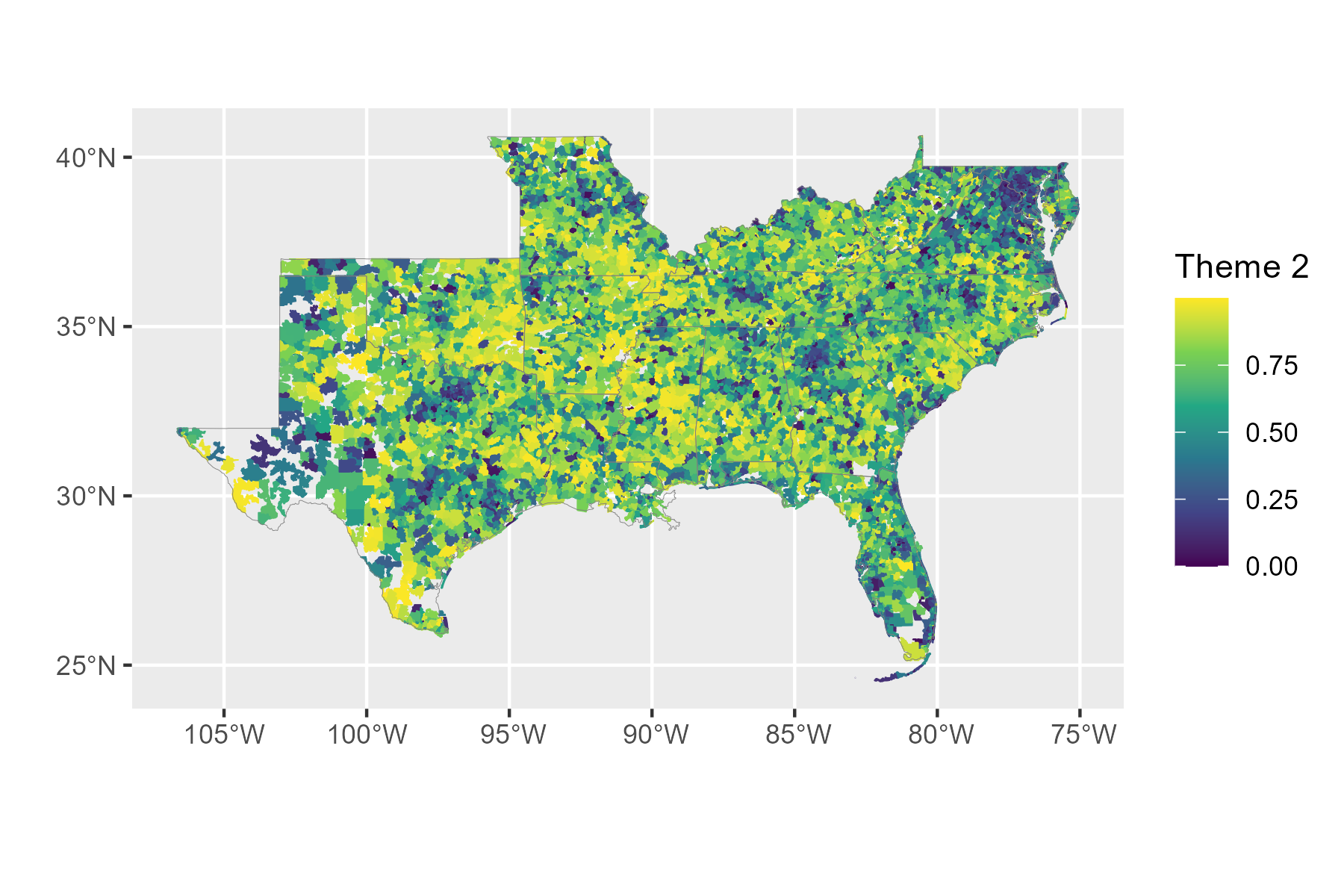}}
    \subfigure[Theme 3]{\includegraphics[trim={.5cm, 1.5cm, 0cm, 1cm}, clip,width=0.49\linewidth]{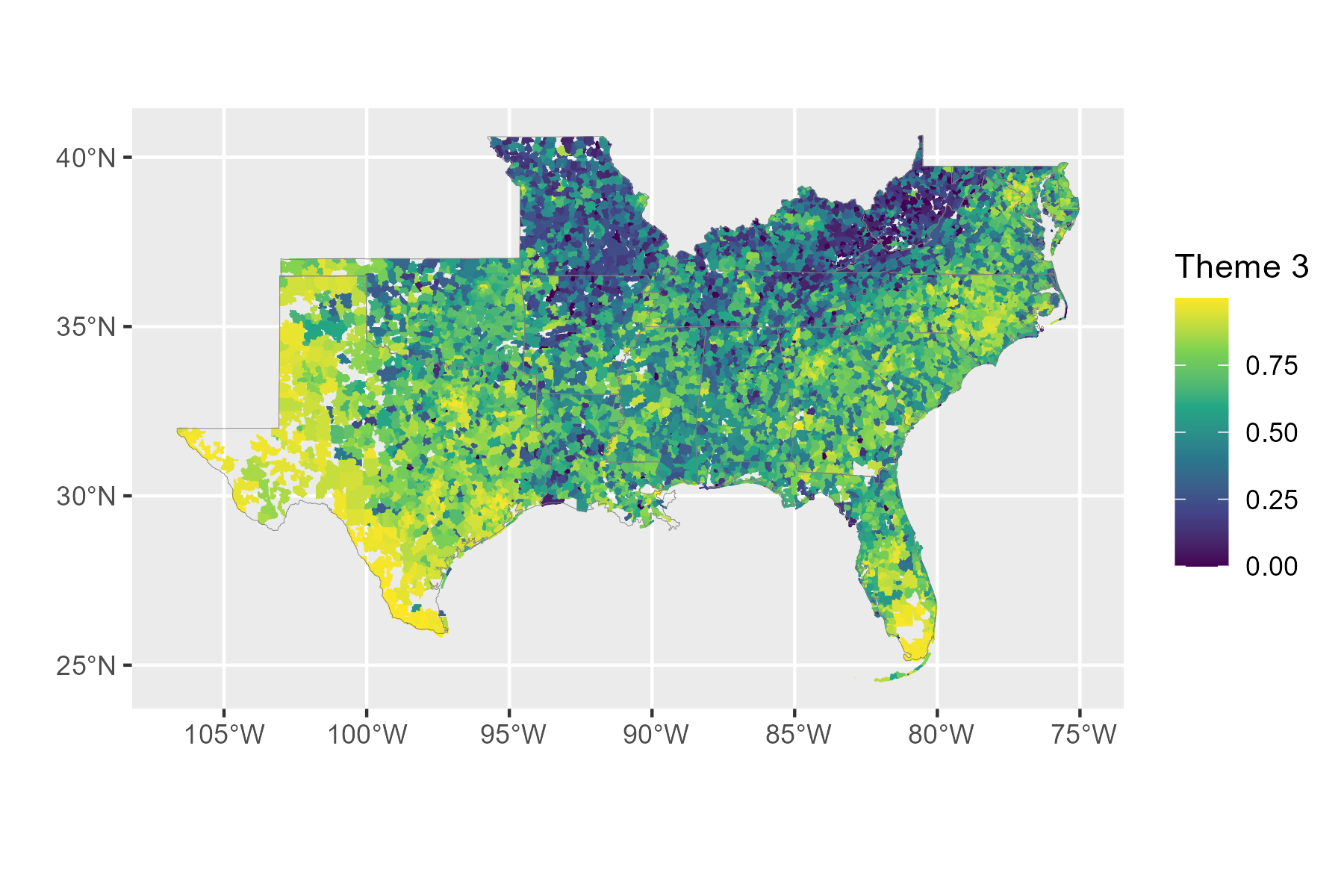}}
    \subfigure[Theme 4]{\includegraphics[trim={.5cm, 1.5cm, 0cm, 1cm}, clip,width=0.49\linewidth]{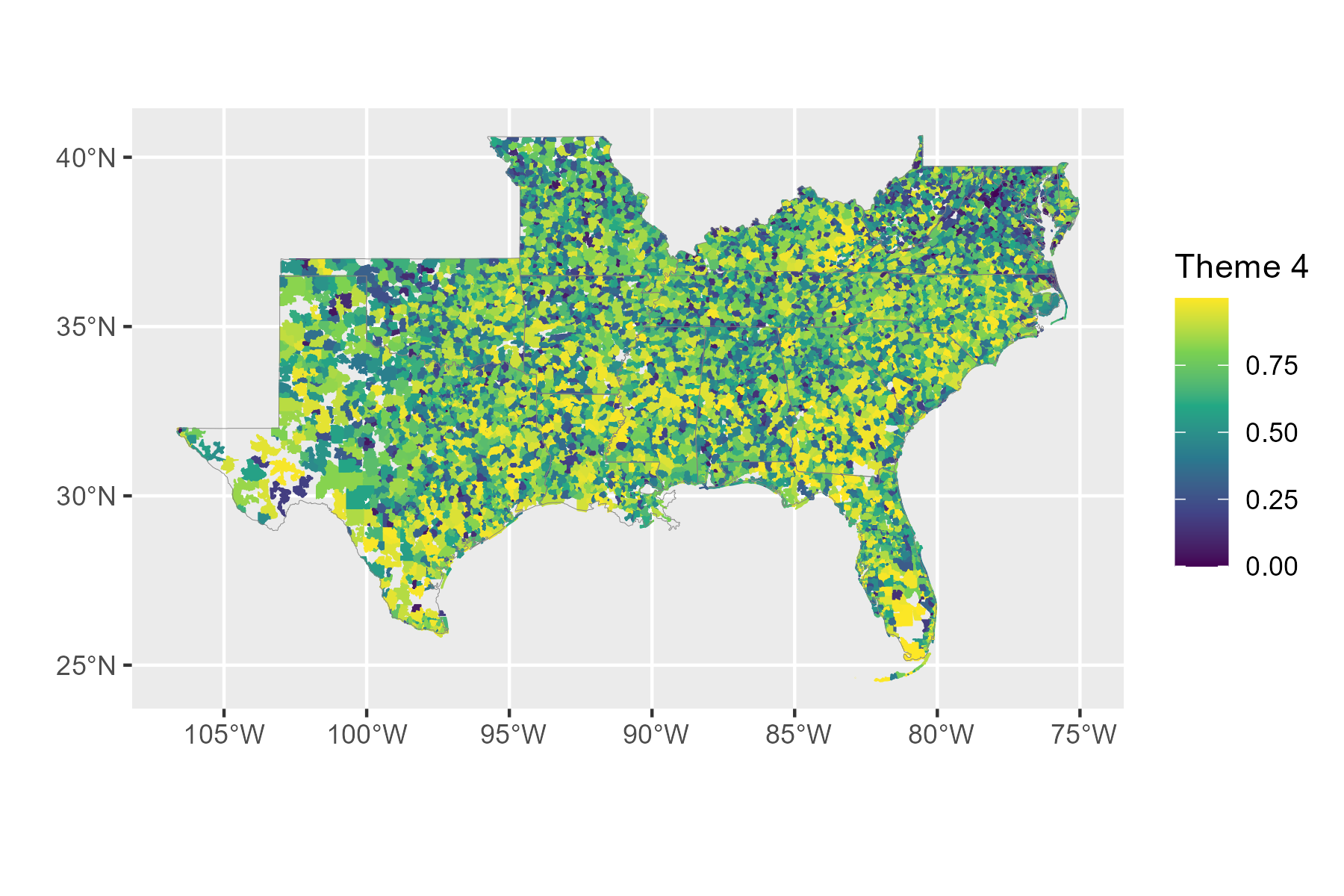}}
    \subfigure[Hypertension]{\includegraphics[width=0.49\linewidth]{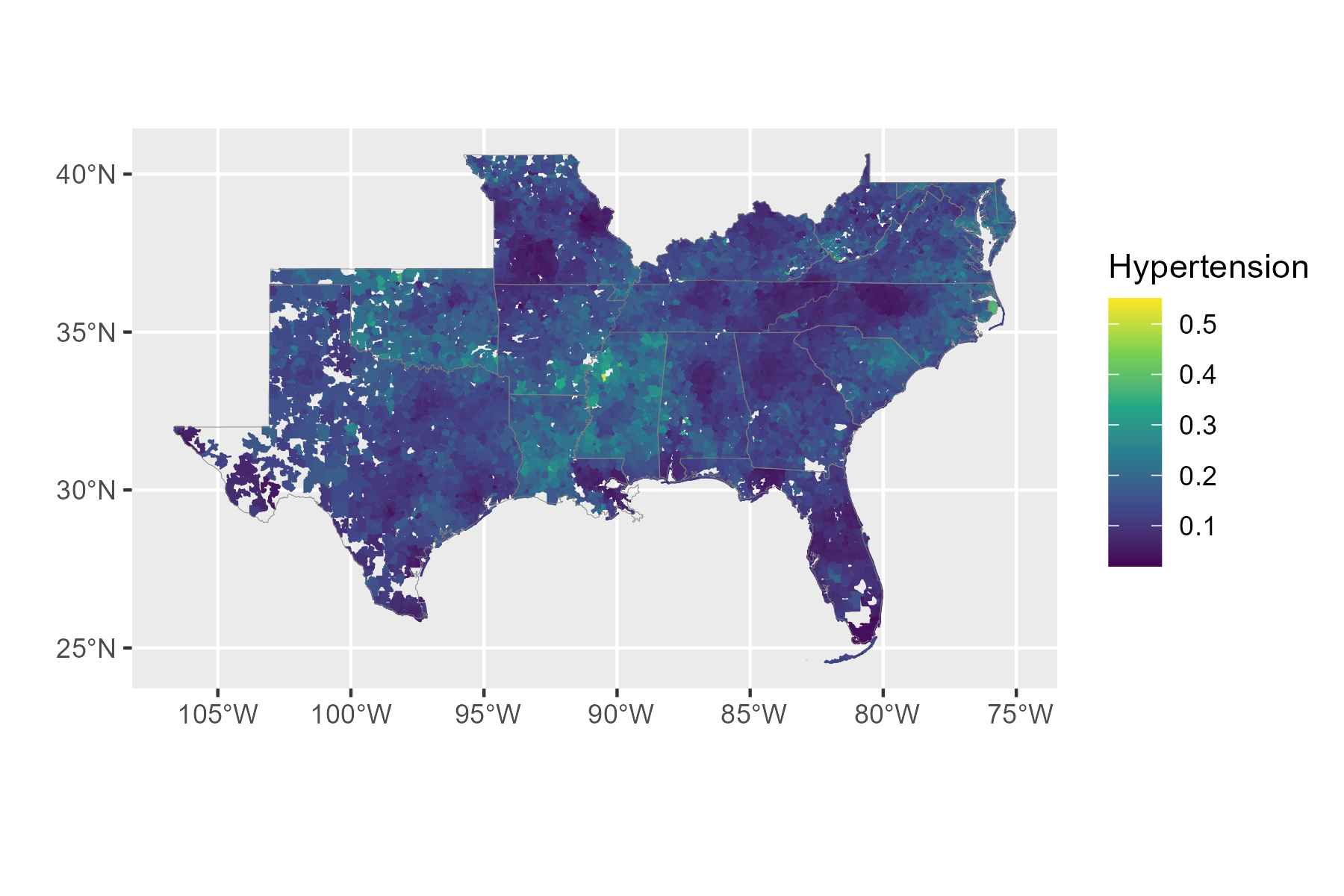}}
    \subfigure[Chronic kidney disease]{\includegraphics[width=0.49\linewidth]{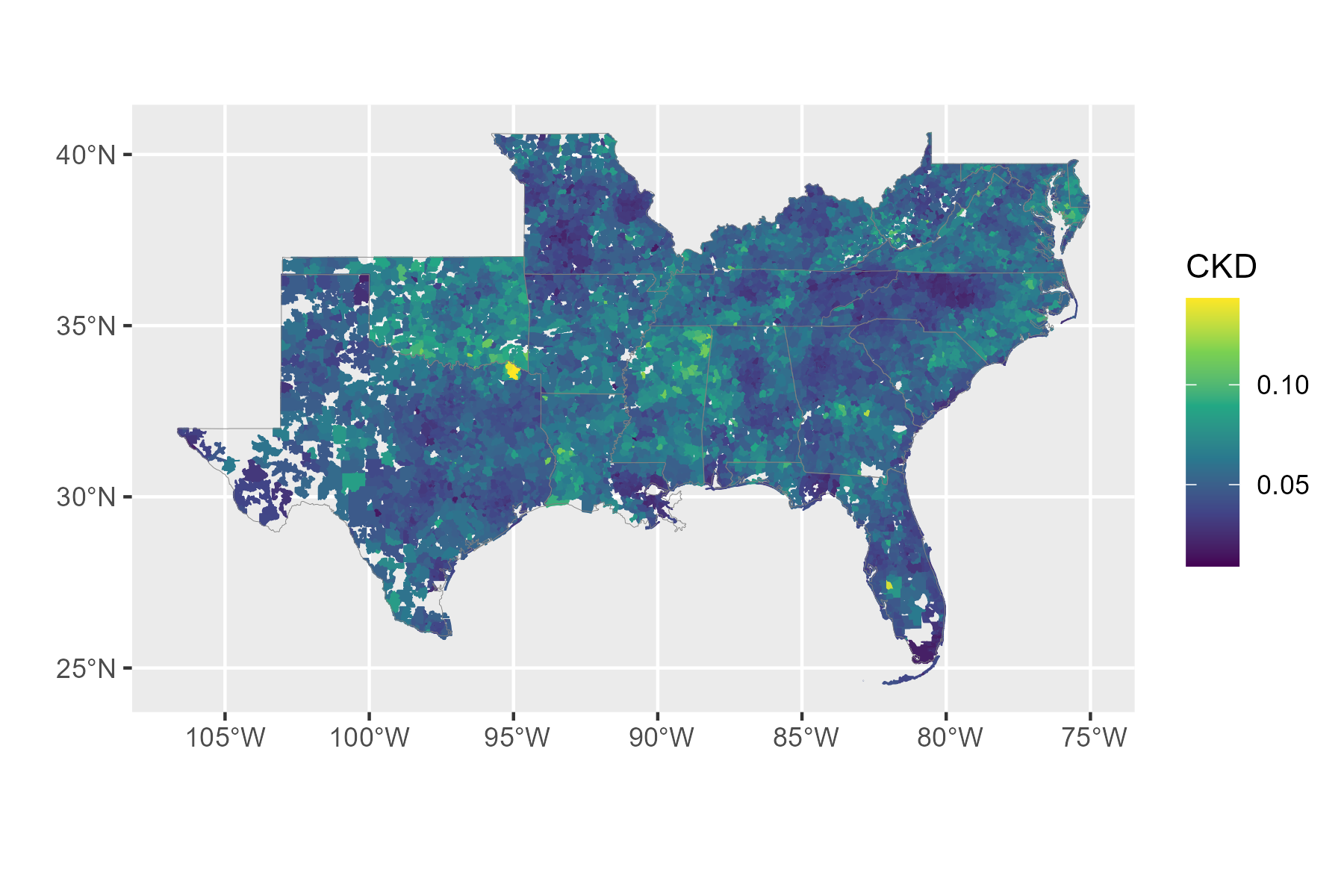}}
    
    \label{f:svi_themes}
\end{figure}

The adjusting covariates include the population size and the indicator of urban/rural ZCTA. Population size data was downloaded from the US Census Bureau's 5-year American Community Survey (ACS-5) \citep{census2023}. The logarithmic scale of the population is used. Urban/rural determination was obtained from 2010 data from the Integrated Public Use Microdata Series (IPUMS) National Historical Geographic Information System (NHGIS) \citep{nhgis}. A ZCTA was considered urban if more than $50\%$ of its population lived in urban areas. The split of rural versus urban areas was $57.2\%$ urban and $42.8\%$ rural. Maps of outcomes, exposures, and covariates, provided in Figures~\ref{f:svi_themes} and \ref{f:map_outcomes}, show strong spatial patterns, and  Tables \ref{t:summary_response} and \ref{t:summary_svidm} show the correlations between exposures and outcomes; these results support our multivariate spatial analysis.
We focus on the South region \citep{cdc_region}, which includes seventeen states (Alabama, North Carolina, Oklahoma, Arkansas, Delaware, Florida, Georgia, Kentucky, Louisiana, Maryland, Missouri, Mississippi, South Carolina, Tennessee, Texas, Virginia, West Virginia) and the District of Columbia. This region is experiencing rapid population growth and is susceptible to an array of natural disasters including hurricanes, storm surge, tornadoes, inland flooding, and extreme heat. After removing ZCTAs with less than $100$ beneficiaries the dataset included $10,149$ ZCTAs.

\section{Statistical methods}\label{s:method}

\subsection{Model in the spatial domain}\label{s:spatialdomain}

Assume the spatial domain is partitioned into $S$ regions, and $a_{st}=1$ indicates that regions $s$ and $t$ are adjacent and $a_{st}=0$ are otherwise.  The data for region $s$ consists of $R$ response variables, $\{ Y_{s1},...,Y_{sR}\}$, $E$ exposure variables, $\{ X_{s1},...,X_{sE}\}$ and $P$ known confounding variables, $\{Z_{s1},...,Z_{sP}\}$. 
In the spatial domain, the data model is: 
\begin{equation}\label{e:Y}
  Y_{sr} = \eta_{0r}+\sum_{p=1}^PZ_{sp} \eta_{pr}  +
  \sum_{e=1}^EX_{se}\beta_{er} + \theta_{sr} + \varepsilon_{sr},
\end{equation}
where $\eta_{pr}$ controls the observed confounder effects, $\beta_{er}$ are the exposure effects of interests, $\theta_{sr}$ are spatially-dependent residuals, and $\varepsilon_{sr}\indep\mbox{Normal}(0,\tau^2_r)$ is error. While the responses are assumed to be normally distributed, the exposures can be all types---continuous, binary, categorical, etc. In the traditional spatial regression model, $\theta_{sr}$ is modeled with a spatial process independent of $X_{se}$. If $\theta_{sr}$ is correlated with $X_{se}$, it is a spatial confounder. If the spatial random effect $\btheta = (\theta_{s1}, \cdots, \theta_{sR})^\top$ for $s = \{1, \cdots, S\}$ is observed, estimation of $\beta_{er}$ is straightforward. Under the potential outcomes framework and the Stable Unit Treatment Value Assumption (SUTVA), if the potential outcomes follow (\ref{e:Y}), $(\beta_{1r}, \cdots, \beta_{Er})^\top$ represents the average treatment effects of the exposures on outcome $r$ (Section \ref{s:causal}). However, when the multivariate, spatial $\btheta$ is unobserved and treated as a spatially-correlated error, the dependence between $\theta_{sr}$ and $X_{se}$ leads to bias in estimating $\beta_{er}$.  

Our approach for mitigating this bias makes use of the Conditional Autoregressive (CAR) model. We study the form of bias in special cases in Section \ref{s:alpha_w}. For an arbitrary random vector $\bV = (V_1,...,V_S)^\top$, the CAR model with Leroux parameterization \citep{leroux2000} is multivariate normal with mean zero and covariance matrix $\sigma_V^2\left\{(1-\lambda_V)\bI_S + \lambda_V\bQ\right\}^{-1}$ where $\sigma^2_V$ controls the variance, $\lambda_V\in(0,1)$ determines the strength of spatial dependence, and $\bQ$ has values $\sum_{t}a_{st}$ on the diagonal and $-a_{st}$ on the off-diagonals. We denote this model by $\bV\sim\mbox{CAR}(\sigma_V^2,\lambda_V)$.

The multivariate spatial random effects $\theta_{sr}$ are modeled using the Linear Model of Coregionalization (LMC) \citep{schmidt2003LMC,wackernagel2003,gelfand2005spatial,gelfand2010mspm,banerjee2014}. The LMC assumes the random effects for the $R$ response variables can be explained by $Q\le R$ latent spatial factors as
\begin{equation}\label{e:LMC}
\theta_{sr} = \sum_{q=1}^Q A_{rq} C_{sq},
\end{equation}
where $A_{rq}$ are the factor loadings and $\bC_q = (C_{1q},...,C_{Sq})^\top \indep\mbox{CAR}(\sigma_q^2,\lambda_C)$ for $q\in\{1,...,Q\}$. 
We take $Q$ to be the same as the number
of responses so that the model is full rank.
This multivariate spatial process includes both spatial dependence via the CAR model and cross dependence via the factor loadings. The cross-correlations have closed forms, i.e., at any location $s$, two different outcomes have correlation 
\[ 
Cor(\theta_{sr}, \theta_{sr'}) = \frac{\sum_{q=1}^Q \sigma^2_q A_{rq} A_{r'q}}{\sqrt{\left(\sum_{q=1}^Q\sigma^2_qA_{rq}A_{rq}\right)\left(\sum_{q=1}^Q\sigma^2_qA_{r'q}A_{r'q}\right)}}. 
\]
To ensure identifiability, we set $A_{qq}=1$ and $A_{rq}=0$ for $q>r$ \citep{gelfand2010mspm}. The remaining terms have prior distributions $A_{rq}\sim\mbox{Normal}(0,0.5^2)$, so that the diagonal terms dominate. 

While the multivariate CAR model in (\ref{e:LMC}) accounts for multivariate spatial residual dependence, it assumes the random effects $\theta_{sr}$ are independent of the exposure variables $X_{se}$. The spatial model without properly accounting for the dependence between $\theta$ and $X$ does not mitigate bias. As a result, the estimates of $\beta_{er}$ can only be interpreted as causal effects when $\theta_{sr}$ and $X_{se}$ are independent, namely, under the assumption of no unmeasured spatial confounding. In Section \ref{s:spectraldomain} we relax this global assumption.

\subsection{Model in the spectral domain}\label{s:spectraldomain}

We propose to estimate $\beta_{er}$ in the spectral domain. By transforming a spatial variable to the spectral domain, we can decompose the spatial information to different scales. The no unmeasured confounder assumption requires $\btheta$ and $\bX$ are independent at each scale, where $\bX = \big( \bX_{1}, \cdots, \bX_{E} \big)$, $\btheta = \big( \btheta_1, \cdots, \btheta_R \big)$, and $\bX_e$ and $\btheta_r$ are vectors of length $S$. We relax this assumption by assuming that global-scale terms are susceptible to confounding whereas local-scale terms are unconfounded, i.e., local unconfoundedness \citep{guan2023spatial}. Our multiscale estimates can thus capture the true effect at scales where there is assumed to be no confounding.

Our spectral approach uses the graph Fourier transform to decompose the precision matrix into independent signals at different spatial scales.  Let $\bQ = \bGamma\bW\bGamma^\top$ be the spectral decomposition of $\bQ$ for eigenvector matrix $\bGamma$ and diagonal matrix $\bW$ with diagonal elements equal to the eigenvalues $w_1\ge...\ge w_S$. For arbitrary vector $\bV\sim\mbox{CAR}(\sigma_V^2,\lambda_V)$, premultiplying by the eigenvectors projects the spatial process into the spectral domain. The spectral process is $\bGamma^\top \bV = \bV^* = (V_1^*,...,V_S^*)^\top$ with $$V_i^*\indep\mbox{Normal}\left\{0,\sigma_V^2/(1-\lambda_V + \lambda_V w_i)\right\}.$$
In the spectral domain, $V_i^*$ are independent across $i$, allowing us to model each scale separately,  account for multiscale confounding, and represent weights of the eigenvectors. Larger weights on the eigenvectors that correspond to small eigenvalues result in smoother spatial process (termed global scale), while larger weights on the eigenvectors corresponding to large eigenvalues lead to rougher spatial process (termed local scale) \citep{reichEffectsResidualSmoothing2006}.  

We specify our model for adjusting for the unmeasured confounders by projecting the model in (\ref{e:Y}) to the spectral domain. For response $r\in\{1,...,R\}$, the data in the spatial domain is $\bY_{r} = (Y_{1r},...,Y_{Sr})^\top$ and the projection to the spectral domain is $\bGamma^\top\bY_{r}=\bY_{r}^* = (Y_{1r}^*,...,Y_{Sr}^*)^\top$, where $Y^*_{ir}$ captures the variation of the outcome $r$ at eigenvalue $w_i$. Similarly, let $Z_{ip}^*$, $X_{ie}^*$ and $C_{iq}^*$ for $p\in\{1,...,P\}$, $q\in\{1,...,Q\}$, and $e\in\{1,...,E\}$ be the projected values of the observed confounding, exposure, and spatial variables. Then (\ref{e:Y}) in the spectral domain is
\begin{eqnarray}\label{e:Ystar}
  Y_{ir}^* &=& \eta_{0r}+\sum_{p=1}^PZ_{ip}^*\eta_{pr} + 
  \sum_{e=1}^EX_{ie}^*\beta_{er} + \theta_{ir}^* + \varepsilon_{ir}^*, \text{ where}\\
  \theta_{ir}^* &=& \sum_{q=1}^QA_{rq} C_{iq}^*\nonumber
\end{eqnarray}
for $i\in\{1,...,S\}$ and $\varepsilon_{ir}^*\indep\mbox{Normal}(0,\tau^2_r)$. 
Identifiability of the regression coefficients in $\beta$ is discussed in Section \ref{s:ident_beta} and that of the other parameters are covered in Section \ref{s:identifiability}.
In the spectral domain, the latent factors are distributed as 
$C_{iq}^*\indep\mbox{Normal}\left\{0,\sigma_q^2/(1-\lambda_C + \lambda_C w_i)\right\}.$ 

\begin{figure}[h]
    \centering
    \caption{{\bf Spatial scales in the South region of the US:} The eigenvalues range from $0$ to $23.12$. Panel (a) shows a local spatial scale that corresponds to eigenvalue $1.78$, and Panel (b) shows a global spatial scale that corresponds to eigenvalue $0.006$.}
    \subfigure[Local scale]{\includegraphics[width=.4\linewidth]{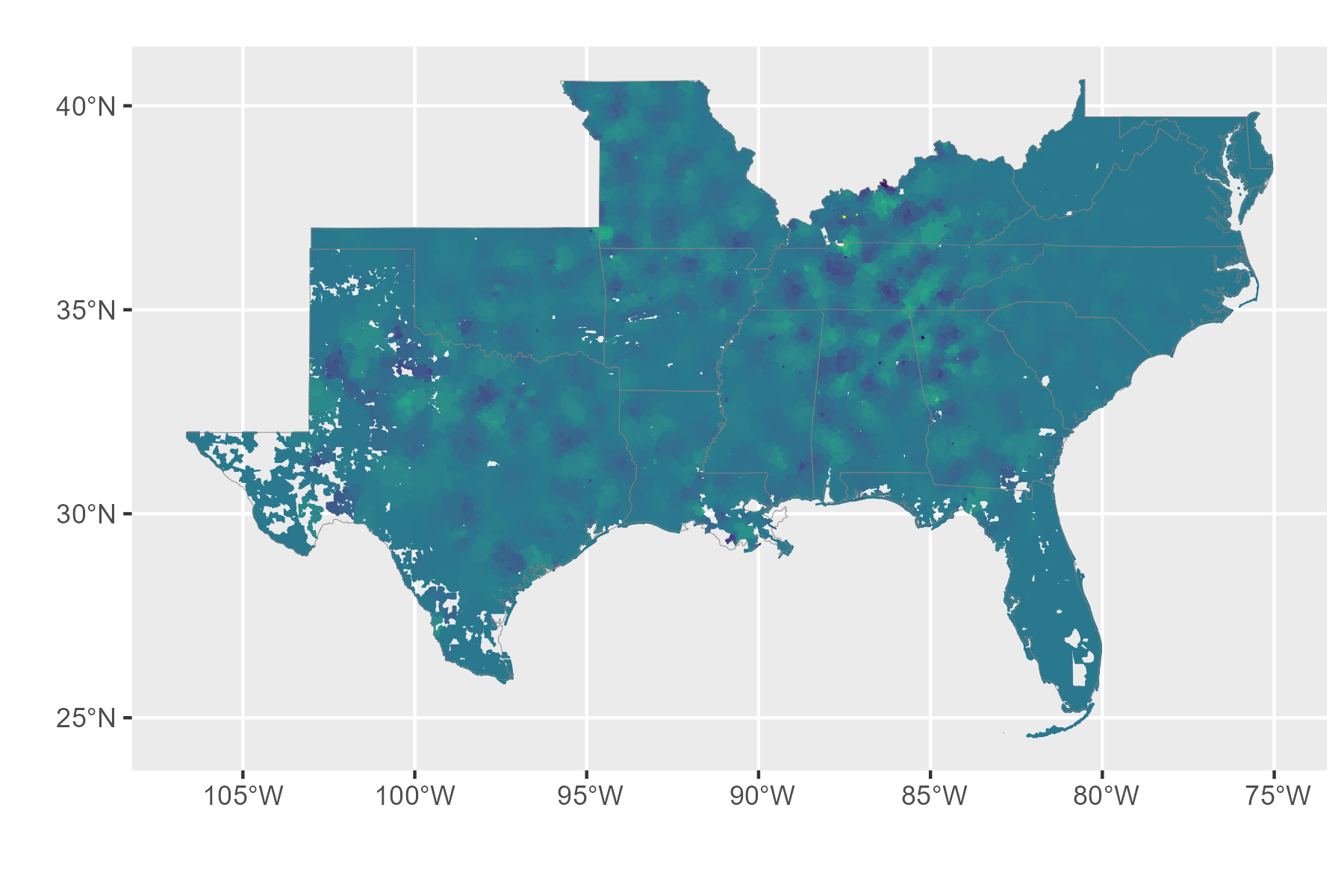}}
    \subfigure[Global scale]{\includegraphics[width=.4\linewidth]{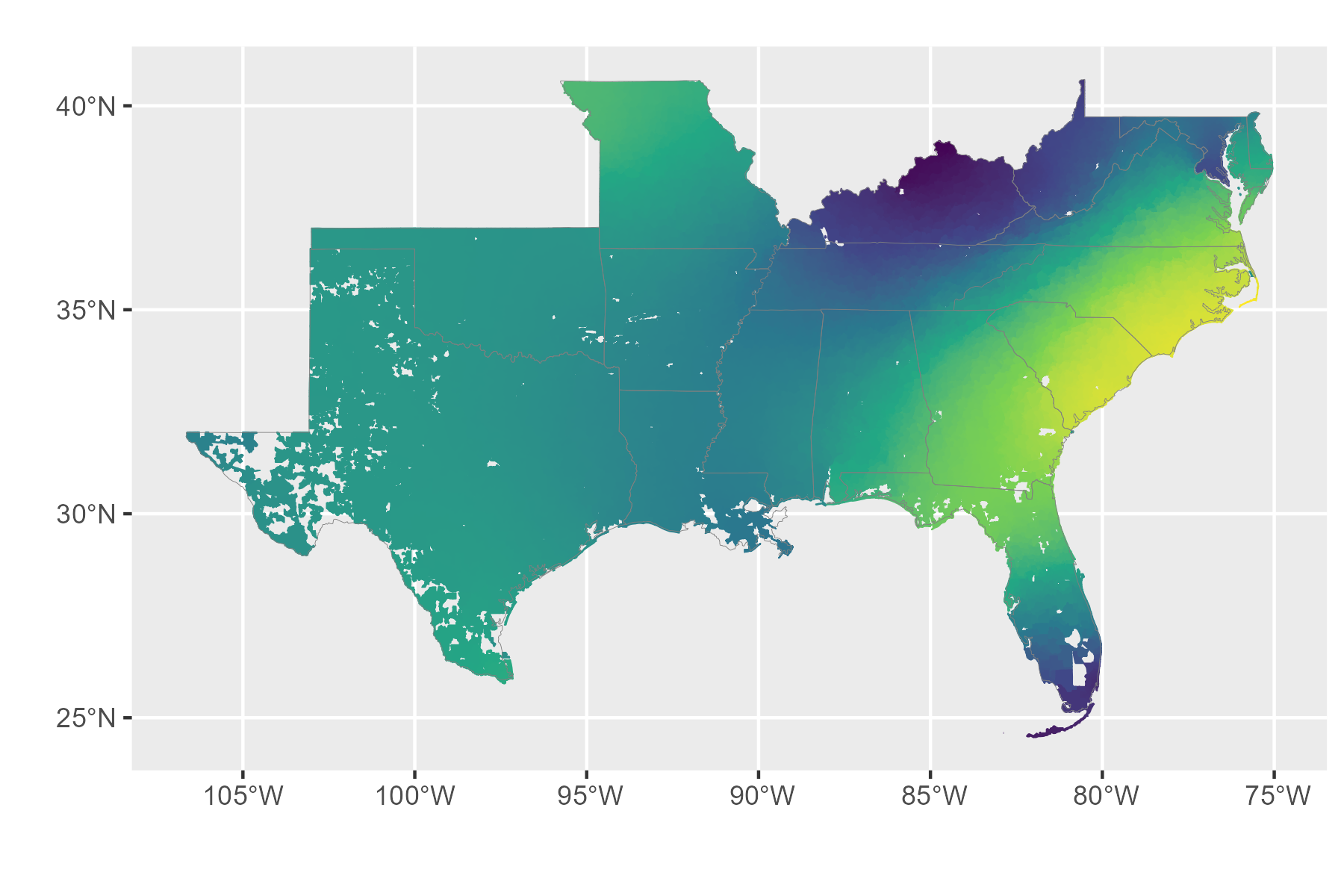}}
    \label{fig:eigenvectors}
\end{figure}

The distinction between confounding in the ``global'' versus ``local'' spatial scales prompts a key assumption of our proposed model. The local unconfoundedness assumption is that terms in (\ref{e:Ystar}) with large $w_i$ are unaffected by spatial confounding.
The matrix $\mathbf{Q}$ in the Leroux parameterization corresponds to a Laplacian matrix defined on an undirected graph defined based on neighborhood structure \cite[Chapter 2][]{Biyikogu2007laplacian}. The Laplacian operator on a rectangular domain with Dirichlet and Neumann boundary condition has eigenfunctions that are harmonic (sine and cosine) functions \citep{grebenkov2013eigen}. For example, Section \ref{s:eigen_scale} provides the closed-form and visualization of the eigenfunction on a 1D path graph. In more general graph structure, the eigen functions are not in closed-form; however, Figure \ref{fig:eigenvectors} shows that the eigenfunctions represent different spatial scales. Global confounding refers to large-spatial scale (low frequency) confounding and vice versa.

\subsection{Identifiability of $\boldsymbol{\beta}$
in spectral domain}
\label{s:ident_beta}

Unlike a standard spatial regression, we model dependence between $X_{ie}^*$ and $\theta_{ie}^*$. We assume a linear confounding relationship between $\theta^*_{ir}$ and $X^*_{ie}$:
$\theta^*_{ir} | X^*_{ie} = \alpha_{ier} X^*_{ie} + \xi_{ir},$
where $\alpha_{ier}$ is the confounding effect of exposure $e$ on response $r$ specific to spatial scale described by eigenvalue $w_i$. 
The spatial error term, $\xi_{ir}$, captures the remaining variation in $\theta^*_{ir}$ after conditioning on $X^*_{ie}$ and is independent of the non-spatial error term, $\varepsilon^*_{ir}$.
The magnitude of $\alpha_{ier}$ determines the level of correlation between $X^*_{ie}$ and $\theta^*_{ir}$. 
Note that a linear confounding relationship is assumed in the projection of (\ref{e:Y}) and (\ref{e:LMC}) to the spectral domain (\ref{e:Ystar}). 
In this scenario, the scale-invariant true effect $\beta_{er}$ is generally inseparable from the confounding bias $\alpha_{ier}$ when scales are not explicitly modeled, since increasing $X^*_{ie}$ by one increases the mean response by $\tilde{\beta}_{ier} = \beta_{er} + \alpha_{ier}$. 

We use a tensor of dimension $S\times E \times R$ to denote multiscale coefficients $\tilde{\bbeta}$, which contains $\tilde{\beta}_{ier}$ at scale $i$, exposure $e$, and response $r$. To accurately estimate the coefficient tensor $\tilde{\bbeta}$, our proposed method allows for a multiscale relationship between $\bX$ and $\btheta$. A spectral analysis of (\ref{e:Ystar}) estimates effects for all $S$ scales. We further assume the strength of correlation diminishes at local scales, i.e. $\alpha_{ier} \to 0$ for large $w_i$. 
This {\bf local unconfoundedness assumption}    is critical in identifying $\tilde{\boldsymbol{\beta}}$.
Thus $\hat{\beta}_{1er}$, the estimate for the multiscale coefficient $\tilde{\beta}_{1er}$, can capture the true $\beta_{er}$ at eigenvalue $w_1$, where the correlation between $X^*_{ie}$ and $\theta^*_{ir}$ goes to zero and 
$$\tilde{\beta}_{1er} = \beta_{er}.$$

\subsection{Bayesian tensor regression} \label{s:btr}

The coefficient $\tilde{\bbeta}$ includes effects across multiple exposures, multiple responses, and multiple scales and is thus structured as a tensor. While decomposition and the computational aspects have been the focus for tensor research \citep{zhang2017paper2,zhang2018paper3, zhang2019optimal}, using tensor for the purpose of statistical inference has seen a recent rise in interest \citep{mai2022DEEM, Llosa2022tensorontensor}. We employ Bayesian tensor regression methods for the regression coefficients. Since we envision $\tilde{\beta}_{ier}$ being smooth across scale $i$ but not exposure $e$ or response $r$, we represent the variation across $i$ using $L$ spline basis functions $B_{i1},...,B_{iL}$ and
$\tilde{\beta}_{ier} = \sum_{l=1}^LB_{il} \gamma_{ler},$
where $B_{il}$ are known B-spline basis functions. The intuition for this model is, first, assuming $\alpha_{ier}$ is a smooth function of $w_i$, the $L$ spline bases are sufficient to capture its relationship with $w_i$. Second, the multiple exposures might influence the outcomes through common underlying biological pathways or mechanisms, or the outcomes may respond in similar ways to different exposures. A low-rank model helps identify these latent structures.

\begin{figure}[h]
    \centering

    \caption{{\bf CP decomposition}: $\bT_1$, $\bT_2$, and $\bT_3$ are matrices, which are formed by columns that are tensor margins: $T_{1k}, T_{2k}, T_{3k}$ for $k \in \{1, \cdots, K\}$. Here $T_{1k} = (T_{11k}, \cdots, T_{1Lk})^\top, T_{2k} = (T_{21k}, \cdots, T_{2Ek})^\top, T_{3k} = (T_{1rk}, \cdots, T_{1Rk})^\top$. The outer product of three tensor margins, each of which a vector of lengths $L$, $E$, and $R$, respectively, is a $L \times E \times R$ tensor. The sum of $K$ such tensors approximates the tensor $\bgamma$.}
    \includegraphics[page=3,trim={5cm 5cm 3.6cm 6cm},clip,width=0.6\linewidth]{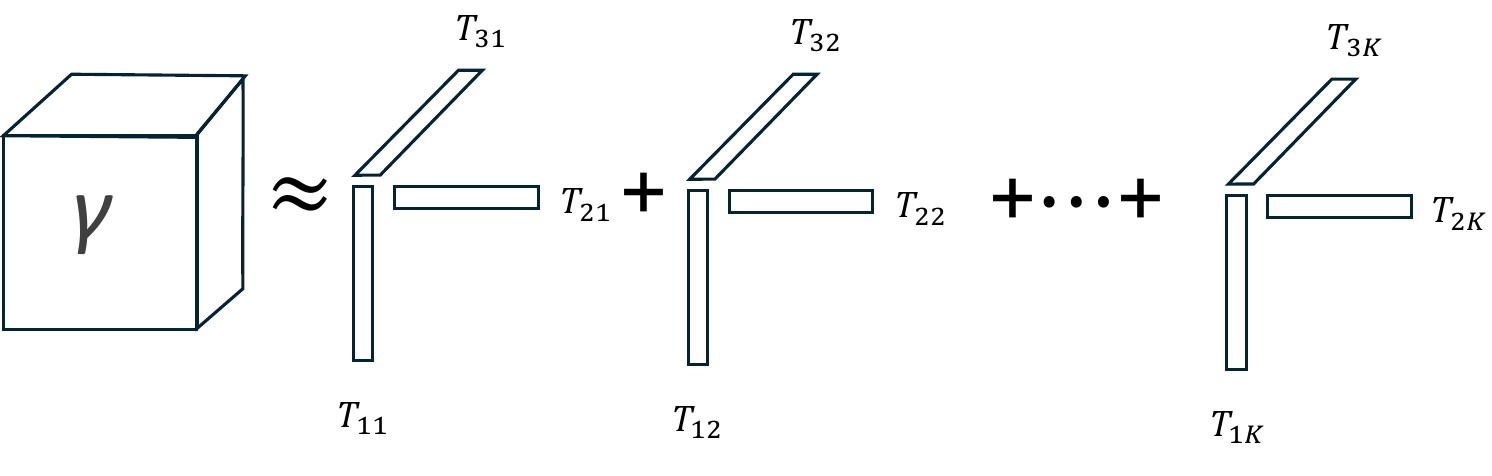}
    \label{fig:cp_decomp}
\end{figure}

The basis coefficients $\gamma_{ler}$ are modeled using the Bayesian tensor regression approach in \cite{guhaniyogi2017bayesian}.  The tensor $\bgamma$ is written as a rank $K$ tensor product of factors in each of the three dimensions, with
\begin{equation} \label{e:tensor_co}
     \gamma_{ler} = \sum_{k=1}^K T_{1lk}T_{2ek}T_{3rk}.
\end{equation} 
$\bT_1$, $\bT_2$, and $\bT_3$ are matrices, whose columns are tensor margins from basis functions, exposures, and outcomes, respectively. More specifically, $\bT_t = (T_{t1}, \cdots, T_{tK}), t = 1, 2, 3$ and tensor margins $T_{tk}$ are defined in Figure \ref{fig:cp_decomp}. Each tensor margin is of the dimension $\mathbb{R}^L$, $\mathbb{R}^E$, and $\mathbb{R}^R$. The latent factors $T_{1lk}$, $T_{2ek}$, and $T_{3rk}$ are given Bayesian shrinkage priors to encourage sparsity, assuming a structure of lower than full rank exists (Section \ref{s:shrinkageprior}). Using canonical polyadic (CP) tensor decomposition \citep{kolda2009tensor}, the sum of $K$ outer products of the tensor margins approximate the tensor coefficient $\bgamma$, as represented in Figure \ref{fig:cp_decomp}. Section \ref{s:identifiability} verifies that the parameters in this model including $\bgamma$ are identifiable. 
The benefits of such decomposition are threefold. It reduces dimensionality and simplifies the model, representing essential patterns with far fewer parameters and avoiding overfitting. Further, this model captures non-linear relationships and interactions between frequency, exposures, and outcomes. Lastly, the low-rank model makes the results more interpretable and generalizable. 

\cite{kolda2009tensor} provide a guideline for tensor ranks: $K = \min \{ LE, ER, LR\}$. In our simulations, we choose a value smaller than the above value that achieves convergence in MCMC chains. To set a range for $L$, we consider the relationship between the total number of parameters in the tensor, $K(L+E+R)$, and the number of response variables $SR$, 
    $0 < L \le \frac{SR}{K} - E - R.$
We fit the model with several $L$ and check trace plots and estimates to choose the optimal $L$, and sensitivity tests are described in Section~\ref{s:A4}.

\subsection{Shrinkage priors for tensor coefficients}\label{s:shrinkageprior}

The tensor regression coefficients in (\ref{e:tensor_co}) are given horseshoe priors to shrink the regression estimates towards zero for unnecessary ranks or null exposures and outcomes. The horseshoe prior is   
$ T_{1lk} \sim N(0, \lambda_{1k}); T_{2ek} \sim N(0, \lambda_{2e}); T_{3rk} \sim N(0, \tau^2 \lambda_{3r} \tau^2_r), $
where $\lambda_{1k}, \lambda_{2e},$ and $\lambda_{3r}$ are the variances and control local shrinkage. A global shrinkage hyperparameter, $\tau$, is embedded in $T_{3rk}$ to shrink the coefficients in all outcomes. The variance component $\tau_r^2$ is the variance of the random errors from $\varepsilon_{ir}^*$ in (\ref{e:Ystar}) and does not affect shrinkage. Following \cite{makalic2016HC}, we parameterize the shrinkage priors hierarchically. Each shrinkage parameter follows an inverse gamma prior, e.g. $\lambda_{jk} | \nu_j \sim \text{InvGamma} \big( \frac{1}{2}, \frac{1}{\nu_j} \big)$ for $j \in \{1,2,3\}$. Each hyper-parameter $\nu_j \sim \text{InvGamma} \big( \frac{1}{2}, 1 \big)$, resulting in the standard half-Cauchy distribution for the square roots of the shrinkage parameters.

\subsection{Simulation study}\label{s:sim}

We apply the methods described in Section \ref{s:method} to simulated data to examine how our MSM method performs with different magnitudes of linear and nonlinear spatial confounding. 
Compared to the naive model that does not account for spatial confounding, MSM has over $40\%$ of reduction in mean MSE (from $0.065$ to $0.039$). The mean coverage rate for MSM is $98.8\%$, close to the nominal rate of $95\%$, as opposed to $85.1\%$ for the naive method. More details are shown in Figure~\ref{fig:allres} and Tables~\ref{t:all_results1} to \ref{t:all_results5}.
The results demonstrate that: 1) our proposed MSM model has low mean squared errors, 2) MSM tends to have coverage rates above the nominal rate, suggesting its conservative nature, 3) when no confounding is present, MSM's performance is similar to the model that does not allow for spatial confounding, and 4) MSM is robust to moderate nonlinear confounding.
Implementation details and results are provided in Section \ref{A:sim}.

\section{Data analysis} \label{s:data_analysis}

In this section, we analyze the SVIDM and chronic conditions data described in Section \ref{s:data} using the proposed MSM.  We first tune and compare models and conduct checks of model fit.  We then use the fitted models in the subsections below to address the study's objectives, outlined in Section~\ref{s:intro} and reiterated below: quantify whether the relationship between SVIDM and chronic conditions varies by spatial scale, and thus establish the presence of spatial confounding;  estimate the causal effect of each component of SVIDM on the incidence of chronic conditions while accounting for spatial confounding; and identify a low-rank structure that distills the many-to-many causal inference problem down to a minimal set of epidemiologically interpretable factors.

\subsection{Selection and validation of the best fit model}
\begin{figure}[h]
    \centering
    \caption{{\bf Residual plot in the spatial domain for diabetes}: No apparent patterns and outliers are present.}
    \includegraphics[width=0.5\linewidth]{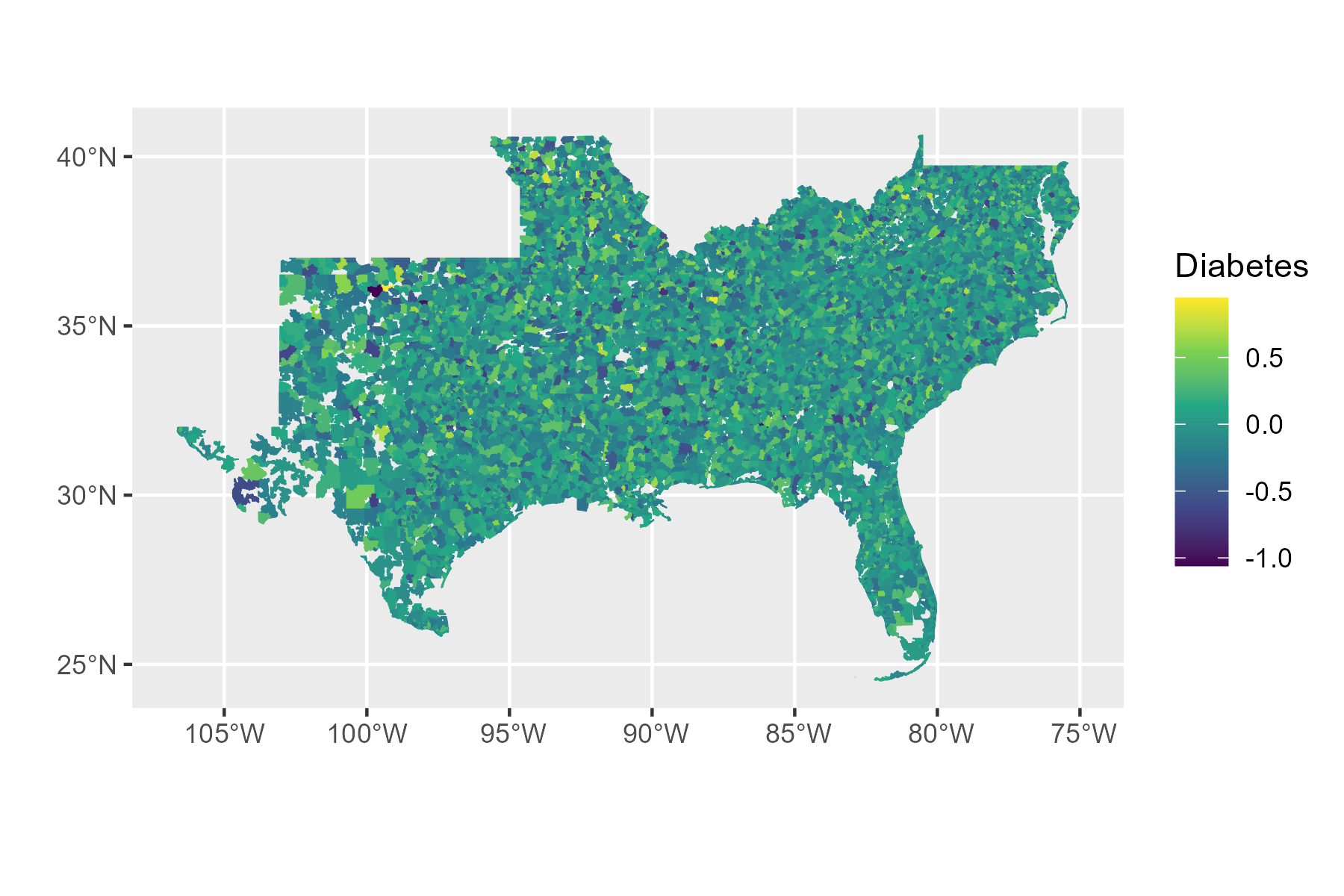}
    \label{fig:res_diabetes}
\end{figure}

To select the best model, we compare model performance of four combinations of the MSM method: $K=5,10$ (tensor rank) and $L=5, 10$ (number of basis functions). 
We split the dataset that contains $10,149$ ZCTAs into a training set ($80\%$) and a test set ($20\%$) and use log scores to select the best model ($K=L=10$). The out-of-sample predictive $R$-squared and coverage verified the performance of the selected model (see Table~\ref{t:pred_cv}). 
All models were run with $20,000$ iterations, $2,000$ burn-ins, and thinning rate $10$. 
Sensitivity study with respect to different $L$ and $K$ is provided in Section~\ref{s:A4}.
The final model was run on the full dataset with $100,000$ iterations, $10,000$ burn-ins, and a thinning rate of $10$. Trace plots confirmed convergence for all coefficients, as shown in Section~\ref{A:dianostic}. 
Residual plots in the spatial domain (Figure~\ref{fig:res_diabetes} for diabetes and Figure~\ref{fig:residuals} for all outcomes) showed no spatial patterns or outliers. In Section~\ref{A:residuals}, the density plot and variogram (Figure~\ref{fig:residuals2}) showed that residuals were normally distributed with mean around zero and have very low spatial correlations. Our analysis assumes a linear relationship with the unmeasured spatial confounders.  This assumption is unverifiable, but our residual analysis in Section~\ref{A:residuals} shows no signs of non-linearity with the observed confounders, and the simulation study in Section~\ref{A:sim} suggests robustness to moderate deviations from linearity.

To highlight the difference between our proposed multiscale method and methods that either lack spatial information or do not allow effects to vary by scale, we compare the results with those from the univariate  spectral models (USM), a naive spatial model (Naive; see Section \ref{s:competitors}) and a non-spatial ordinary least squares (OLS). The USM model used the same settings ($K$, $L$, iterations, burn-ins, and thinning rate) as MSM.

\begin{table}[h]
    \centering
    \small
    \caption{{\bf Predictive validity assessment with $80\%$ training data and $20\%$ test data}: The predictive $R^2$ shows the squared correlation between predictive and true responses, and coverage shows the percentage of observed responses in the $95\%$ credible intervals.}
    \begin{tabular}{c|cc}
       Model & Predictive $R^2$ & Coverage  \\
       \hline
       MSM ($K=5,L=5)$  & 0.216 & 0.950 \\
       MSM ($K=5,L=10)$ & 0.248 & 0.950 \\
       MSM ($K=10,L=5)$ & 0.154 & 0.951 \\
       MSM ($K=10,L=10)$ & 0.278 & 0.951 \\
       Naive ($L=1$) & 0.065 & 0.950 \\
    \end{tabular}
    
    \label{t:pred_cv}
\end{table}

\subsection{The summary effect estimates across spatial scales}

\begin{figure}[h]
    \centering
    \caption{{\bf Posterior estimates by spatial scale}: The posterior means for all frequencies and the $95\%$ credible intervals are shown below. Larger eigenvalues correspond to local scales, and smaller eigenvalues global scales. The four themes for the disaster resilience index are economic resilience (Theme 1), household composition (Theme 2), fraction of minority (Theme 3), and housing type and transportation (Theme 4). For each Theme, higher scores indicate less resilience.} 
    \includegraphics[width=.9\linewidth]{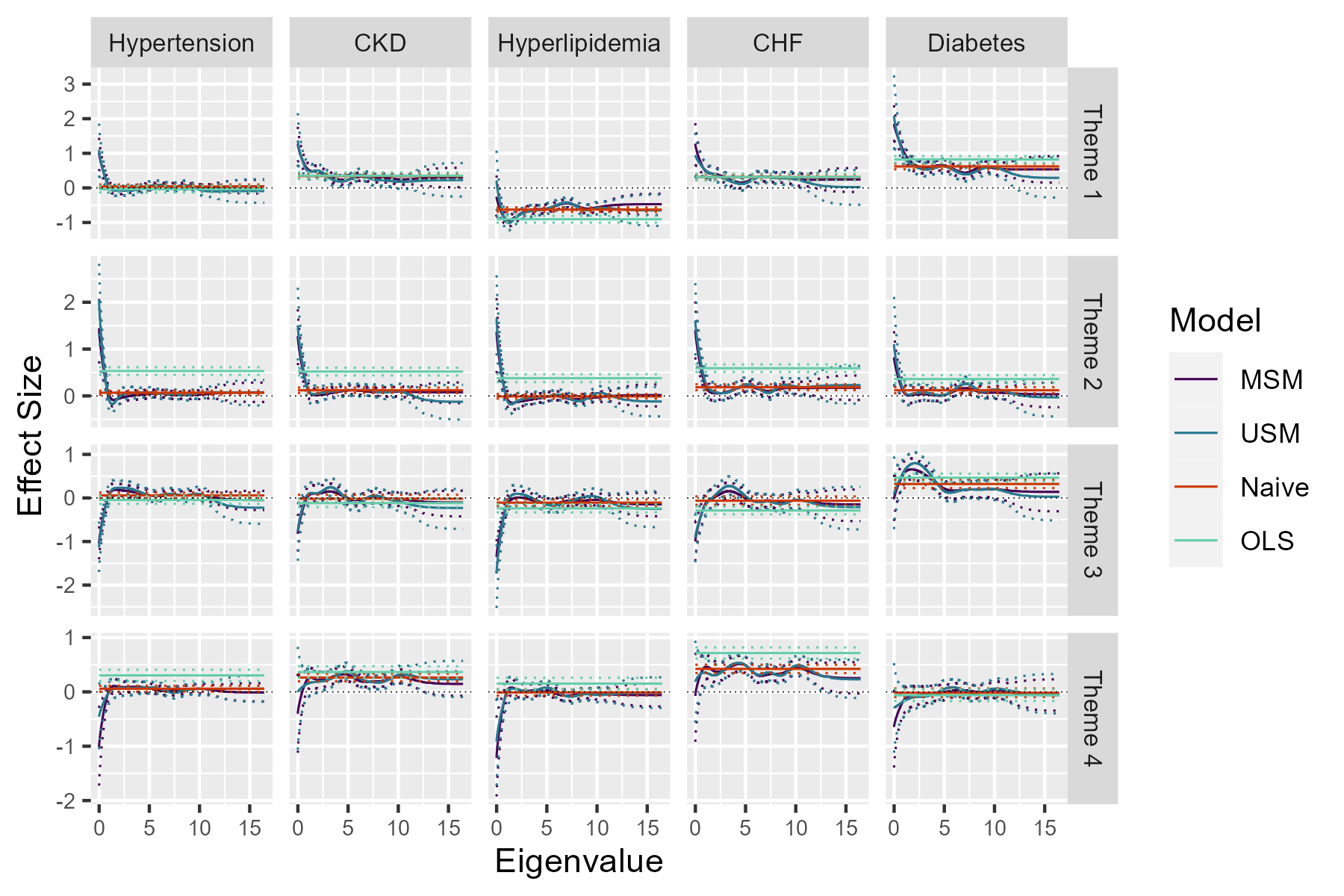}    \label{f:coef_by_freq_svi}
\end{figure}

The relationship between SVIDM and chronic conditions varied by spatial scale, strongly indicating that the results are sensitive to spatial confounding.  The size and uncertainty associated with the coefficient estimates at all eigenvalues (spectral frequencies) are given in Figure \ref{f:coef_by_freq_svi}. Focusing on the MSM model across outcomes and SVIDM themes, we note that the global spatial scale estimates---which correspond to smaller eigenvalues---differ markedly from local scale estimates with larger eigenvalues. In most treatment-outcome pairs, the global scale estimate of the effect coefficient (eigenvalues$<2$) was biased away from the null. This implies that the models that do not account for unmeasured confounding at larger spatial scales are biased away from the null. The effect of socioeconomic measures on kidney disease, hyperlipidemia, heart failure, and diabetes did not reduce to null (at larger eigenvalues) but were markedly reduced in comparison to the global effects at lower eigenvalues. In contrast, the effect of Theme 4 scores describing housing type and transportation for kidney disease was reduced to null at a global scale (smaller eigenvalues).

The discrepancy between local and global estimates can be quantified by the posterior probabilities of the mean effect at the lowest frequency being greater than those at the highest frequency (Table \ref{t:coef_vary}). Treatment-outcome pairs corresponding to Theme 1, 2 and 4 have very different estimates at the most local and most global spatial scales, strongly suggesting the presence of spatial confounding. In contrast, Theme 3 coefficient estimates for racial and ethnic minority measures from the most local and most global scales have the largest percentages of overlap and stand out as not having overall trends for the outcomes.

\label{S:spatial_units}
The multiscale coefficients also highlight how spatial scale of analysis could affect the result. As explained in Section \ref{s:eigen_scale}, derived from the eigen decomposition of the adjacency matrix's precision matrix, eigenvalues correspond to spatial scales: the larger the eigenvalues, the more local the scales. Though eigenvalues cannot be uniquely translated into specific physical distances, a rough correspondence can be provided for illustrative purposes. Among the pairwise distances of ZCTAs, the $10^{th}$ percentile is around 283 kilometers, and the $90^{th}$ percentile is around 1638 kilometers. For context, the mean width of a state is around 400 kilometers. These numbers and the curves in Figure \ref{f:coef_by_freq_svi} reveal that very different conclusions can be drawn, depending on whether data come from the census tract, ZCTA, county, state, or region level. If the information only comes from the regional level, for example, no local information is available, and the results can be biased away from null.


\begin{table}[h]
    \caption{{\bf Probabilities of coefficient estimates at the lowest frequency greater than those at the highest frequency}: Values close to one (in red) or zero (in blue) mean that coefficient estimates vary by spatial scales, which suggests likely spatial confounding. As shown below, strong spatial confounding is suspected for most exposure/outcome pairs.}
    \centering
    \small
    \begin{tabular}{c|ccccc}
         & Hypertension & CKD & Hyperlipidemia & CHF & Diabetes  \\
         \hline
         Theme 1 & \color{red} 1.00 & \color{red} 1.00 & 0.68 & \color{red} 1.00 & \color{red} 1.00 \\
         Theme 2 & \color{red} 1.00 & \color{red} 1.00 & \color{red} 1.00 & \color{red} 1.00 & \color{red} 0.99 \\
         Theme 3 & \color{blue} 0.00 & \color{blue} 0.00 & \color{blue} 0.00 & \color{blue} 0.00 & 0.39 \\
         Theme 4 & 0.08 & 0.12 & \color{blue} 0.01 & 0.22 & 0.13 \\
    \end{tabular}
    \label{t:coef_vary}
\end{table}

 \subsection{Causal effect estimates}

\begin{figure}[h]
    \centering
    \caption{{\bf Forest plot for estimated coefficients}: MSM is our multivariate spectral model, and USM is the univariate version. The naive model is spatial but does not allow coefficients to vary by scales. The OLS model is non-spatial and treats each location independently. MSM, more efficient than USM, has larger uncertainty than Naive and OLS but provides unbiased estimates, assuming local unconfoundedness.}
    \includegraphics[width=.75\linewidth]{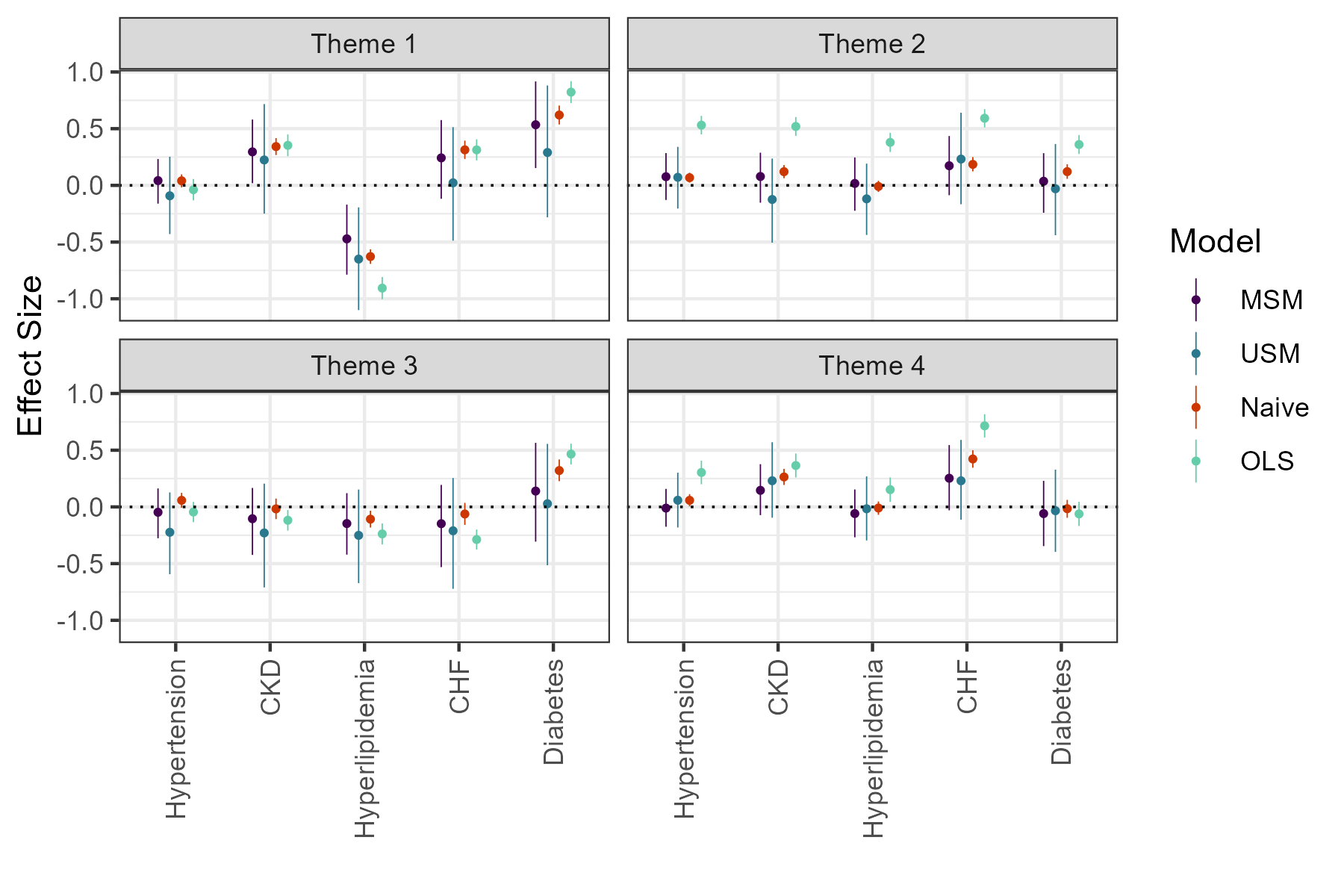}    
    \label{fig:forest}
\end{figure}

Despite the evidence of confounding at the global spatial scale, a number of outcome-treatment pairs have a significant causal effect at the local scale, 
indicated by the corresponding posterior probabilities of coefficients being different from zero are close to $0$ or $1$ (the pairs in red or blue in Table~\ref{t:prob_pos}). 
Economic resilience (Theme 1) demonstrates a positive effect on diabetes, chronic kidney disease, and congestive heart failure but a negative effect on hyperlipidemia. Furthermore, vulnerable household characteristics (Theme 2), including age extremes, disability, and single-parent structures, has a 90\% probability of having a positive effect on diabetes rates.  Housing type and transportation (Theme 4) shows positive effects on chronic kidney disease and congestive heart failure. The effects are not universal, however, as other pairs have less extreme posterior probabilities.
To additionally demonstrate the utility of the multiscale model, we contrast MSM coefficients with those from the Univariate (USM), Naive, and OLS models. The  multivariate (MSM) and univariate (USM) coefficient estimates were similar in magnitude across all outcomes, but the MSM model had advantageously smaller credible intervals, especially for local spatial scales/large eigenvalues (Figure~\ref{fig:forest}).  

\begin{table}[h]
    \caption{{\bf Posterior probability of effect size being different from zero}: Positive coefficients suggest that higher scores are associated with higher incidence of chronic health conditions (worse health outcomes), and negative coefficients suggest otherwise. Posterior probability over 0.90 (in red) and under 0.10 (in blue) suggest strong evidence for causal effect to be different from zero; positive or negative effect respectively. }
    
\centering
    \small
    \begin{tabular}{c|ccccc}
         & Hypertension & CKD & Hyperlipidemia & CHF & Diabetes  \\
         \hline
         Theme 1 & 0.63 & \color{red} 0.97 & \color{blue} 0.00 & \color{red}0.90 & \color{red} 1.00 \\
         Theme 2 & 0.73 & 0.76 & 0.54 & \color{red}0.90 & 0.61 \\
         Theme 3 & 0.34 & 0.26 & 0.11 & 0.23 & 0.77 \\
         Theme 4 & 0.56 & \color{red}0.92 & 0.35 & \color{red} \color{red}0.96 & 0.36 \\
    \end{tabular}
    \label{t:prob_pos}
\end{table}

Figure \ref{f:coef_by_freq_svi} shows how the global spatial scales can nudge the estimates towards biased for Naive and OLS. Although the credible intervals of MSM and Naive generally overlap, the naive model takes in all information and its credible intervals might be overly narrow, resulting in false positive results. The nudge toward bias is particularly evident for measures in Theme 2 (household composition and disability) where OLS estimates best compare to the global MSM estimates. 
As shown in Figure \ref{fig:allres}, the naive model has the lowest mean standard deviation, but the coverage rate can be lower than nominal levels, especially when spatial confounding is strong. Similarly, OLS takes in information from all locations, but unlike the naive model, it does not consider spatial locations. Its results also have smaller uncertainty and can be influenced by spatial confounding to a larger degree, which is particularly evident in plots for household composition and disability. 

A forest plot in Figure \ref{fig:forest} shows the coefficient estimates with $95\%$ credible intervals for MSM, USM, Naive, and OLS. This is essentially the same as Figure \ref{f:coef_by_freq_svi}, except the curves are sliced for the largest eigenvalues (local effects). 
For MSM, the intervals that do not cross the zero line are the same as those colored blue or red in Table \ref{t:prob_pos}.
The estimates from Naive and OLS show how adjusting for spatial relationships moves the estimates closer to null, but the naive model, nonetheless, can still be over-confident, as shown in Figure \ref{fig:allres}. 
As discussed above, MSM and USM have the largest uncertainty because the spectral model breaks down the estimates by spatial scales, and the estimates come from the most local spatial scale. 
The separation of signals filters out confounded information and calibrates uncertainty, as reflected in the wider intervals. 
Further, the comparison between MSM and USM highlights the efficiency gained by jointly modeling all exposures and outcomes. All credible intervals from MSM are narrower than those from USM, and significant associations—such as for theme 1 and CKD—can be overlooked due to the too-wide credible interval of USM.

The difference among these four models points to important implications for real world data. For epidemiological studies, data analysis—--even when spatial information is available—--univariate linear models are widely used. Figure \ref{fig:forest} shows how much results from univariate OLS can deviate from those from spatial and spectral models. Further, the naive spatial method can be overly confident, when unmeasured confounding is present. The results thus reflect the danger of basing policies on potentially false positive results, if OLS and naive spatial models are the only ones used. Spectral models can account for unmeasured confounding present in global spatial scales; moreover, if no confounding exists, spectral methods will give results that agree with the naive model. In other words, the adjustment provided by MSM will not result in false conclusions if no confounding exists. The multivariate model also provides efficiency gain, especially for correlated outcomes, commonly seen in epidemiological studies. 

\subsection{Sensitivity analysis for bias}
\label{S:sensitivity}

Our analysis relies on the unverifiable assumption of local unconfoundedness. Using methods motivated by \cite{10.1111/rssb.12348}, we conduct a sensitivity analysis to gauge the strength of a missing local confounder that would alter our conclusions.  We assume there are $N_U = \max\{R,E\}$ latent confounders, and let $\rho$ determine the strength of confounding.  Specifically, we assume confounding in the spectral domain has the form 
\begin{align*}
    \mathbf{X}^* &= \left(\sqrt{1-\rho^2} \mathbf{U}_X + \rho\mathbf{U}_C\right)\bB_X^T \\
    \mathbf{\Theta}^* &= \left(\sqrt{1-\rho^2} \mathbf{U}_\Theta + \rho\mathbf{U}_C\right)\bB_\Theta^T,
\end{align*}
where $\mathbf{U}_{X}$, $\mathbf{U}_{\Theta}$, and $\mathbf{U}_{C}$ are mutually independent $S\times N_U$ matrices with independent standard normal elements.  To preserve the cross-correlation between exposures and spatial random effects, 
$\bB_X$ and $\bB_\Theta$ are selected so that $\bB_\Theta\bB_\Theta^T$ and $\bB_X\bB_X^T$ match the sample covariance of the posterior mean of the spatial random effects, ${\hat \Theta}^*$, and observed exposure variables, $\bX_0^*$, respectively, with columns of zeros added so that both $\bB_X$ and $\bB_\Theta$ have $N_U$ columns. In this formulation, $\rho$ is both the correlation between the latent confounder $\bU_C\bB_X^T$ and exposure $\bX^*$ and the latent confounder $\bU_C\bB_\Theta^T$ and spatial random effect $\Theta^*$.    The resulting bias for exposure $e$ and response $r$ is
\begin{equation*}
\alpha_{ier} = \frac{\text{Cov}(\Theta^*_{ir}, X^*_{ie})}{\text{Var}(X^*_{ie})} = \rho^2\frac{ (\mathbf{B}_\Theta^\top \mathbf{B}_X)_{re}}{(\mathbf{B}_X^\top \mathbf{B}_X)_{ee}}.
\end{equation*}
    Derivations are provided in Section~\ref{A:sensitivity}. To focus on the local spatial scale, we compute $\mathbf{B}_\Theta$ and $\mathbf{B}_X$ using only the ${\hat \Theta}$ and $\bX_0^*$ for the one hundred most local terms.  

\begin{figure}[h]
    \centering
    \caption{{\bf Sensitivity analysis}: The red solid and dotted lines indicate the posterior means and $95\%$ credible intervals, respectively, from MSM. The black curve is the absolute bias that would be induced by a missing local confounding variable plotted against its correlation $\rho$ with the exposure and response, and the vertical lines indicate correlation between the exposure and two observed confounders from the one hundred most local spatial scales}
    \includegraphics[trim={0.1cm 0.1cm 1.2cm 0.7cm},clip, width=0.6\linewidth]{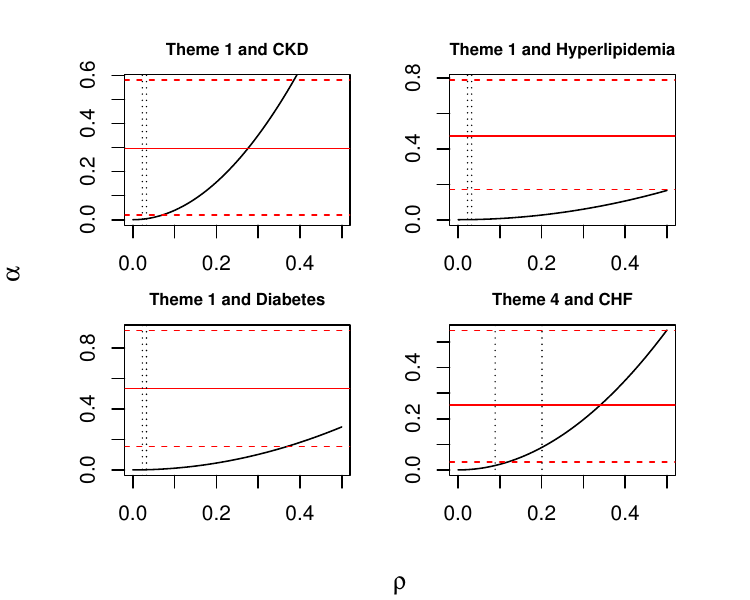}
    \label{fig:sen_bias}
\end{figure}

Figure \ref{fig:sen_bias} shows how bias increases with $\rho$ for four exposure-outcome combinations that have significant effect in MSM. The critical value of $\rho$ is the value for which the bias equals the lower boundary of the credible intervals (dashed line) of the effect estimate, as shifting the interval by this value of the bias would place zero in the credible interval.  For the combinations Theme 1 and CKD and Theme 4 and CHF, the effects remain significant only for  $\rho<0.069$ and $\rho<0.117$, respectively. For Theme 1 and Hyperlipidemia and Theme 1 and Diabetes, the effects remain significant for greater values of $\rho$ ($0.507$ and $0.368$, respectively). As a benchmark to interpret these results, the absolute correlation over the largest one hundred eigenvalues between the observed confounders and exposures  ranges between $0.02$ and $0.20$.
Thus, the causal conclusions are robust for the associations between Theme 1 and Hyperlipidemia and Theme 1 and Diabetes but fragile for those between Theme 1 and CKD and Theme 4 and CHF.

\subsection{Summarizing the tensor structure}\label{S:tensor_latent}

\begin{figure}[h]
    \centering
    \caption{{\bf Tensor estimates}: The curves show one rank-2 CP decomposition of the posterior means of the low-rank tensor that is of the dimension $L\times E \times R$, as presented in (\ref{e:tensor_co}).For basis functions (left panel), the x-axis shows the ten basis functions that represent weights from the most global (1) to the most local (10). The x-axes for exposures (middle panel) and outcomes (right panel) include the four themes of the disaster resilience index and the five health conditions. Each curve represents a tensor margin and shows the relationship among basis functions, exposures, and outcomes.}  
    \includegraphics[trim={0.2cm 0.5cm .1cm 1cm},clip, width=.6\linewidth]{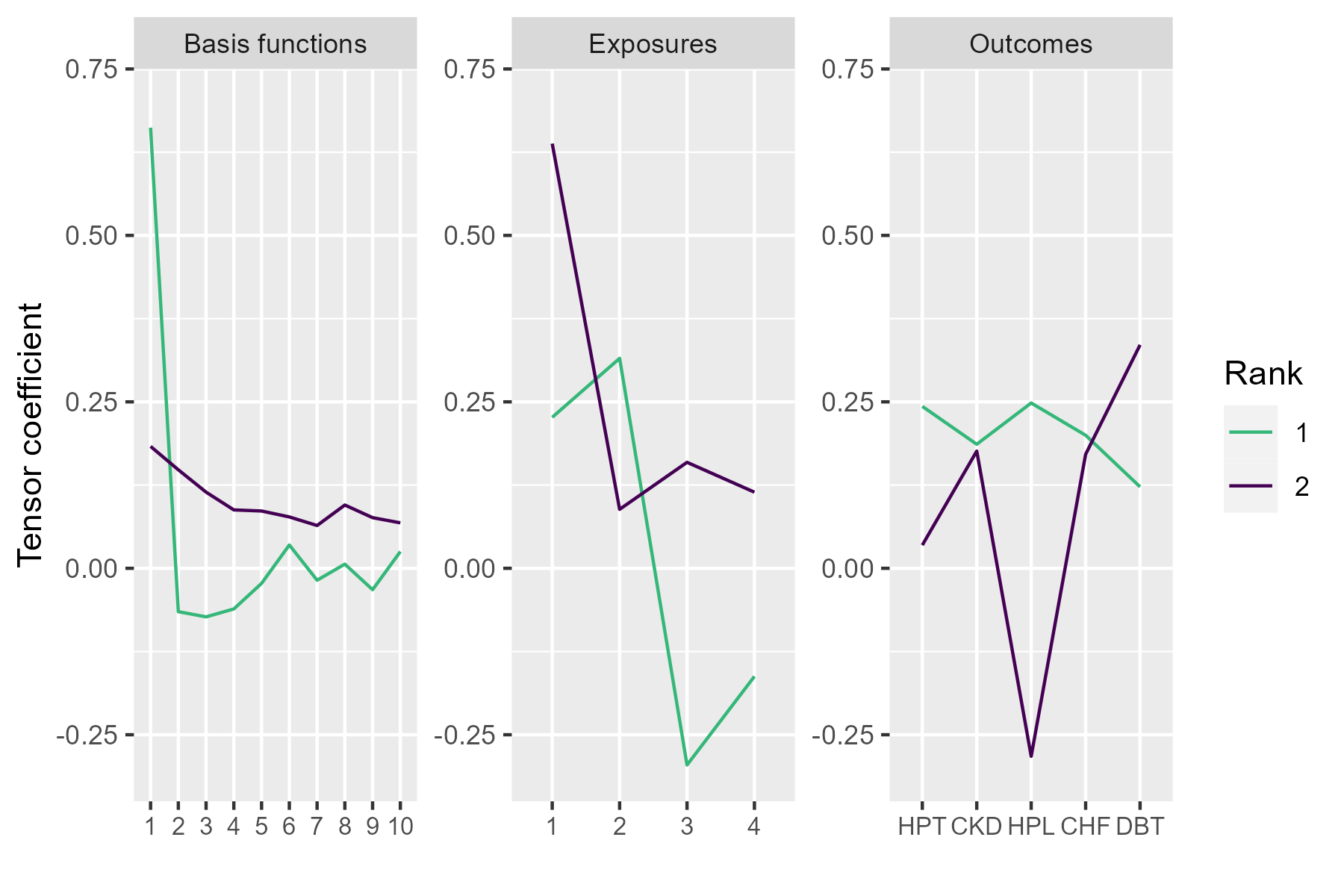}

    \label{f:tensor_estimates}
\end{figure}

Lastly, to answer our motivating question regarding the latent structure between variables, we use the tensor estimates to decompose the health effects.
To provide a succinct representation of this structure, we use the {\tt R} package {\tt rTensor} \citep{rTensor} to find a rank-2 canonical polyadic tensor decomposition of the posterior mean of the tensor $\hat{\boldsymbol{\gamma}}$ in (\ref{e:tensor_co}). Each curve in Figure \ref{f:tensor_estimates} represents a tensor margin (see Figure \ref{fig:cp_decomp}). For basis functions, term with index near one capture global features and indices near $L=10$ capture local features. For tensor margins representing basis functions, the first factor (green) captures global trends. For tensor margins representing exposures, the first factor is positive for Theme 1 (lack of economic resilience) and Theme 2 (household-level adaptive capacity) and negative for Theme 3 (fraction of minority) and Theme 4 (housing type and transportation). 
The green tensor margins for basis function and exposure together explain why the coefficient estimates for global scales are positive for Themes 1 and 2 and negative for Themes 3 and 4 (Figure \ref{f:coef_by_freq_svi}).
Factor 2 (purple) dominates for the local-scale trends of interest and Theme 1 has the highest loading and thus explains the strongest effects in Theme 1, shown in Figure \ref{fig:forest}.
For this factor and thus the final estimates in Figure \ref{fig:forest}, the responses have similar loadings except for hyperlipidemia, which is also reflected in Figure \ref{fig:forest}. The tensor margins for outcomes indicate that the exposure effects for chronic kidney disease and congestive heart failure are largely similar, while the effects for other outcomes differ. The complexity in data with multiple exposures and multiple outcomes makes it difficult to parse complex structures. By using a tensor structure, MSM offers a way to construct interpretable latent structures, enabling understanding of the relationship among different dimensions.

\section{Discussion}\label{s:discussion}

We introduce a multivariate spectral model (MSM) for areal data that estimates the effects of multiple exposures on multiple outcomes in the presence of spatial confounding. The model relaxes the no unmeasured confounder assumption and permits global confounding. As the simulation study demonstrates, when the conditions hold, the MSM method has low MSE and achieves conservative coverage rates that corrects over-confidence---reflected in the too-narrow credible intervals---from single-scale methods. Nonetheless, being a parametric model, the proposed MSM has its limitations in the pre-specified model. A semiparametric or nonparametric framework could provide more flexibility and robustness in the case that the model is misspecified. Further, unmeasured local confounding is unverifiable, and our proposed method does not account for spillover and interference. Spillover effects are possible if neighboring, economically vulnerable communities influence local health outcomes. For instance, residents may travel to adjacent areas to access public resources, thereby diminishing the supply available to local residents. We leave these limitations for future research.  Finally, we note that our results and interpretations are limited to the Southeastern US and require external validation before generalizing.

Data analysis shows the associations between disaster resilience index in four themes and the incidence of five chronic diseases. The $95\%$ credible intervals show that ZCTA areas lacking economic resilience (high scores for Theme 1) are more likely to have higher incident rates of chronic kidney disease and diabetes but lower rates of hyperlipidemia, while lacking disaster resilience in terms of housing type and transportation (high scores for Theme 4) are associated with higher incident rates of congestive heart failure. 
Besides being able to provide unbiased causal estimate effects with adjustment for unmeasured spatial confounders, MSM offers efficiency gain, provides interpretable findings, and encourages the consideration of the role of spatial scales in the effect estimates.



\section*{Funding}

Research reported in this publication was partially supported by the National Heart, Lung, And Blood Institute of the National Institutes of Health under Award Number T32HL079896, National Institutes of Health grants R01ES031651-01 and R01ES036270-01A1, and National Science Foundation grant DMS2152887. The content is solely the responsibility of the authors and does not necessarily represent the official views of the National Institutes of Health.
This project was supported by an appointment (of S.E.C.) to the Research Participation Program at the Center for Public Health and Environmental Assessment, U.S. Environmental Protection Agency (EPA), administered by the Oak Ridge Institute for Science and Education (ORISE) through an interagency agreement between the U.S. Department of Energy and the U.S. EPA. 

\section*{Disclosure statement}

The authors have no conflicts of interest. Although this work has been reviewed for publication by the U.S. EPA, it does not necessarily reflect the views and policies of the agency. Mention of trade names or commercial products does not constitute endorsement or recommendation for use.

\section*{Data Availability Statement}

All available data and code are provided in our public git repository:  \url{https://github.com/snprim/multivar_spectral_confounder}. 

\section*{Acknowledgements}

We thank Dr. Wei-Lun Tsai for her support and meaningful discussions during the
development of this manuscript, in particular her work on incidence rates and numerous
spatial indicators of environmental exposure. We also thank the reviewers and editors for
their insightful feedback on earlier versions of the manuscript.

\singlespacing
\bibliography{bibliography.bib}





\spacingset{1.8}
\appendix

\newpage\begin{center}
    {\LARGE\bf Supplementary Materials for ``A Spectral Confounder Adjustment for Spatial Regression with Multiple Exposures and Outcomes''}
\end{center}

\section{Spatial causal framework}\label{s:causal}
Under the potential outcome framework \citep{rubin1974}, we let 
\[ Y_{ir}(\mathbf{x}_{i}) = \sum_{e=1}^E \beta_{ier} x_{ie} + \sum_{p=1}^P \alpha_{pr} Z_{ip} + \theta_{ir} + \epsilon_{ir}, \]
where $Y_{ir}(\mathbf{x}_{i})$ is the potential outcome at location $i$ when the levels of $E$ exposures for the location $i$ are $\mathbf{x}_i = (x_{i1}, \cdots, x_{iE})^T$. $Z_{ip}$ is covariate $p$ at location $i$, $\alpha_{pr}$ is the coefficient of the covariate $p$ associated with the outcome $r$, $\theta_{ir}$ is a confounder for outcome $r$ at location $i$, and $\epsilon_{ir}$ is random noise and distributed by N$(0, \tau_r^2)$.

Since our exposure is continuous, our estimand is $\beta_{er} = E[Y_{ir}(x_{i1}, \cdots, x_{ie}+1, \cdots, x_{iE}) - Y_{ir}(x_{i1}, \cdots, x_{ie}, \cdots, x_{iE})]$, the average treatment effect (ATE) of exposure $e$ on response $r$. Since we can only observe one outcome at each location, the potential outcome framework requires several assumptions to ensure identifiability. One assumption is the stable unit treatment value assumption (SUTVA), which means only one version for each level of the treatment exists and the potential outcome of one unit is not affected by that of another unit \citep{rubin1980}. In the spatial context, the second half of SUTVA means the effect does not spill over to neighboring areas. Further, the assumption of consistency states that the observed outcome is the same as the potential outcome determined by the observed treatment, i.e. $Y_{ir} = Y_{ir}(\mathbf{x}_{i})$. 

The next assumption, latent ignorability, states that, given the treatment $\mathbf{x}_{i}$ and the confounder $\theta_{ir}$, the potential outcome $Y_{ir}(\mathbf{x}_{i})$ is independent of the treatment variable, i.e., $\bX \perp \bY (\bx) | \bZ,\btheta$ for all $\bx \in \text{supp}(\bX)$, where 
\begin{align*}
\bX &= \big( \begin{smallmatrix}
    \mathbf{x}_{1} & \cdots & \mathbf{x}_{S}
\end{smallmatrix} \big)^\top,& 
\mathbf{x}_s &= \big( \begin{smallmatrix}
    x_{s1} & \cdots & x_{sE}
\end{smallmatrix} \big), \\
\bY &= \big( \begin{smallmatrix}
    \bY_{i1} & \cdots & \bY_{iR}
\end{smallmatrix} \big), &
\bY_{ir} &= \big( \begin{smallmatrix}
    Y_{1r} & \cdots & Y_{Sr}
\end{smallmatrix} \big)^\top, \\
\btheta &= \big( \begin{smallmatrix}
    \btheta_{1} & \cdots & \btheta_{R}
\end{smallmatrix} \big), &
\btheta_{r} &= \big( \begin{smallmatrix}
    \theta_{1r} & \cdots & \theta_{Sr}
\end{smallmatrix} \big)^\top.
\end{align*} 


\section{Understanding confounding in special cases} \label{s:alpha_w}

Our multiscale spectral method for identifying the exposure effects relies on the local unconfoundedness assumption that $\alpha_{ier} \to 0$ as $w_i\to\infty$. In this section, we give two special cases to relate this assumption in the spectral domain to more intuitive properties of a spatial model: we first show the connection between $w_i$ and spatial scale through an example of a ring graph (Section \ref{s:eigen_scale}) and then derive and examine $\alpha_{ier}$ for a model with an explicit generative model for the missing spatial confounders (Section \ref{s:LMC}). 

\subsection{Connecting eigenvalues and scales in the case of the ring graph} \label{s:eigen_scale}

\begin{figure}
    \centering
    \caption{{\bf Spatial scale and eigenvalues}: Plots of the eigenvectors ($\Gamma$) for three frequencies ($k$) and eigenvalues $(w)$ for the ring graph with $S=50$ nodes.}
\includegraphics[page=1,width=0.3\linewidth]{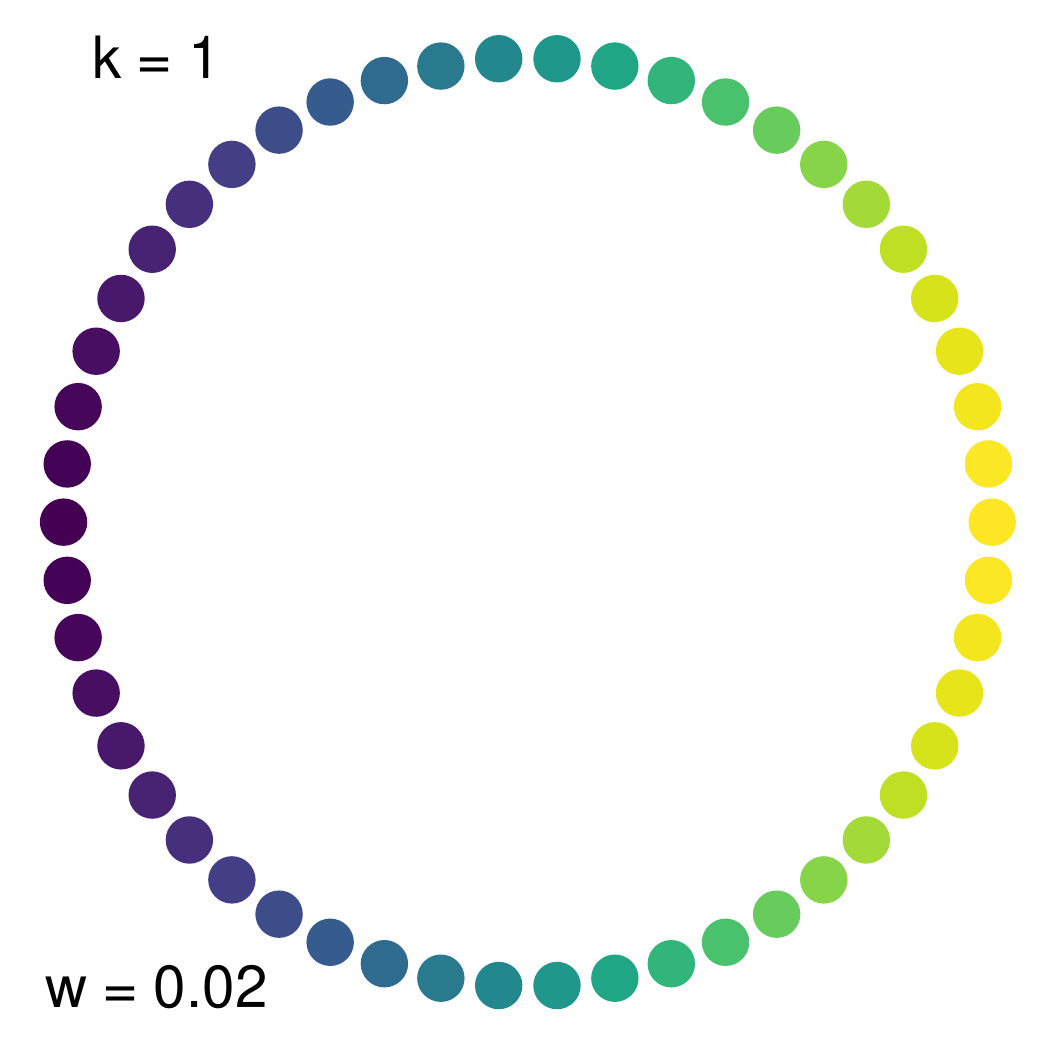}
\includegraphics[page=2,width=0.3\linewidth]{images/ring.pdf}
\includegraphics[page=3,width=0.3\linewidth]{images/ring.pdf}
    \label{f:rings}
\end{figure}

For regions on a regular grid,
the eigenvalues and eigenvectors of the adjacency matrix have explicit forms, which lays bare the relationship between the eigenvalues and the spatial scale \citep{vandenheuvel2001}. The simplest case is the ring graph with $S$ vertices $s\in\{1,...,S\}$ (Figure \ref{f:rings}). The first $S/2$ eigenvectors and eigenvalues are \citep{rabikka2022}
\begin{equation}\label{e:eigenvalues}
    \Gamma_{is} = \cos(2\pi k_i s / S) \mbox{\ \ \ and \ \ \ }
    w_i = 2 - 2 \cos(2\pi k_i / S) \end{equation} 
for frequency $k_i = i/2, i = 1,...,S$ (the remaining $S/2$ eigenpairs have the same form but with a period shift).
The data in the spectral domain is $Y_i^* = \sum_{s=1}^S\Gamma_{is}Y_s$, so the eigenvector on the left of Figure \ref{f:rings} with the smallest eigenvalue is effectively the east/west gradient, which we refer to as a global trend.  In contrast, the eigenvalue on the right has the largest eigenvalue and corresponds to an average of local differences.  Although the connections are not always explicit, we refer to eigenvalues by their scale and in particular focus on the case where $w_i\to\infty$ and the corresponding eigenvectors measure increasingly local differences. 

\subsection{Confounding in a linear, multivariate Gaussian model} \label{s:LMC}

In this example, we examine the confounding that arises from a particular model with unmeasured confounding variables to gain insights about the settings that lead to the local unconfoundedness assumption being satisfied. In this particular model, confounding results from $n_u$ latent factors (with unit variance for identifiability) $\bU_1,...,\bU_{n_u} \iid \mbox{CAR}(1, \lambda_U)$ and the exposure and spatial random effects are generated similarly with $\bM_{11},...,\bM_{1E} \iid \mbox{CAR}(1, \lambda_X)$ 
 and $\bM_{21},...,\bM_{2R} \sim \mbox{CAR}(1, \lambda_\theta)$.  Let  $\bU = (\bU_1,...,\bU_{n_u})$, $\bM_1 = (\bM_{11},...,\bM_{1E})$ and $\bM_2 = (\bM_{21},...,\bM_{2R})$.  The data are then $\bX = \bU\bB_1+\bM_1\bS_1^{1/2}$, $\btheta = \bU\bB_2+\bM_2\bS_2^{1/2}$ and $\bY = \bX\bbeta + \btheta + \bepsilon$, where $\bB_1$ and $\bB_2$ control the strength of confounding, $\bS_1$ and $\bS_2$ are the covariance of the exposures and responses, and $\bbeta$ is the true effect matrix of interest.   

It can be shown (Section \ref{s:A23}) that the matrix of bias parameters $\alpha_{ier}$ is
\begin{equation} \label{e:alpha_omega}
    \balpha_i = \bB_2^\top \bB_1 \left(\bB_1^\top \bB_1 + \frac{1-\lambda_U + \lambda_U w_i}{1-\lambda_X + \lambda_X w_i} \mathbf{S}_1 \right)^{-1}. 
\end{equation}
If $\bB_1$ or $\bB_2$ is zero, then $\balpha_i=0$ since $\btheta$ and $\bX$ are independent and there is no confounding.  If $\bB_2^\top\bB_1\ne 0$ and there is confounding, then $\balpha_i\to 0$ as $w_i\to \infty$ if and only if $\lambda_U>\lambda_X$.  That is, the local confounding assumption is satisfied if the missing confounders have stronger spatial dependence than the covariates, which agrees with \cite{paciorekImportanceScaleSpatialconfounding2010} and others.

\section{Identifiability}\label{s:identifiability}

\subsection{Tensor coefficients $\bgamma$} \label{s:tensor_coeff}

Regarding tensor coefficients, when $\balpha_i$ goes towards zero, the multiscale regression coefficients $\tilde{\bbeta} = \bgamma \times_n \bB$ is identifiable, where   $\times_n$ denotes n-mode product \citep{kolda2009tensor} and $\bB$ is of dimension $S \times L$ and contains $L$ B-spline basis functions evaluated at knots. Since $\bB$ is given, $\bgamma$ is identifiable, but the individual tensor margins, as discussed in Section \ref{s:btr} of the main document, are not identifiable. As stated in  \cite{guhaniyogi2017bayesian}, identifiability of tensor margins (columns in $T_{1}, T_{2}, T_{3}$, see equation    (\ref{e:tensor_co}) in the main text) comes with three restrictions: (1) scale indeterminacy, (2) permutation indeterminacy, and (3) orthogonal transformation indeterminacy. They found that the non-identifiability of tensor margins does not affect convergence of the tensor coefficients in their simulation studies, and thus imposing identifiability restrictions on tensor margins is unnecessary.  

\subsection{Other parameters} \label{A:iden_others}

The distribution of our transformed data for response $r$ is 
\begin{equation} \label{e:likelihood}
    \bY_{r}^* | \bX^* \overset{ind}{\sim} N \bigg( \bX^* \big(\mathbf{1}_S \bbeta_r + \balpha_r \big), \tau_r^2 \bI_S \bigg),
\end{equation} 
where $\bY^*_{r} = \big( \bY^*_{r}(w_1), \cdots, \bY^*_{r} (w_S)  \big)^\top$, $\balpha_r = \big( \balpha_r(w_1), \cdots, \balpha_r(w_S) \big)^\top$, $\bbeta_r = \big( \beta_{1r}, \cdots, \beta_{Er} \big)$, $\mathbf{1}_S$ is a length $S$ vector of $1$, and $\bX^*$ defined as in Section \ref{s:alpha_w}. The parameters of interest are $\Theta = ( \bbeta_r, \tau^2_r, \balpha_r).$ To show that the parameters are identifiable, we need to show that $P_1(\bY^*_{r} | \Theta^{(1)})$ = $P_2(\bY^*_{r} | \Theta^{(2)})$ if and only if $\Theta^{(1)} = \Theta^{(2)}$, where $P(\bY^*_{r} | \Theta)$ is the likelihood function of (\ref{e:likelihood}) in the main text. Thus, 
\begin{align*}
    & (2\pi \tau_r^{2 (1)})^{-S/2}\exp \bigg\{ \bigg( \bY^*_r - \bX^* \big( \mathbf{1}_S \bbeta_r^{(1)} + \balpha_r^{(1)} \big) \bigg)^\top \bigg( \bY^*_r - \bX^* \big( \mathbf{1}_S \bbeta_r^{(1)} + \balpha_r^{(1)} \big)  \bigg) \bigg\} \\
    =&  (2\pi \tau_r^{2 (2)})^{-S/2} \exp \bigg\{ \bigg( \bY^*_r - \bX^* \big( \mathbf{1}_S \bbeta_r^{(2)} + \balpha_r^{(2)} \big)  \bigg)^\top \bigg( \bY^*_r - \bX^* \big(\mathbf{1}_S \bbeta_r^{(2)} + \balpha_r^{(2)} \big) \bigg) \bigg\} \\
    \Leftrightarrow & \tau^{2(1)}_r = \tau^{2(2)}_r \text{ and } \big( \mathbf{1}_S \bbeta_r^{(1)} + \balpha_r^{(1)} \big) = \big( \mathbf{1}_S \bbeta_r^{(2)} + \balpha_r^{(2)} \big). 
\end{align*} 
When $\balpha_r \to 0$ in local scales, as described in Section \ref{s:alpha_w}, $\bbeta_r^{(1)} = \bbeta_r^{(2)}$. Since $\bbeta + \balpha_i = \tilde{\bbeta}_i$, $\tilde{\bbeta}_i$ is identifiable as shown in Section \ref{s:tensor_coeff}. The other direction is straightforward. When $\bbeta_r^{(1)} = \bbeta_r^{(2)}$ and $\tau^{2(1)}_r = \tau^{2(2)}_r$, $P_1(\bY^*_r | \Theta^{(1)})$ = $P_2(\bY^*_r | \Theta^{(2)})$ as long as $\balpha_r \to 0$ for large eigenvalues.

\section{Computational details}\label{s:A1}

The MCMC sampler is comprised of Gibbs and Metropolis-Hasting sampling. For Gibbs sampling, the full-conditional posterior distributions are provided in Section \ref{s:full_conditional}. For Metropolis-Hasting sampling, our algorithm keeps the acceptance rate around $40\%$. Each MCMC simulation is run for $5,000$ iterations with $1,000$ burn-ins. We use different values of $L$ and $K$ and set $L=10$ and $K=5$ with verification that the MCMC chains achieve good convergence by examining trace plots and effective sample sizes. 

\section{Derivations}\label{s:A2}

\subsection{Full conditional distribution}\label{s:full_conditional}

The likelihood function is
\[ Y^*_{ir} \sim N \bigg( \sum_{e=1}^{E} X^*_{ie} \bigg( \sum_{l=1}^L B_{il} \Big( \sum_{k=1}^K T_{1lk} T_{2ek} T_{3rk} \Big) \bigg)  + \bA_r \bC_i^{*\top}, \tau_r^2 \bigg). \]
The full conditional distributions are given below. 

\subsubsection{Update $T_{1lk}$}
The full conditional depends on
\begin{align*}
   L_{lk} &= \sum_{r=1}^R \bigg( \sum_{i=1}^S \Big(  \sum_{e=1}^{E} X^*_{ie} B_{il} T_{2ek} T_{3rk} \Big)^2 /\tau^2_r \bigg), \\
   A_{lk} &= L_{lk} + \frac{1}{\lambda_{1k}}, \\
   h^{(l)}_{i,k,l,r} &= Y^*_{ir} - \underbrace{\sum_{e=1}^{E} X^*_{ie} \sum_{z \ne l} B_{iz} T_{1zk} T_{2ek} T_{3rk}}_{\text{same rank, different basis function}} - \underbrace{\sum_{e=1}^{E} X^*_{ie} \sum_{l=1}^L B_{il} \sum_{j \ne k} T_{1lj} T_{2ej} T_{3rj}}_{\text{different rank, all basis functions}} - \bA_r \bC_i^{*\top}, \\
   B_{lk} &= \sum_{r=1}^R \bigg( \sum_{i=1}^S h^{(l)}_{i, k, l, r} \bigg) / \tau^2_r.
\end{align*}
Then, $T_{1lk} | \text{rest} \sim N(B_{lk}/A_{lk}, 1/A_{lk})$.

\subsubsection{Update $T_{2ek}$} 
The full conditional depends on
\begin{align*}
    E_{ek} &= \sum_{r=1}^R \bigg( \sum_{i=1}^S X^*_{ie} \sum_{l=1}^L B_{il} T_{1lk} T_{3rk} \bigg)^2 /\tau^2_r \bigg), \\
   A_{ek} &= E_{ek} + \frac{1}{\lambda_{2e}}, \\
   h^{(e)}_{i,k,e,r} &= Y^*_{ir} - \underbrace{\sum_{z\ne e} X^*_{iz} \sum_{l=1}^L B_{il} T_{1lk} T_{2zk} T_{3rk}}_{\text{same rank, different exposure}} - \underbrace{\sum_{e=1}^{E} X^*_{ie} \sum_{l=1}^L B_{il} \sum_{j \ne k} T_{1lj} T_{2ej} T_{3rj}}_{\text{different rank, all exposures}} - \bA_r \bC_i^{*\top}, \\
   B_{ek} &= \sum_{r=1}^R \bigg( \sum_{i=1}^S h^{(e)}_{i,k,e,r} \bigg) / \tau^2_r. 
\end{align*}
Then, $T_{2ek} | \text{rest} \sim N(B_{ek}/A_{ek}, 1/A_{ek})$.

\subsubsection{Update $T_{3rk}$}
The full conditional depends on
\begin{align*} 
    R_{rk} &= \sum_{i=1}^S \Big(  \sum_{e=1}^{E} X^*_{ie} \sum_{l=1}^L B_{il} T_{1lk} T_{2ek} \Big)^2 /\tau^2_r, \\
    A_{rk} &= R_{rk} + \frac{1}{\lambda_{3r} \tau^2 \tau^2_r}, \\
    h^{(r)}_{i,k,r} &= Y^*_{ir} - \underbrace{\sum_{e=1}^{E} X^*_{ie} \sum_{l=1}^L B_{il} \sum_{j \ne k} T_{1lj} T_{2ej} T_{3rj}}_{\text{different rank, same response}} - \bA_r \bC_i^{*\top},  \\
    B_{rk} &= \bigg( \sum_{i=1}^S h^{(r)}_{i,k,r} \bigg) / \tau^2_r. 
    \end{align*}
Then, $T_{3rk} | \text{rest} \sim N(B_{rk}/A_{rk}, 1/A_{rk})$.

\subsubsection{Update $\tau^2_r$}
The prior distribution is
$ \tau_r^2 \sim \mbox{InvGamma} (a, b).$
Thus,
\[ \tau_r^2 | \text{rest} \sim \mbox{InvGamma} \bigg( \frac{S + K}{2}+a, \frac{\sum_{i=1}^S (Y^*_{ir} - \bX^*_i\bbeta_r - \bA_r \bC_i^{*T})^2}{2} + \frac{\sum_{k=1}^K T_{3rk}^2}{2 \lambda_{3r} \tau^2} + b \bigg). \]

\subsubsection{Update $\sigma^2_q$}
The prior distribution is $\sigma^2_q \sim \mbox{InvGamma(c,d)}$. Then,
\[ \sigma^2_q|\text{rest} \sim \mbox{InvGamma} \bigg( \frac{S}{2} + c, \sum_{i=1}^S \frac{\bC_q^{*2} (1-
\lambda_C + \lambda_C w_i)}{2} + d \bigg) \]

\subsubsection{Update $\bA_r$}
The prior distribution is $A_{rq} \sim \mbox{N}(0,0.5^2)$. Thus,
\[ A_{rq} | \text{rest} \sim \mbox{N} \bigg( \Big( \bC_q^{*T} \bSigma_r^{-1} \bC_q^* + \frac{1}{0.5^2} \Big)^{-1} \Big( \bC_q^{*T} \bSigma_r^{-1} (\bY^*_r - \bX^* \bbeta_r) \Big), \Big( \bC_q^{*T} \bSigma_r^{-1} \bC_q^* + \frac{1}{0.5^2} \Big)^{-1} \bigg).\]
We update one row at a time: 
\[ \bA_{r} | \text{rest} \sim \mbox{N} \bigg( \Big( \bC^{*T} \bSigma_r^{-1} \bC^* + \frac{1}{0.5^2} \bI_Q \Big)^{-1} \Big( \bC^{*T} \bSigma_r^{-1} (\bY^*_r - \bX^* \bbeta_r) \Big), \Big( \bC^{*T} \bSigma_r^{-1} \bC^* + \frac{1}{0.5^2} \bI_Q \Big)^{-1} \bigg).\]
Although all columns are used for the full conditionals, we only update the cells corresponding in the lower triangle.

\subsubsection{Update $\bC_q$} 
The prior distribution is $\bC_q \sim \mbox{CAR}(\sigma^2_q, \lambda_U)$. Then,
\[\bC^{*T}_q | \text{rest} \sim \mbox{N} \bigg\{ \bD_{C_q}^{-1} \bigg( \sum_{r=1}^R  A_{rq} \bSigma_r^{-1} \Big( \bY^*_{r} - \bX^*\bbeta_r - \sum_{p \ne q} \bC^*_p A_{rp} \Big) \bigg), \bD_{C_q}^{-1} \bigg\},\]
where $\bD_{C_q}$ is a $S \times S$ diagonal matrix with the diagonal components $\bigg( \sum_{r=1}^R \frac{A^2_{rq}}{\tau^2_r} +  \frac{1-\lambda_U + \lambda_U w_i}{\sigma^2_q} \bigg)$ for $i \in \{ 1, 2, \cdots, S\}$.

\subsection{Implementation of Spatial+}
\label{s:sptialplus}

In a univariate case, \cite{Urdangarin2024simplified} decompose the covariates $\bX$ into two groups of linear combinations of eigenvectors, $\bX = \mathbf{F} + \mathbf{F}'$ with $\mathbf{F} = a_1 \bD_1 + \cdots + a_{n-(k+1) \bD_{n-(k+1)}}$ and $\mathbf{F}' = a_{n-k} \bD_{n-k} + \cdots + a_n \bD_n$, where $\bD_i$ are eigenvectors and $a_i$ are constants. To connect this model to our implementation, we rewrite $\mathbf{F}$ and $\mathbf{F}'$: $\mathbf{F} = \bGamma_1 A_1, \mathbf{F}' = \bGamma_2 A_2$ where the eigenvector matrix $\bGamma = \begin{bmatrix} \bGamma_1 & \bGamma_2 \end{bmatrix}$, which corresponds to large and small eigenvalues, and $A_1$ and $A_2$ contain the coefficients of the corresponding linear combinations. Their model uses $\mathbf{F}=\bGamma_1 A_1$, as the covariates in the spatial domain:
\[ \bX^* = \bGamma^\top \bX = \begin{bmatrix}
    \bGamma_1^\top \\ \bGamma_2^\top \end{bmatrix}
    \bigg( \bGamma_1 A_1 + \bGamma_2 A_2 \bigg) = 
    \begin{bmatrix}
        \bGamma_1^\top \bGamma_1 A_1 + \bGamma_1^\top \bGamma_2 A_2 \\
        \bGamma_2^\top \bGamma_1 A_1 + \bGamma_2^\top \bGamma_2 A_2
    \end{bmatrix} = \begin{bmatrix}
        A_1 \\ A_2
    \end{bmatrix},
\]
which is equivalent to the covariates in our spectral model, $A_1$.

\subsection{Derivations of bias from unmeasured confounding}\label{s:A23}

The matrix $\bU$ introduces confounding to $\bX$ and $\btheta$. The derivation here does not include extra smoothing introduced in Section \ref{s:sim_gen} in the main text and focuses on demonstrating how correlation, and thus bias, decreases at highest eigenvalues when $\lambda_U > \lambda_X$.
\[ \bU \sim \mbox{CAR} (1, \lambda_U) \]
\[ \bX | \bU = \bU \bB_1 + \bM_1 \mathbf{S}_1^{1/2}, \text{where } \bM_1 \sim \mbox{CAR}(1, \lambda_X) \text{ and } \mathbf{S}_{1,ij} = \rho_1^{|i-j|} \]
\[ \boldsymbol{\theta} | \bU = \bU \bB_2 + \bM_2 \mathbf{S}_2^{1/2}, \text{where } \bM_2 \sim \mbox{CAR} (1, \lambda_\theta) \text{ and } \mathbf{S}_{2,ij} = \rho_2^{|i-j|}. \]
In the spatial domain, 
\[ \bY = \bX \boldsymbol{\beta} + \boldsymbol{\theta} + \boldsymbol{\epsilon}. \]
To project the data into the spectral domain, we pre-multiply $\bY$, $\bX$, and $\boldsymbol{\theta}$ by the eigenvector matrix $\bGamma$. 
\[ \underbrace{\bGamma^\top Y}_{\bY^*} = \underbrace{\bGamma^\top \bX}_{\bX^*} \bbeta + \underbrace{\bGamma^\top \btheta}_{\btheta^*} + \underbrace{\bGamma^\top \bepsilon}_{\bepsilon^*} \]
\[ \bU^*_i = \bGamma^\top \bU_{i} \sim \mbox{N} (0, \frac{1}{1-\lambda_U + \lambda_U w_i} \bI_{n_u})\]
\[ \bX_i^* | \bU_i^* = \bU_i^* \bB_1 + \bM_{1i}^* \mathbf{S}_1^{1/2}, \text{where } \bM_{1i}^* \mathbf{S}_1^{1/2} \sim \mbox{N} \bigg( 0, \frac{1}{1-\lambda_X + \lambda_X w_i} \mathbf{S}_1 \bigg)  \]
and
\[ \btheta_i^* | \bU_i^* = \bU_i^* \bB_2 + \bM^*_{2i} \mathbf{S}_2^{1/2}, \text{where } \bM^*_{2i} \mathbf{S}_2^{1/2} \sim \mbox{N} \bigg( 0, \frac{1}{1-\lambda_\theta + \lambda_\theta w_i} \mathbf{S}_2 \bigg) \]
Now we can derive expectations and covariances: 
\[ E(\bX^*_i | \bU^*_i) = E(\btheta^*_i | \bU^*_i) = 0 \Rightarrow E(\bX^*_i) = E(\btheta^*_i) = 0 \]
\begin{align*}
   Var(\bX_i^*) &= E\{ Var(\bX^*_i | \bU^*_i) \} + Var\{ E(\bX_i^* | \bU_i^*)\} \\
   &= \frac{1}{1-\lambda_U + \lambda_U w_i} \bB_1^\top \bB_1 + \frac{1}{1-\lambda_X + \lambda_X w_i} \mathbf{S}_1.
\end{align*}
Similarly,
\begin{align*}
    Var(\btheta_i^*) &= E\{ Var(\btheta^*_i | \bU^*_i) \} + Var\{ E(\btheta_i^* | \bU_i^*)\} \\ 
    &= \frac{1}{1-\lambda_U + \lambda_U w_i} \bB_2^\top\bB_2 + \frac{1}{1-\lambda_\theta + \lambda_\theta w_i} \mathbf{S}_2.
\end{align*} 
Then, 
\[ Cov(\bX_i^*, \btheta^*_i) = \bB_1^\top Cov(\bU^*_i) \bB_2^\top = \frac{1}{1-\lambda_U + \lambda_U w_i} \bB_1^\top \bB_2\]
Thus the joint distribution of $\btheta^*_i$ and $\bX^*_i$ is: 
\[ \begin{pmatrix} \btheta^*_i \\ \bX^*_i \end{pmatrix} \sim \mbox{MVN} \begin{pmatrix} \begin{bmatrix} 0 \\ 0 \end{bmatrix}, 
\begin{bmatrix}
Var(\btheta^*_i  )
&
\frac{1}{1-\lambda_U + \lambda_U w_i} \bB_2^\top \bB_1
\\
\frac{1}{1-\lambda_U + \lambda_U w_i} \bB_1^\top \bB_2 
&
Var(\bX^*_i)
\end{bmatrix} 
\end{pmatrix} \]
    Thus, by the property of multivariate normal distribution, 
    \begin{align*}
        E(\btheta^*_i | \bX^*_i) &= \bX_i^* \bigg( \frac{1}{1-\lambda_U + \lambda_U w_i} \bB_2^\top \bB_1 
    Var(\bX^*_i)^{-1} \bigg) \\
       &= \bX_i^* \bigg\{\frac{1}{1-\lambda_U + \lambda_U w_i} \bB_2^\top \bB_1 \bigg( \frac{1}{1-\lambda_U + \lambda_U w_i} \bB_1^\top\bB_1 + \frac{1}{1-\lambda_X + \lambda_X w_i} \mathbf{S}_1 \bigg)^{-1}\bigg\}^\top \\
       &= \bX_i^*  \underbrace{ \bigg\{\bB_2^\top \bB_1 \bigg( \bB_1^\top\bB_1 + \frac{1-\lambda_U + \lambda_U w_i}{1-\lambda_X + \lambda_X w_i} \mathbf{S}_1 \bigg)^{-1}\bigg\}^\top}_{=:\balpha_i}. 
    \end{align*}
    
    The model can be written as $\bY^*_i | \bX^*_i = \bX^*_i \bbeta + \btheta^*_i | \bX^*_i + \bepsilon^*_i = \bX^*_i \bbeta + \bX^*_i \balpha_i + \bepsilon^*_i$.

\section{Simulation study}\label{A:sim}

We apply the methods described in Section \ref{s:method} to simulated data to examine how our method performs with different strengths or patterns of spatial confounding. We used four data generation settings as described in Section \ref{s:sim_gen}. For each of these four settings as well as two additional settings with no confounding and nonlinear confounding, we generated $N=500$ datasets and compared the performance across five models listed in Section \ref{s:sim:methods}. Computation details are provided in Section \ref{s:A1}. This simulation study does not include observed covariates, as the focus here is to demonstrate the effectiveness of our models in adjusting for unmeasured confounding and reducing bias in regression coefficient estimates for the exposures.

\subsection{Data generation}\label{s:sim_gen}


We generated data in $S$ spatial locations arranged in a $\sqrt{S} \times \sqrt{S}$ square grid. The response matrix $\bY$ is generated from fixed covariates $\bX$ and random effects $\boldsymbol{\theta}$: 
\begin{equation} \label{e:dgY}
    \bY = \bX \boldsymbol{\beta} + \boldsymbol{\theta} + \boldsymbol{\epsilon}, \text{ where } \epsilon_{sr} \overset{iid}{\sim} \mbox{N}(0, \tau^2_r) \text{ for } s = 1, \cdots, S \text{ and } r = 1, \cdots, R.
\end{equation}  
\renewcommand\arraystretch{1} 
Confounding is induced by $n_u$ common factors $\bU = \beta_{XZ} \bG \bV$ where $\beta_{XZ}$ controls the overall confounding, $\bV \sim \mbox{CAR} (\sigma^2_V, \lambda_V )$, and smoothing controlled by the kernel matrix $\bG$ with $G_{ij} = g_{ij} / (\sum_{l=1}^S g_{il})$, $\log(g_{ij}) = -(\|s_i - s_j \|/\phi)^2$, and $s_i \in \{1, \cdots, S\}$ is the location of the $i^{th}$ grid cell \citep{guan2023spatial}. The bandwidth $\phi$ controls the amount of smoothing, with larger $\phi$ giving smoother $\bU$.  

Given $\bU$, $\bX$ and $\boldsymbol{\theta}$ are generated with independent components $\bM_1 \mathbf{S}_1^{1/2}$ and $\bM_2 \mathbf{S}_2^{1/2}$. 
\begin{equation} \label{e:dgx}
    \bX | \bU = \bU \bB_1 + \bM_1 \mathbf{S}_1^{1/2}, \text{where } \bM_{1e} \sim \mbox{CAR}(\sigma^2_X, \lambda_X ) \text{ for } e \in \{1, \cdots, E\} \text{ and } \mathbf{S}_{1,ij} = \rho_1^{|i-j|}
\end{equation}  
and
\begin{equation} \label{e:dgtheta}
    \btheta | \bU = \bU \bB_2 + \bM_2 \mathbf{S}_2^{1/2}, \text{where } \bM_{2r} \sim \mbox{CAR} (\sigma^2_\theta, \lambda_\theta ) \text{ for } r \in \{1, \cdots, R\} \text{ and } \mathbf{S}_{2,ij} = \rho_2^{|i-j|},    
\end{equation}
where both $\bB_1$ (dimension $n_u \times E$) and $\bB_2$ (dimension $n_u \times R$) are non-negative and determine the latent structure between exposure and response. Emulating the relationships commonly found in mixtures and correlated health outcomes, $\mathbf{S}_1$ (dimension $E \times E$) and $\mathbf{S}_2$ (dimension $R \times R$) generate correlation among exposures and responses, respectively.  

\begin{figure}
    \centering
    \caption{{\bf Correlation versus eigenvalue}:  Correlation between $X$ and $\theta$ decreases with the increase of eigenvalues. The rate of decrease varies with parameters of extra smoothing $\phi$ and $\beta_{XZ}$, with $\phi$ controlling the rate of changes in confounding and $\beta_{XZ}$ controlling overall level of confounding.}
    \includegraphics[width=.7\linewidth]{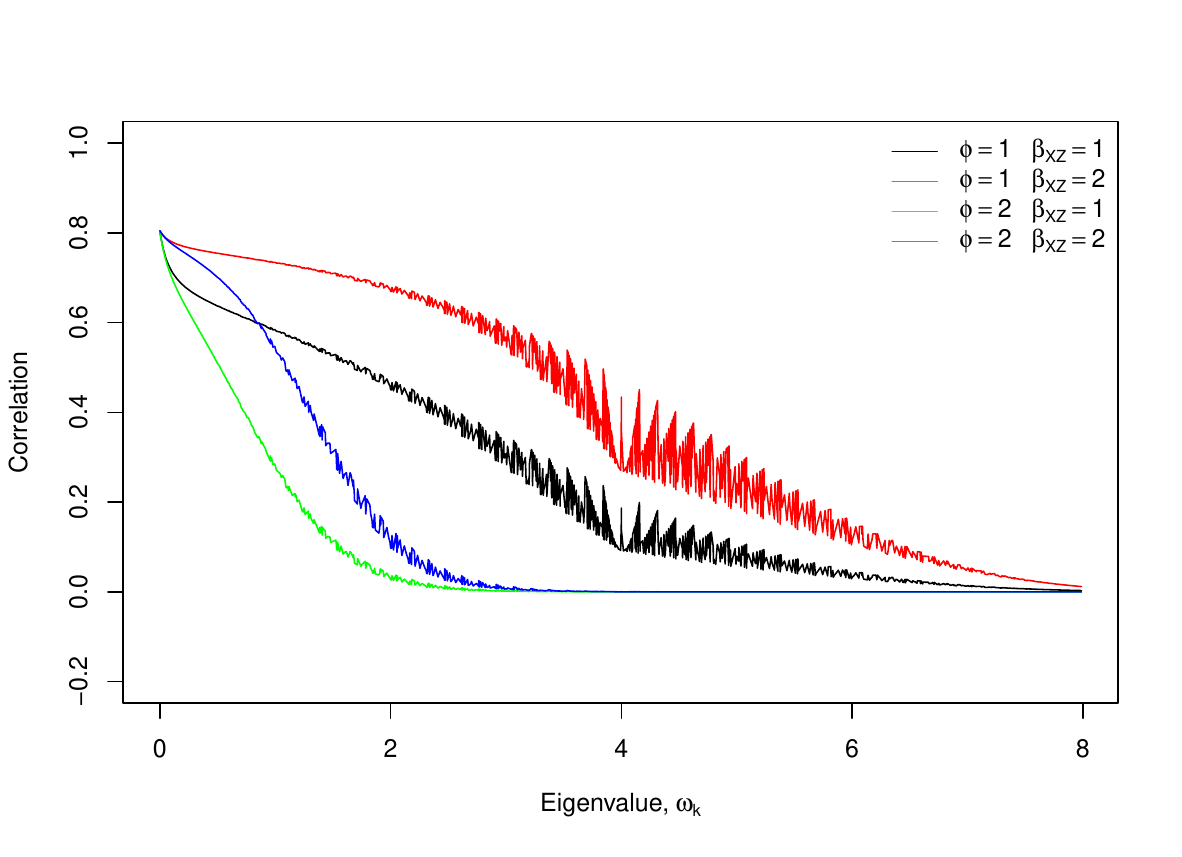}
    \label{f:corr-decrease}
\end{figure}

We first consider Setting 0 with $(\phi    ,\beta_{XZ}) = (1,0)$ and thus no confounding.
The next four data-generation settings use the combinations of the two hyperparameters $(\phi, \beta_{XZ})=(1,1),(1,2),(2,1),(2,2)$ for settings 1-4, respectively. As demonstrated in Figure \ref{f:corr-decrease}, when $\phi=2$, confounding is strong only for low eigenvalues; when $\phi = 1$, confounding remains strong for all but the highest eigenvalues. Also shown in Figure \ref{f:corr-decrease} is how quickly the correlation between $X_{ie}$ and $\theta_{ir}$ decreases as the eigenvalue $w_i$ increases to the maximum value in a fixed grid. As the grid becomes finer, the bias lessens. The correlation decreases much faster when $\phi=2$ compared to $\phi=1$. The difference affects model performance and demonstrates that our model produces the smallest mean squared error in regression coefficient estimates in the presence of unmeasured spatial confounding. 

All scenarios set $\lambda_U = 0.9999, \lambda_X = \lambda_\theta = 0.9, \rho_1 = \rho_2 = 0.9, S = 400, E = 9, R = 5, n_u = 10$. The variances are set at $\sigma^2_X=3$, $\sigma^2_\theta=2$, $\sigma^2_U = 16$ for $\bU$, and $\tau^2_r = 4$ for all responses. Values in $\bB_1$ and $\bB_2$ are generated from uniform distribution between $0$ and $1$. All settings use the same set of coefficients, $\bbeta$. Out of the $10$ coefficients (for nine exposures and an intercept), five are non-zero and five are zero. Note that $\beta$, though full-rank, has an inherently low-rank structure. Two of the five singular values ($1.86, 0.49, 0.12, 0.04, 0.02$) are much greater than the rest, accounting for $92.67\%$ of variation. 

\[ \bbeta = \begin{bmatrix}
        0.3 & 0.4 & 0.2 & 0.1 & 0.5 & 0 & 0 & 0 & 0 & 0 \\
        -0.3 & -0.2 & -0.3 & -0.4 & 0.1 & 0 & 0 & 0 & 0 & 0 \\
        0.2 & 0.4 & 0.3 & 0.2 & 0.3 & 0 & 0 & 0 & 0 & 0 \\
        0.3 & 0.5 & 0.4 & 0.3 & 0.4 & 0 & 0 & 0 & 0 & 0 \\
        0.5 & 0.7 & 0.6 & 0.4 & 0.6 & 0 & 0 & 0 & 0 & 0 \\ 
    \end{bmatrix}^\top.
\]

\subsection{Competing methods and metrics}\label{s:sim:methods}

\subsubsection{Competing methods}
\label{s:competitors}

The first method is our Multivariate Spectral Model (``MSM'') with horseshoe prior for the tensor margin. Method 2 is a univariate version of MSM with each response analyzed separately and is labelled ``USM.'' The MSM and USM models both set $L=10$ and $K=5$; sensitivity to these values is examined in Supplementary Materials Section \ref{s:A4}. Method 3 is the constant coefficient version of our model, with $\tilde{\bbeta}_i \equiv \bbeta$ for all $i$, incapable of accounting for different levels of spatial confounding at different scales and labeled the ``naive'' method. Methods 4 and 5 adapt the simplified spatial+ method by \cite{Urdangarin2024simplified}, who essentially fit (\ref{e:Ystar}) using a subset of the data that corresponds to local scales (Supplementary Materials Section \ref{s:sptialplus}). We follow \citeauthor{Urdangarin2024simplified}'s recommendation and keep $80\%$ (``Spatial+80'') and $40\%$ (``Spatial+40'') of the eigenvectors in the spectral domain for analysis. 

\subsubsection{Metrics}

For the MSM and USM models, we use posterior mean $\hat{\beta}_{1er}$ as the estimate of $\beta_{er}$. We evaluate model performance by four metrics: mean squared error (MSE), mean absolute bias (MAB), coverage of 95\% intervals, and mean posterior standard deviation (SD). Each metric is averaged over exposure and response. For example, 
\[ \text{MSE} = \frac{1}{ERN} \sum_{e=1}^E \sum_{r=1}^R \sum_{n=1}^{N} (\hat{\beta}^{(n)}_{1er} - \beta_{er})^2.\]
For MAB, we calculate the mean absolute bias by taking the absolute values of the mean bias across replicates and then average over exposures and outcomes: 
\[ \text{MAB} = \frac{1}{ERN} \sum_{e=1}^E \sum_{r=1}^R \bigg\vert \sum_{n=1}^{N} \hat{\beta}^{(n)}_{1er} - \beta_{er} \bigg\vert. \]

\subsection{Results}\label{s:sim:results}

\begin{figure}
    \centering
    \caption{{\bf Simulation results}: MSM is our multivariate spectral model, USM is the univariate spectral model, and the naive model does not allow coefficients to vary by scales. Based on \cite{Urdangarin2024simplified}, Spatial+ models retain $80\%$ and $40\%$ of the eigenvectors for Spatial+80 and Spatial+40, No confounding exists in Setting 0. Settings 1-4 set the hyperparameters in the data-generation process $(\phi, \beta_{XZ})$ to be $(1,1),(1,2),(2,1),(2,2)$, respectively.}\includegraphics[width=1\linewidth]{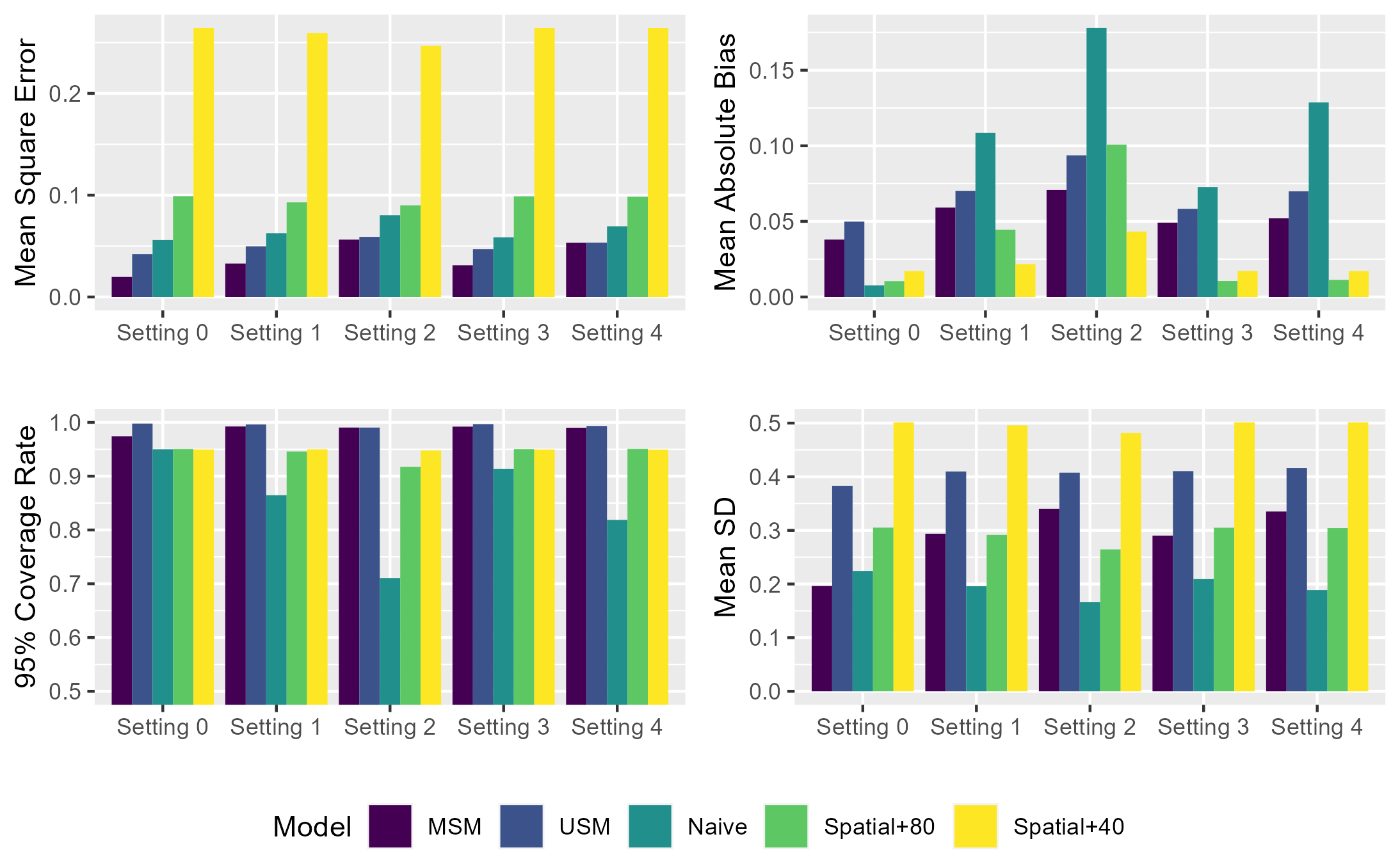}
    
    \label{fig:allres}
\end{figure}

A summary of performance metrics is provided in Figure 
\ref{fig:allres}. MSM typically has the lowest MSE. Note that MSM has lower MSE compared to USM due to its use of a tensor structure that borrows strength across outcomes. In terms of MAB, MSM is still better than USM and the naive method. The low MAB and high MSE of Spatial+80 and Spatial+40 are noteworthy. The high MSE can be attributed to biases of the methods; since the biases differ among different exposure/outcome pairs, the overall much smaller credible intervals lower MAB. This phenomenon can be verified by the coverage rates around or lower than the nominal rates of Spatial+80 and Spatial+40. In terms of coverage rates, spectral models have nominal or higher coverage rates, suggesting some efficiency loss from using a fraction of the decomposed information. This loss is perhaps unavoidable, since MSM assumes that the most local scale provides the most unbiased estimates. If the assumption is that a range of local scales is unconfounded, some efficiency loss can be avoided, but this is beyond the scope of the current study. 
Compared to USM, MSM has the main strength of efficiency; borrowing across responses reduces mean SD.
Lastly, MSM's standard deviations are higher than the naive method and Spatial+80 but much lower than that of the univariate spectral model, indicating improved efficiency achieved by the multivariate structure through information sharing among outcomes. 

Notably, the naive model has especially low $95\%$ coverage rate and high MAB for Setting 2, in which the correlation between $\bX$ and $\btheta$ decreases at the slowest rate (Figure \ref{f:corr-decrease}). This setting accentuates the importance of considering how confounding varies by scale and adjusting the range of scales on which the estimates should be based. The naive model does not discard information from any scales and thus has detrimental performance for settings where confounding decreases slowly. The correlation drops off at a fastest rate for Setting 3; therefore, the naive model performs better in this setting than in the other three settings. A similar pattern can be observed for Spatial+80 and, to a lesser degree, Spatial+40, again emphasizing the need to match confounding patterns to a model's adjustment for confounding. Supplementary Materials Section \ref{s:A3} provides the results separately for true coefficients that are zero and non-zero and the corresponding standard errors.

It is clear from the simulation results that Spatial+ is a competitive method; however, it is sensitive to the number of terms included in the covariates. More specifically, while Spatial+80 is competitive in Settings 1-4, Spatial+40 is not. Furthermore, in Setting 0, which does not have spatial confounding, Spatial+80 has much higher variance on average compared to MSM. In summary, MSM is more robust and adaptable to different levels of confounding. Further, if the spatial confounding is known to be present and the goal is inference for causal effects, the multivariate spectral model has competitive  MSE, low MAB, and reliable coverage rates. Lastly, if the effects are inherently low-ranked, the multivariate model has greater efficiency compared to univariate models.   

\subsection{Nonlinear confounding}
\label{A:nonlinear}

\begin{figure}
    \centering
    \caption{{\bf Nonlinear confounding:} Performance of the proposed method (MSM) and naive methods that does not account of spatial confounding as a function of the nonlinearity parameter, $k$. }
    \includegraphics[width=0.8\linewidth]{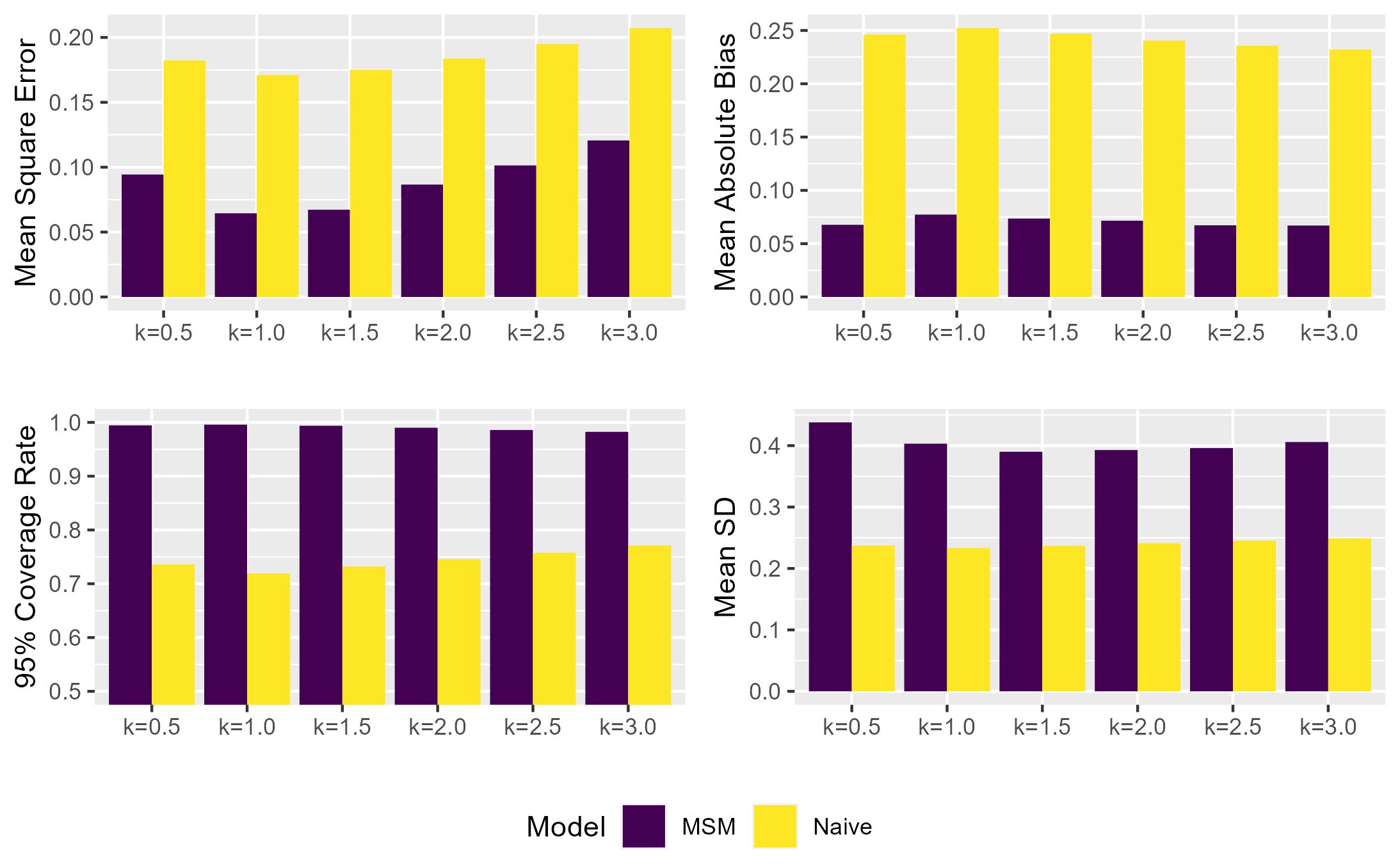}
    
    \label{fig:nonlinear}
\end{figure}

To examine how MSM performs when $\mathbf{X}$ and $\boldsymbol{\theta}$ are nonlinearly confounded, we generated $\mathbf{X}$ as described in (\ref{e:dgx}). For $\boldsymbol{\theta}$, instead of (\ref{e:dgtheta}), we let $\btheta | \bU = \frac{b}{\sigma_{(UB_2)^k}} (\bU \bB_2)^k + \bM_2 \mathbf{S}_2^{1/2}$, where $\sigma_{(UB_2)^k}$ denotes the standard deviation of $(\bU \bB_2)^k$. The standardization step prevents the variance $\sigma_{(UB_2)^k}$ from dominating that of the outcomes. The magnitude of nonlinear confounding is controlled by $b$, set as $8$ in our demonstration. The nonlinearity intensifies as $k$ increases. As shown in Figure \ref{fig:nonlinear}, MSM outperforms the naive model in MSE, MAB, and coverage rate. As expected, the MSE is the lowest at $k=1$, where confounding is linear. As nonlinearity becomes more severe, both models have higher MSE, but MSM still has much lower MSE. Even when $k=3$, MSM has a coverage rate above the nominal rate, suggesting that MSM is robust to moderate nonlinear confounding.

\section{Additional simulation study results}\label{s:A3}

Tables \ref{t:all_results1}, \ref{t:all_results2}, \ref{t:all_results3}, \ref{t:all_results4}, \ref{t:all_results5} provides five performance metrics---MSE, MAB, standard deviation, coverage rate for $95\%$ credible interval, and credible interval length---separately for coefficients $(\hat{\bbeta})$ that are nonzero and zero.  

As shown in the results presented in Section~\ref{s:sim:results}, MSM typically has low MSE and MAB. When the results are separated by whether coefficients are zero or nonzero, a slight difference is noted. MSM consistently outperforms other models for zero coefficients. While its MSE and MAB are still low for nonzero coefficients, the strengths are not as apparent across settings. More specifically, MSM's MAB are higher than those of Spatial+40 for nonzero coefficients. While MSM has overall reliable performance, it is especially strong in determining exposures with null effects. 

\begin{table}
\begin{center}
\caption{{\bf Simulation results}: The first two columns indicate the values for hyperparameters for data generation. Results here are separate for nonzero versus zero coefficients. The five models are Mutivariate Spectral Model (MSM), Univaraite Spectral Model (USM), a model that do not separate estimates by scale (Naive), and Simplified Spatial+ that retains $80\%$ and $40\%$ of the eigenvectors. The numbers in parentheses are standard errors.}
\label{t:all_results1}

(a) Mean Square Error
\begin{tabular}{lll|rrrrr} \\
$\phi$ & $\beta_{XZ}$ & coefficents & MSM & USM & Naive & Spatial+80 & Spatial+40 \\
\hline
 1 & 0 & nonzero & 0.04 (0.01) & 0.08 (0.01) & 0.05 (0.00) & 0.08 (0.01) & 0.23 (0.01)    \\
 1 & 1 & nonzero & 0.05 (0.00) & 0.08 (0.01) & 0.05 (0.00) & 0.08 (0.01) & 0.22 (0.01) \\ 
 1 & 2 & nonzero & 0.07 (0.00) & 0.09 (0.01) & 0.07 (0.00) & 0.08 (0.01) & 0.21 (0.01) \\
 2 & 1 & nonzero & 0.05 (0.01) & 0.08 (0.01) & 0.05 (0.00) & 0.08 (0.00) & 0.23 (0.01) \\ 
 2 & 2 & nonzero & 0.07 (0.00) & 0.09 (0.01) & 0.06 (0.00) & 0.08 (0.01) & 0.23 (0.01) \\[.05in]
 1 & 0 & zero & 0.01 (0.01) & 0.01 (0.00) & 0.06 (0.00) & 0.11 (0.01) & 0.29 (0.02) \\
 1 & 1 & zero & 0.02 (0.00) & 0.02 (0.00) & 0.07 (0.00) & 0.10 (0.01) & 0.29 (0.02) \\ 
 1 & 2 & zero & 0.05 (0.00) & 0.03 (0.00) & 0.09 (0.00) & 0.10 (0.01) & 0.27 (0.02) \\
 2 & 1 & zero & 0.02 (0.01) & 0.02 (0.00) & 0.06 (0.00) & 0.11 (0.01) & 0.29 (0.02) \\ 
 2 & 2 & zero & 0.04 (0.00) & 0.03 (0.00) & 0.08 (0.00) & 0.11 (0.01) & 0.29 (0.02) \\ 
\end{tabular}
\end{center}
\end{table}

\begin{table}
\begin{center}
\caption{{\bf Simulation results (continued)}}
\label{t:all_results2}

(b) Mean Absolute Bias
\begin{tabular}{lll|rrrrr} \\
$\phi$ & $\beta_{XZ}$ & coefficents & MSM & USM & Naive & Spatial+80 & Spatial+40 \\
\hline
 1 & 0 & nonzero & 0.08 (0.01) & 0.10 (0.01) & 0.01 (0.01) & 0.01 (0.01) & 0.01 (0.02) \\
 1 & 1 & nonzero & 0.10 (0.01) & 0.11 (0.01) & 0.09 (0.01) & 0.03 (0.01) & 0.01 (0.02) \\ 
 1 & 2 & nonzero & 0.11 (0.01) & 0.13 (0.01) & 0.16 (0.01) & 0.09 (0.01) & 0.03 (0.02) \\
 2 & 1 & nonzero & 0.09 (0.01) & 0.10 (0.01) & 0.06 (0.01) & 0.01 (0.01) & 0.01 (0.02) \\ 
 2 & 2 & nonzero & 0.10 (0.01) & 0.11 (0.01) & 0.11 (0.01) & 0.01 (0.01) & 0.01 (0.02) \\[.05in] 
 1 & 0 & zero & 0.01 (0.01) & 0.01 (0.01) & 0.01 (0.01) & 0.01 (0.01) & 0.02 (0.02) \\
 1 & 1 & zero & 0.03 (0.00) & 0.04 (0.00) & 0.12 (0.01) & 0.05 (0.01) & 0.03 (0.02) \\ 
 1 & 2 & zero & 0.04 (0.01) & 0.06 (0.01) & 0.19 (0.01) & 0.11 (0.01) & 0.06 (0.02) \\
 2 & 1 & zero & 0.01 (0.01) & 0.02 (0.01) & 0.08 (0.01) & 0.01 (0.01) & 0.02 (0.02) \\ 
 2 & 2 & zero & 0.01 (0.01) & 0.04 (0.01) & 0.14 (0.01) & 0.02 (0.01) & 0.02 (0.02)  
\end{tabular}
\end{center}
\end{table}

\begin{table}
\begin{center}
\caption{{\bf Simulation results (continued)}}
\label{t:all_results3}

(c) Mean Posterior Standard Deviation
\begin{tabular}{lll|rrrrr} \\
$\phi$ & $\beta_{XZ}$ & coefficents & MSM & USM & Naive & Spatial+80 & Spatial+40 \\
\hline
 1 & 0 & nonzero & 0.23 (0.01) & 0.46 (0.01) & 0.20 (0.00) & 0.27 (0.00) & 0.45 (0.00) \\
 1 & 1 & nonzero & 0.32 (0.00) & 0.48 (0.01) & 0.18 (0.00) & 0.26 (0.00) & 0.45 (0.00) \\ 
 1 & 2 & nonzero & 0.35 (0.01) & 0.46 (0.01) & 0.15 (0.00) & 0.24 (0.00) & 0.43 (0.00) \\
 2 & 1 & nonzero & 0.32 (0.01) & 0.48 (0.01) & 0.19 (0.00) & 0.27 (0.00) & 0.45 (0.00) \\ 
 2 & 2 & nonzero & 0.35 (0.01) & 0.48 (0.01) & 0.17 (0.00) & 0.27 (0.00) & 0.45 (0.00) \\[.05in] 
 1 & 0 & zero & 0.17 (0.01) & 0.32 (0.01) & 0.24 (0.00) & 0.33 (0.00) & 0.54 (0.00) \\
 1 & 1 & zero & 0.28 (0.00) & 0.36 (0.01) & 0.21 (0.00) & 0.31 (0.00) & 0.54 (0.00) \\ 
 1 & 2 & zero & 0.33 (0.01) & 0.36 (0.01) & 0.18 (0.00) & 0.28 (0.00) & 0.52 (0.00) \\
 2 & 1 & zero & 0.27 (0.01) & 0.36 (0.01) & 0.22 (0.00) & 0.33 (0.00) & 0.54 (0.00) \\ 
 2 & 2 & zero & 0.32 (0.01) & 0.37 (0.01) & 0.20     (0.00) & 0.33 (0.00) & 0.54 (0.00) \\ 
\end{tabular}
\end{center}
\end{table}

\begin{table}
\begin{center}
\caption{{\bf Simulation results (continued)}}
\label{t:all_results4}

(d) Empirical Coverage 
\begin{tabular}{lll|rrrrr} \\
$\phi$ & $\beta_{XZ}$ & coefficents & MSM & USM & Naive & Spatial+80 & Spatial+40 \\
\hline
 1 & 0 & nonzero & 0.94 (0.00) & 1.00 (0.00) & 0.95 (0.02) & 0.95 (0.01) & 0.95 (0.01) \\
 1 & 1 & nonzero & 0.98 (0.01) & 0.99 (0.00) & 0.86 (0.01) & 0.95 (0.01) & 0.95 (0.01) \\ 
 1 & 2 & nonzero & 0.98 (0.00) & 0.98 (0.00) & 0.70 (0.01) & 0.92 (0.01) & 0.95 (0.01) \\
 2 & 1 & nonzero & 0.98 (0.00) & 0.99 (0.00) & 0.91 (0.02) & 0.95 (0.01) & 0.95 (0.01) \\ 
 2 & 2 & nonzero & 0.98 (0.00) & 0.98 (0.00) & 0.82 (0.01) & 0.95 (0.01) & 0.95 (0.01) \\[.05in]
 1 & 0 & zero & 1.00 (0.00) & 1.00 (0.00) & 0.95 (0.02) & 0.95 (0.01) & 0.95 (0.01) \\
 1 & 1 & zero & 1.00 (0.00) & 1.00 (0.00) & 0.87 (0.01) & 0.94 (0.01) & 0.95 (0.01) \\ 
 1 & 2 & zero & 1.00 (0.00) & 1.00 (0.00) & 0.72 (0.01) & 0.92 (0.01) & 0.95 (0.01) \\
 2 & 1 & zero & 1.00 (0.00) & 1.00 (0.00) & 0.91 (0.02) & 0.95 (0.01) & 0.95 (0.01) \\ 
 2 & 2 & zero & 1.00 (0.00) & 1.00 (0.00) & 0.82 (0.01) & 0.95 (0.01) & 0.95 (0.01) \\ 
\end{tabular}
\end{center}
\end{table}

\begin{table}
\begin{center}
\caption{{\bf Simulation results (continued)}}
\label{t:all_results5}

(e) Mean Width of $95\%$ Credible Intervals
\begin{tabular}{lll|rrrrr} \\
$\phi$ & $\beta_{XZ}$ & coefficents & MSM & USM & Naive & Spatial+80 & Spatial+40 \\
\hline
 1 & 0 & nonzero & 0.92 (0.03) & 1.87 (0.03) & 0.79 (0.00) & 1.07 (0.00) & 1.77 (0.01) \\
 1 & 1 & nonzero & 1.28 (0.02) & 1.94 (0.03) & 0.70 (0.00) & 1.03 (0.00) & 1.75 (0.01) \\ 
 1 & 2 & nonzero & 1.41 (0.02) & 1.89 (0.03) & 0.60 (0.00) & 0.94 (0.00) & 1.71 (0.01) \\
 2 & 1 & nonzero & 1.28 (0.03) & 1.95 (0.03) & 0.75 (0.00) & 1.07 (0.00) & 1.77 (0.01) \\ 
 2 & 2 & nonzero & 1.41 (0.02) & 1.95 (0.03) & 0.68 (0.00) & 1.07 (0.00) & 1.77 (0.01) \\[.05in]
 1 & 0 & zero & 0.66 (0.03) & 1.35 (0.04) & 0.95 (0.00) & 1.29 (0.00) & 2.12 (0.01) \\
 1 & 1 & zero & 1.13 (0.02) & 1.50 (0.03) & 0.82 (0.00) & 1.23 (0.00) & 2.10 (0.01) \\ 
 1 & 2 & zero & 1.35 (0.03) & 1.50 (0.03) & 0.69 (0.00) & 1.11 (0.00) & 2.03 (0.01) \\
 2 & 1 & zero & 1.11 (0.03) & 1.50 (0.03) & 0.88 (0.00) & 1.29 (0.00) & 2.12 (0.01) \\ 
 2 & 2 & zero & 1.32 (0.03) & 1.53 (0.03) & 0.79 (0.00) & 1.29 (0.00) & 2.12 (0.01) \\ 
 \end{tabular}
\end{center}
\end{table}

\section{Sensitivity analysis for varying $L$ and $K$}\label{s:A4}

Our approach allows for flexibility in two parameters: the number of spline basis functions ($L$) and the rank of the tensor decomposition ($K$). Table \ref{t:compLK} shows the MSE and coverage rates for several $L$ and $K$ for $N=20$ replicates from simulation Setting 2. MSE is not sensitive to different $L$ and $K$, but coverage rates might be too low if $K$ is set too low. Our recommendation is setting $L$ and $K$ to be around and slightly larger than the expected complexity and make sure $K$ is not too small. Setting $K$ to be too large, however, risks poor convergence and slow computation.  

\begin{table}[ht]
    \centering
    \caption{{\bf MSE and coverage comparison} with different L and K for Setting 2 $(\phi=1, \beta_{XZ}=2)$, 20 replicates }
    \begin{tabular}
    {c|ccc|ccc} 
    MSE & \multicolumn{3}{|c|}{nonzero coefficients} & \multicolumn{3}{|c}{zero coefficients} \\
    \hline
    & L=5 &  L=10 & L=20 & L=5 & L=10 & L=20 \\
    \hline
    K=2 & 0.06 (0.02) & 0.06 (0.02) & 0.07 (0.02) & 0.02 (0.01) & 0.03 (0.01) & 0.02 (0.01) \\
    K=5 & 0.05 (0.01) & 0.06 (0.02) & 0.06 (0.02) & 0.03 (0.01) & 0.03 (0.01) & 0.02 (0.01) \\
    K=10 & 0.05 (0.01) & 0.05 (0.01) & 0.06 (0.02) & 0.03 (0.01) & 0.03 (0.01) & 0.02 (0.01) \\
    \hline
    Coverage & \multicolumn{3}{|c|}{nonzero coefficients} & \multicolumn{3}{|c}{zero coefficients} \\
    \hline
    & L=5 &  L=10 & L=20 & L=5 & L=10 & L=20 \\
    \hline
    K=2 & 0.82 (0.07) & 0.87 (0.06) & 0.91 (0.05) & 0.95 (0.04) & 0.98 (0.02) & 0.99 (0.00) \\
    K=5 & 0.99 (0.01) & 0.99 (0.01) & 0.99 (0.01) & 0.99 (0.01) & 1.00 (0.00) & 1.00 (0.00) \\
    K=10 & 1.00 (0.00) & 1.00 (0.00) & 1.00 (0.00) & 1.00 (0.00) & 1.00 (0.00) & 1.00 (0.00) \\
    \hline
    \end{tabular}
    \label{t:compLK}
\end{table}

\section{Maps of health conditions} \label{s:A5}

\begin{figure}
    \centering
    \caption{{\bf Maps of health outcomes}: 
    Four of the five chronic health outcomes in southern states are in (a)-(d). (The map for hypertension is in the main text.) For all conditions, the raw percentages are between 0.003 and 0.55.rural/urban determination in (e), and population size (number of people) in (f).}
    \subfigure[Chronic kidney disease]{\includegraphics[width=0.49\linewidth]{images/CKD2.png}}
    \subfigure[Hyperlipidemia]{\includegraphics[width=0.49\linewidth]{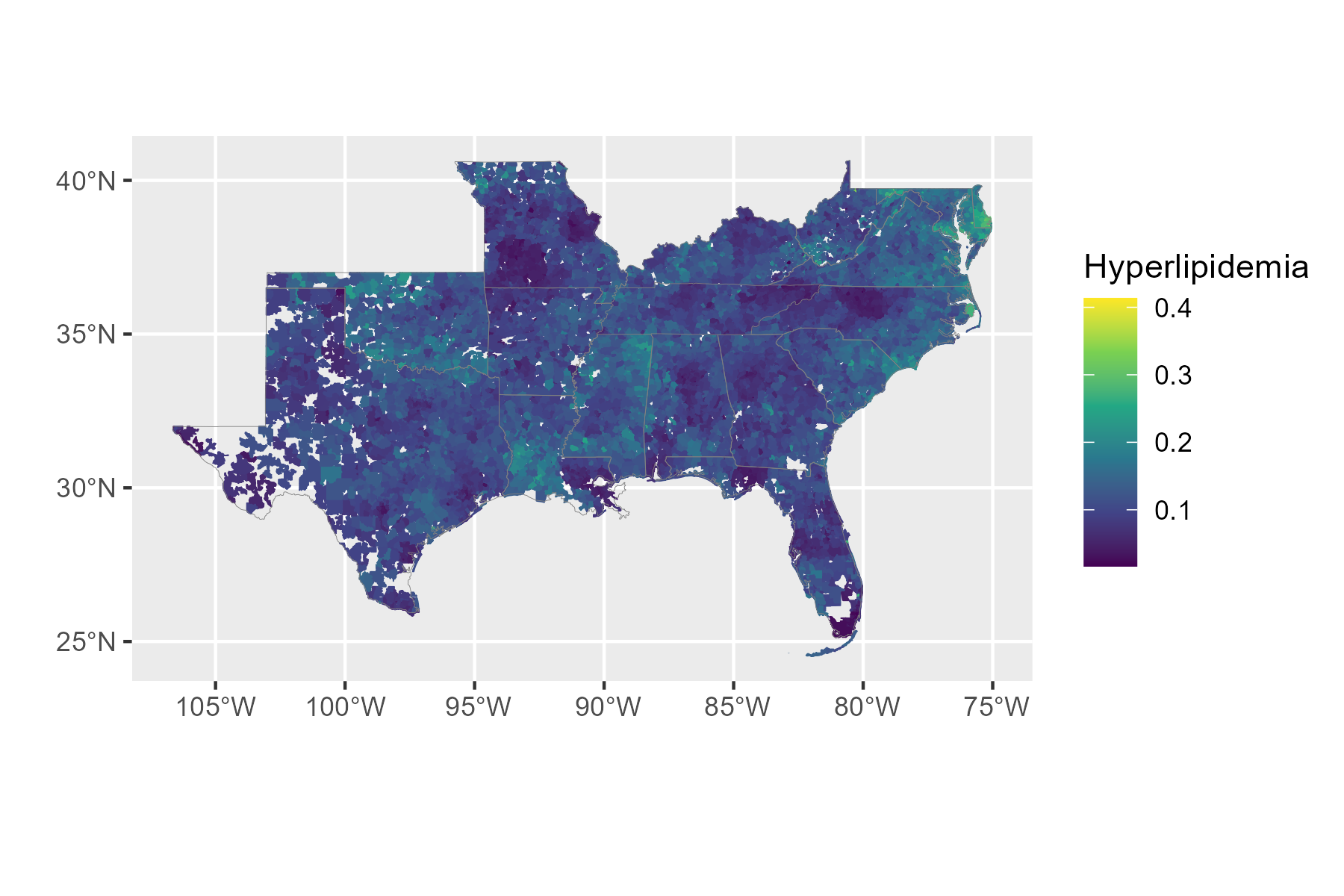}}
    \subfigure[Congestive heart failure]{\includegraphics[width=0.49\linewidth]{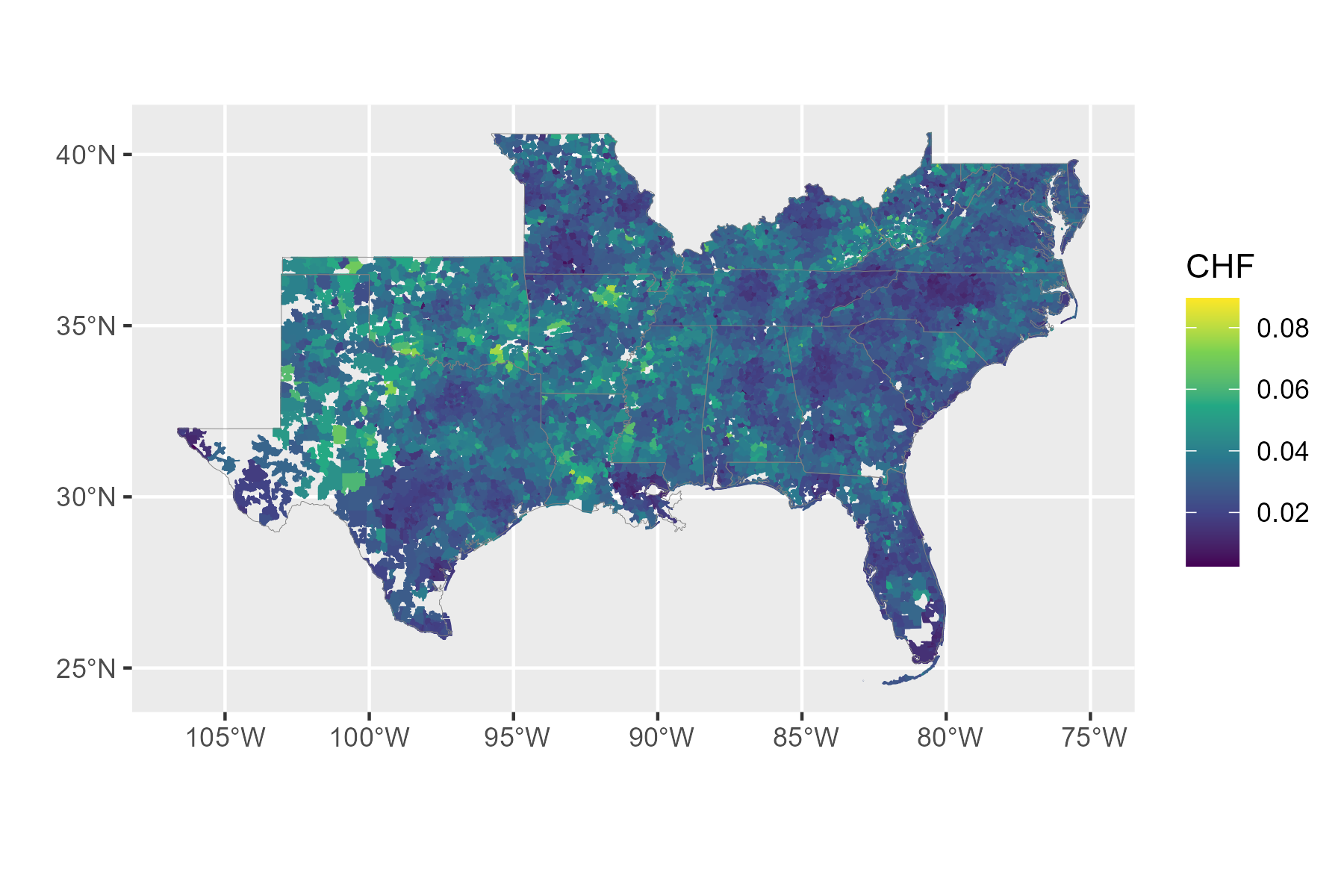}}
    \subfigure[Diabetes]
    {\includegraphics[width=0.49\linewidth]{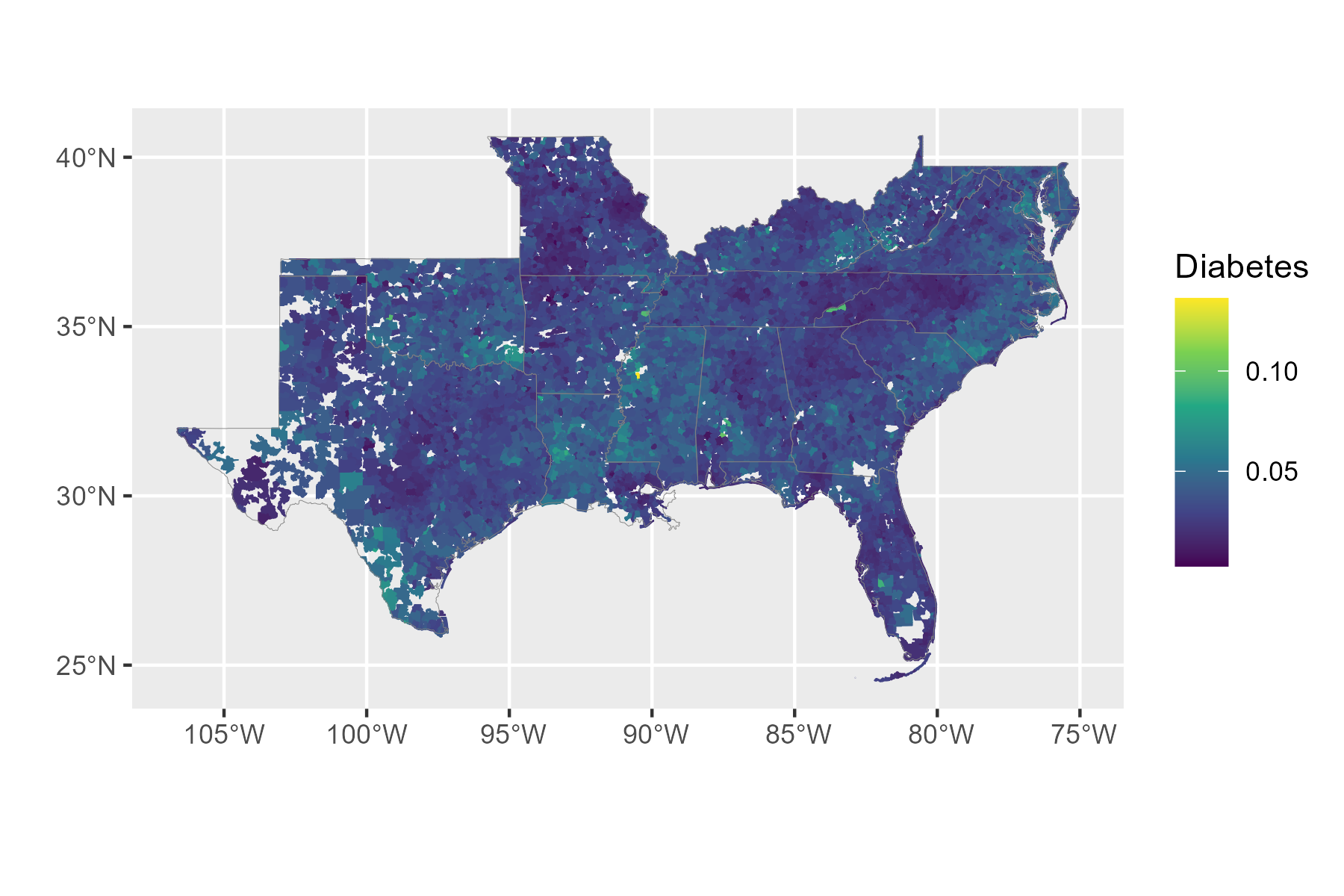}}
    \label{f:map_outcomes}
    \subfigure[Urbanicity]
    {\includegraphics[trim={.5cm, 1.5cm, 0cm, 1cm}, clip,width=0.49\linewidth]{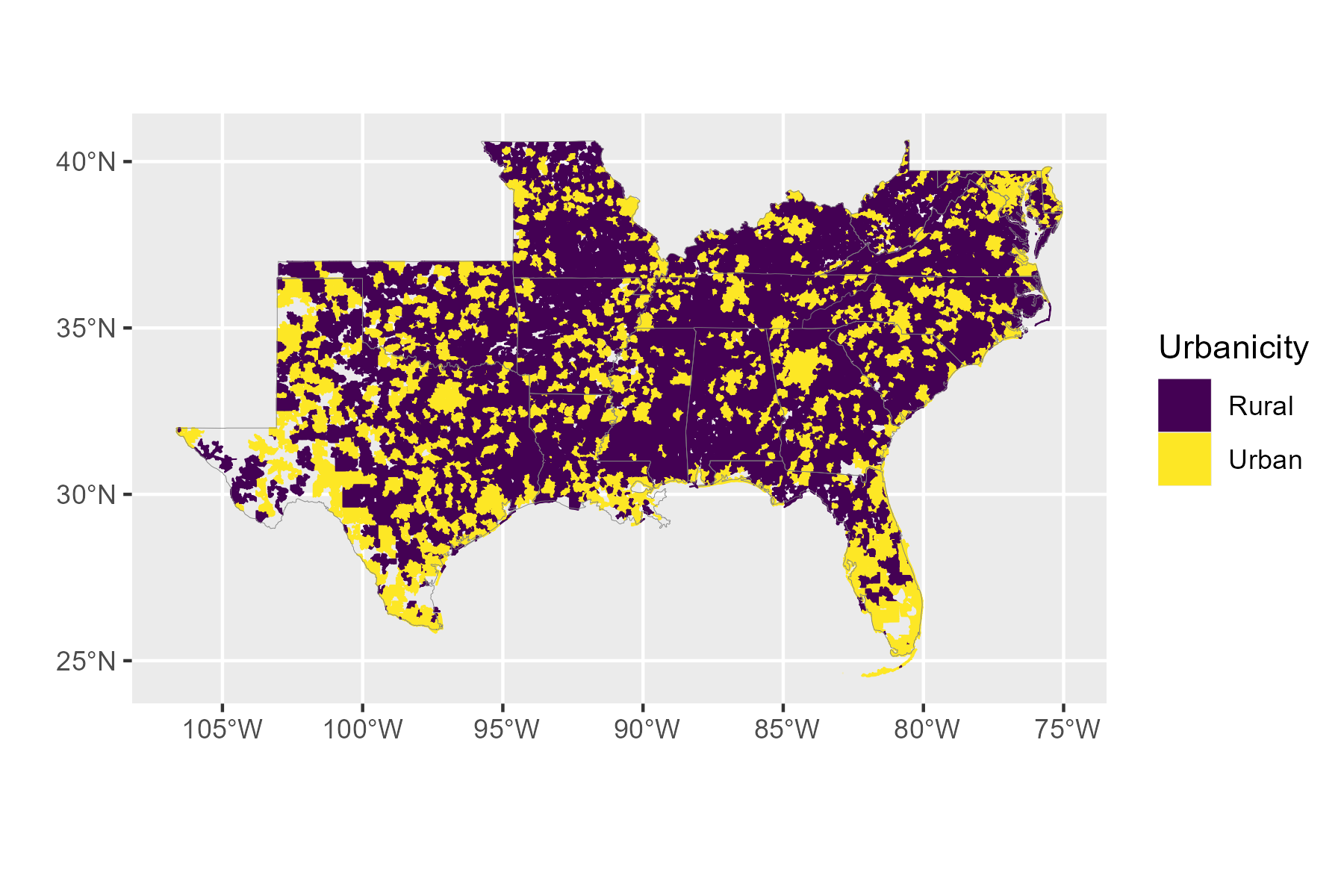}}
    \subfigure[Population size (log scale)]
    {\includegraphics[trim={.5cm, 1.5cm, 0cm, 1cm}, clip,width=0.49\linewidth]{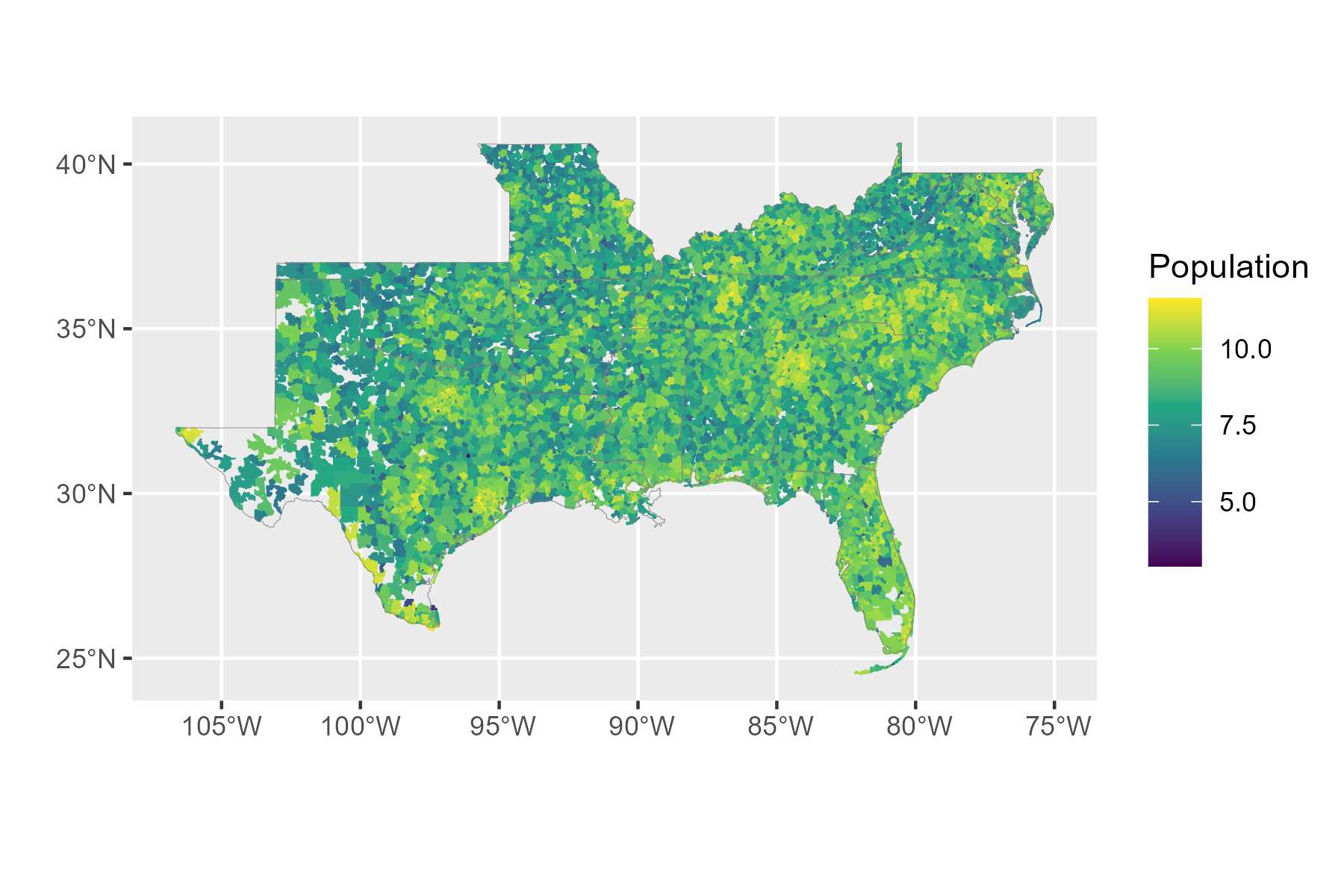}}
\end{figure}

Figure \ref{f:map_outcomes} plots the health outcome data (incident of five chronic conditions) and shows that some spatial trends exist in the outcomes.

\section{MCMC diagnostics for the data analysis}
\label{A:dianostic}

\begin{figure}
    \centering
    \caption{{\bf Trace plots from our proposed MSM model for the data analysis: } The rows are Themes 1-4 of SVI, and the column show the health outcomes (hypertension, chronic kidney disease, hyperlipidemia, congestive heart failure, and diabetes).}
    \includegraphics[width=1\linewidth]{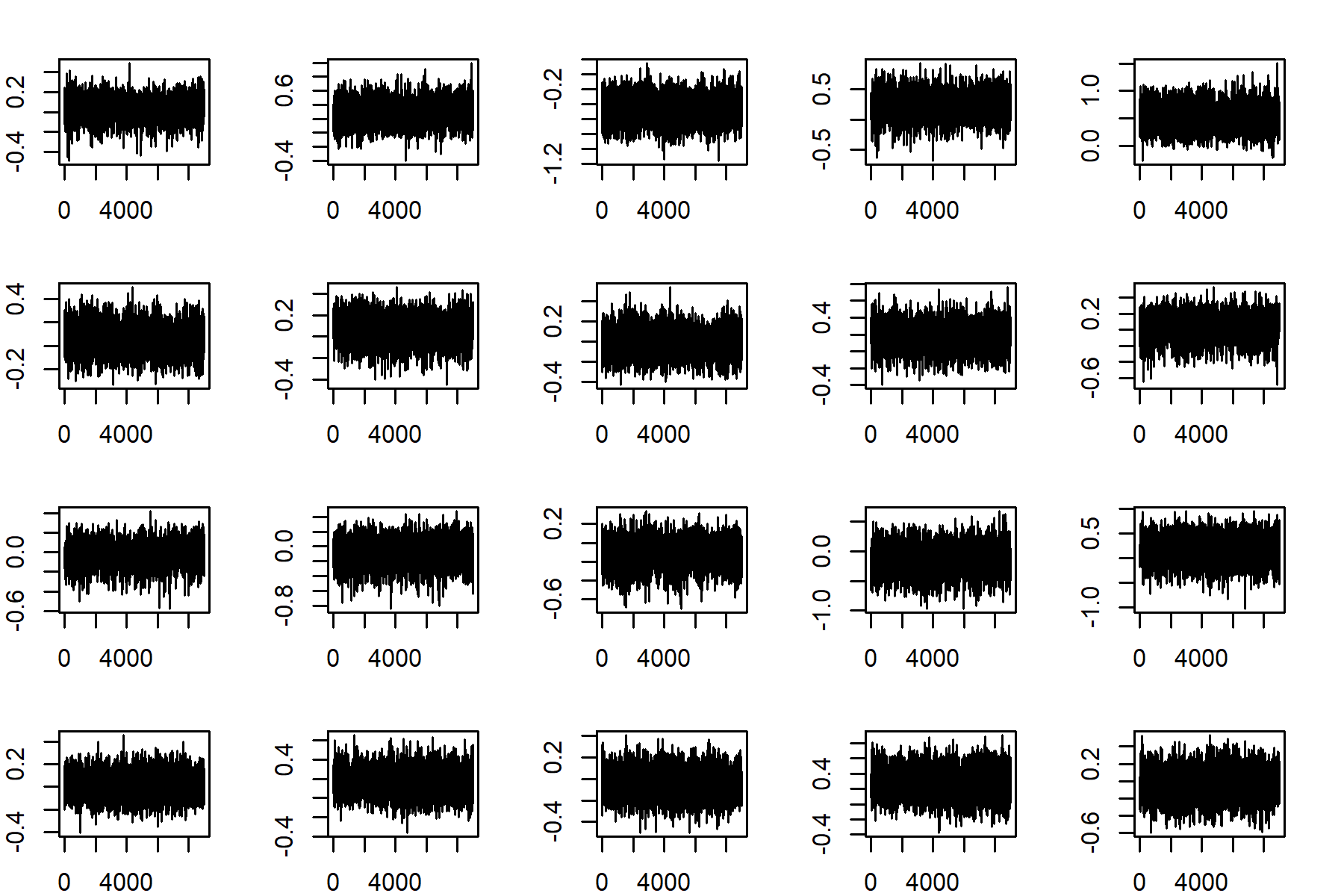}
    \label{fig:trace_final}
\end{figure}

Our final data analysis was run with $100,000$ iterations and $10,000$ burn-ins with a thinning rate of $10$. Convergence was confirmed by visual inspection of trace plots (Figure \ref{fig:trace_final}) and effective sample size for the exposure effects $\beta_{1er}$ (Table \ref{t:ESS}).

\begin{table}[]
    \centering
    \caption{{\bf Effective sample size for the data analysis: } The result was obtained from the final run of data analysis and confirmed convergence.}
    \begin{tabular}{c|ccccc}
         &  Hypertension & CKD & Hyperlipidemia & CHF & Diabetes\\
         \hline
         Theme 1 & 1173 & 2059 & 1285 & 1564 & 1797 \\
         Theme 2 & 1800 & 2630 & 3017 & 1756 & 2988 \\
         Theme 3 & 1633 & 1381 & 1272 & 2180 & 2280 \\
         Theme 4 & 3377 & 2471 & 2768 & 2309 & 2534 \\
    \end{tabular}
    
    \label{t:ESS}
\end{table}

\section{Residual analyses} \label{A:residuals}

\begin{figure}[H]
    \centering
    \caption{{\bf Residual plots in the spatial domain for five health outcomes}}
    \subfigure[Hypertension]{\includegraphics[width=0.49\linewidth]{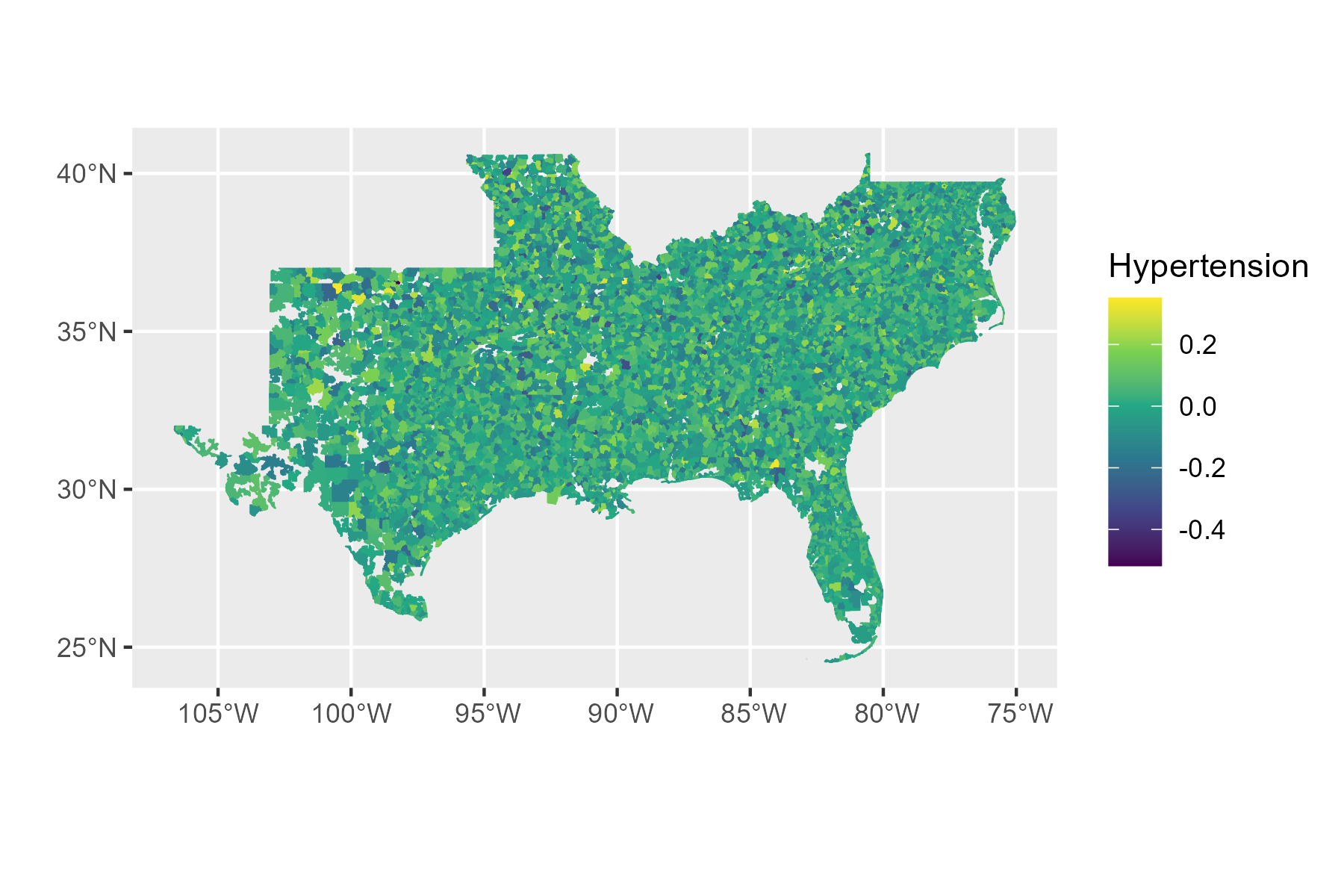}}
    \subfigure[Chronic kidney disease]{\includegraphics[width=0.49\linewidth]{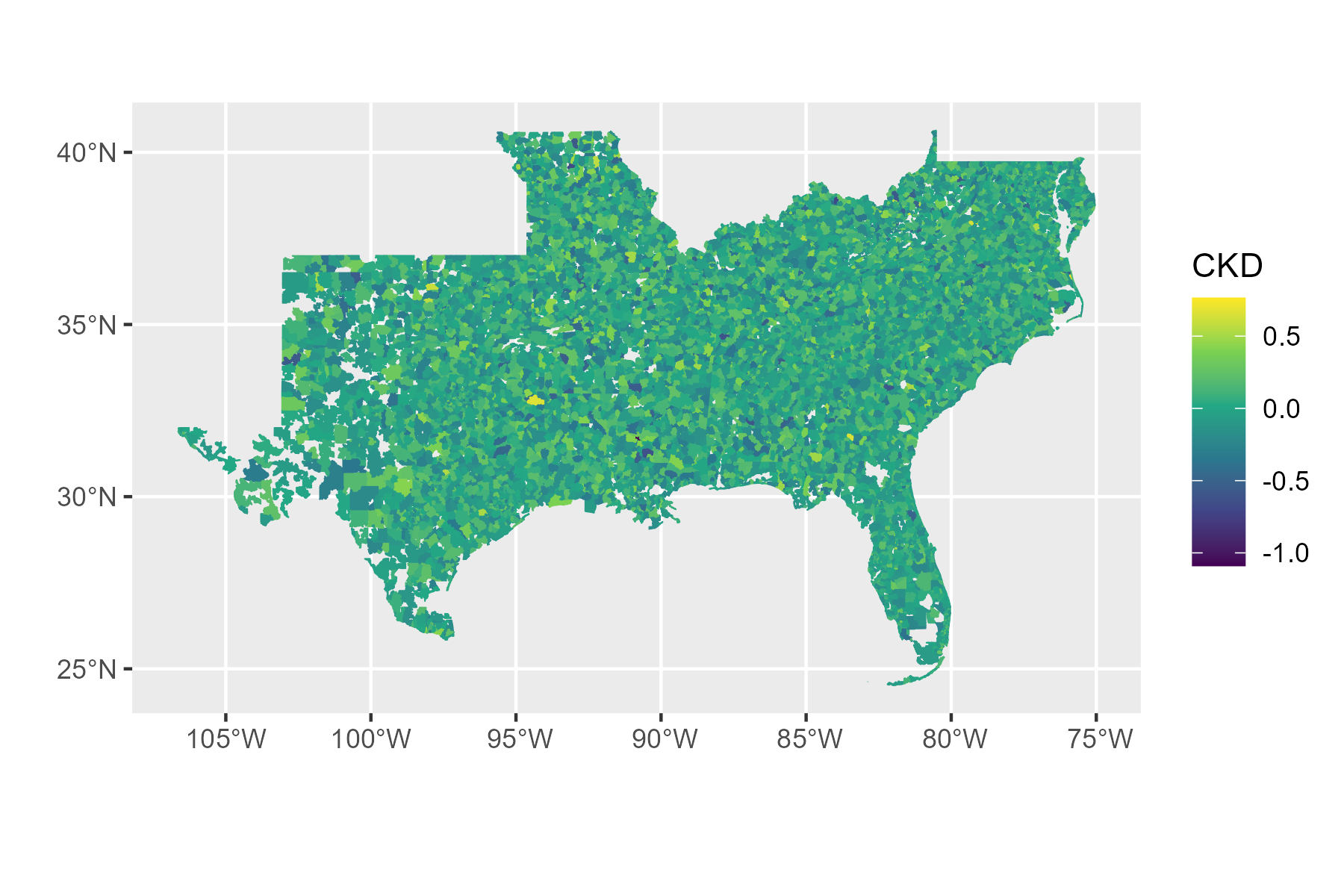}}
    \subfigure[Hyperlipidemia]{\includegraphics[width=0.49\linewidth]{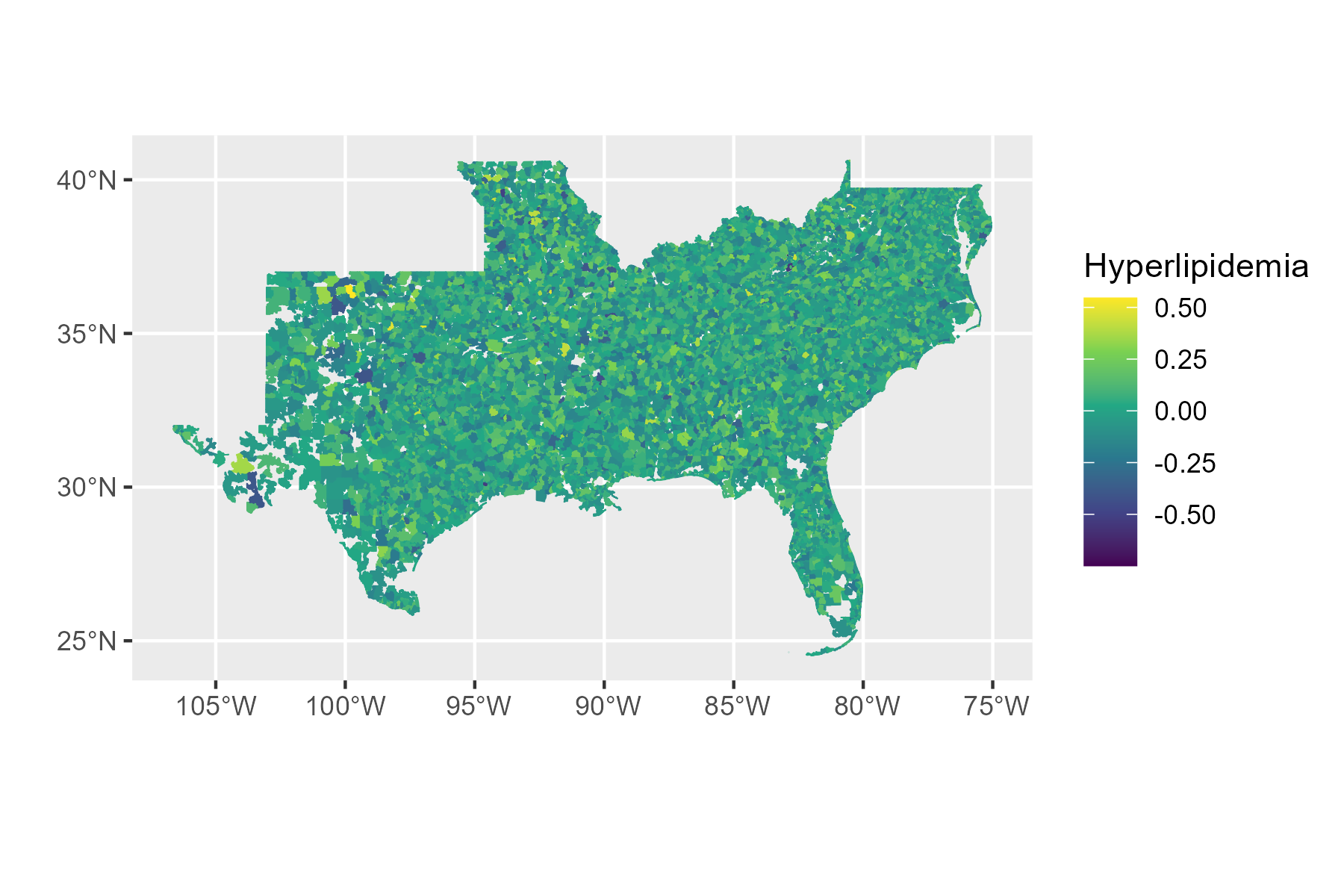}}
    \subfigure[Congestive heart failure]{\includegraphics[width=0.49\linewidth]{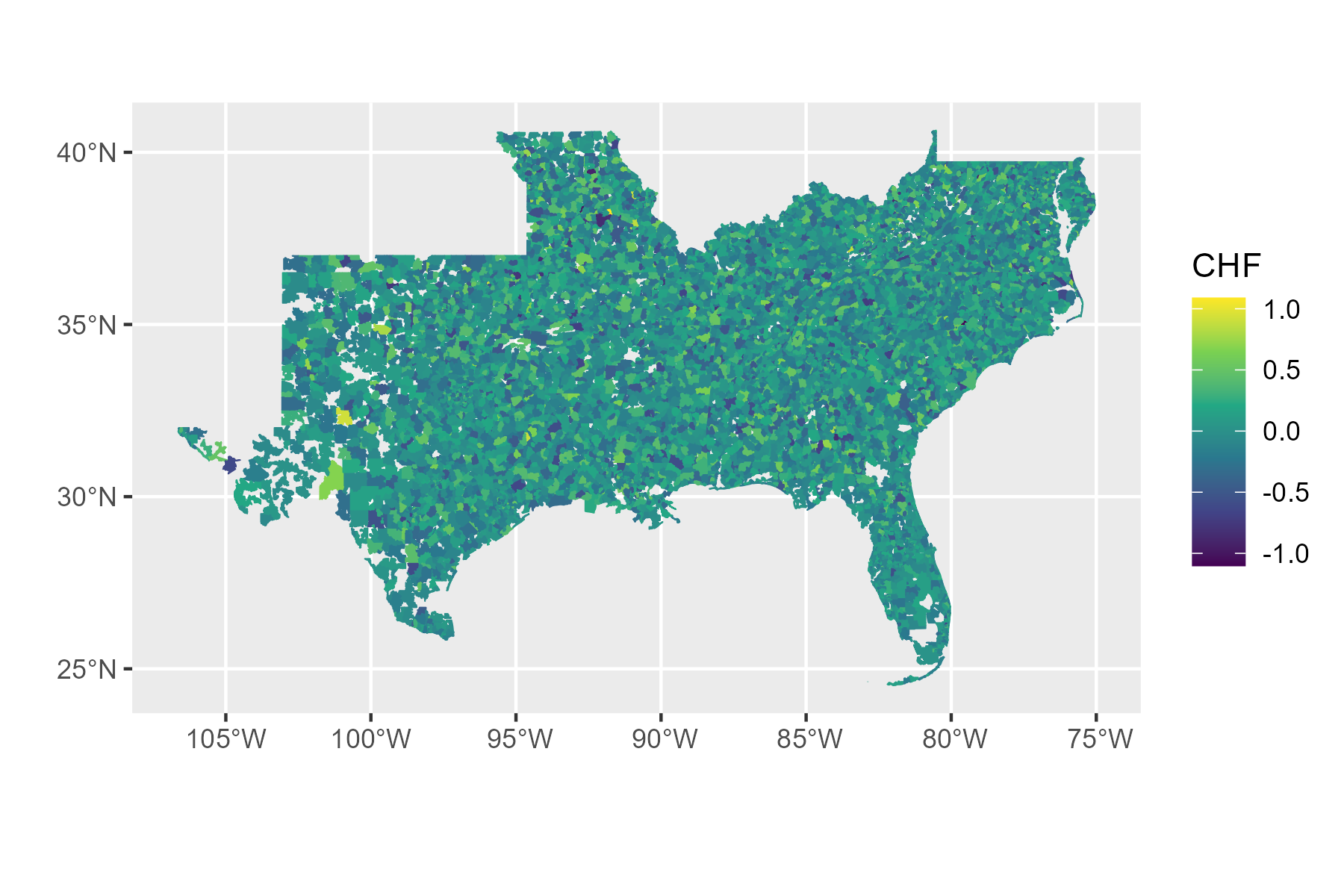}}
    \subfigure[Diabetes]
    {\includegraphics[width=0.49\linewidth]{images/residualsDiabetes.png}}
    \label{fig:residuals}
\end{figure}

\begin{figure}[H]
    \centering
    \caption{{\bf Residual analysis}: Density plots (left panel) show the distribution of residuals, and variograms (right panel) show spatial correlations.}
    \subfigure[Density plots of the residuals]{\includegraphics[width=0.49\linewidth]{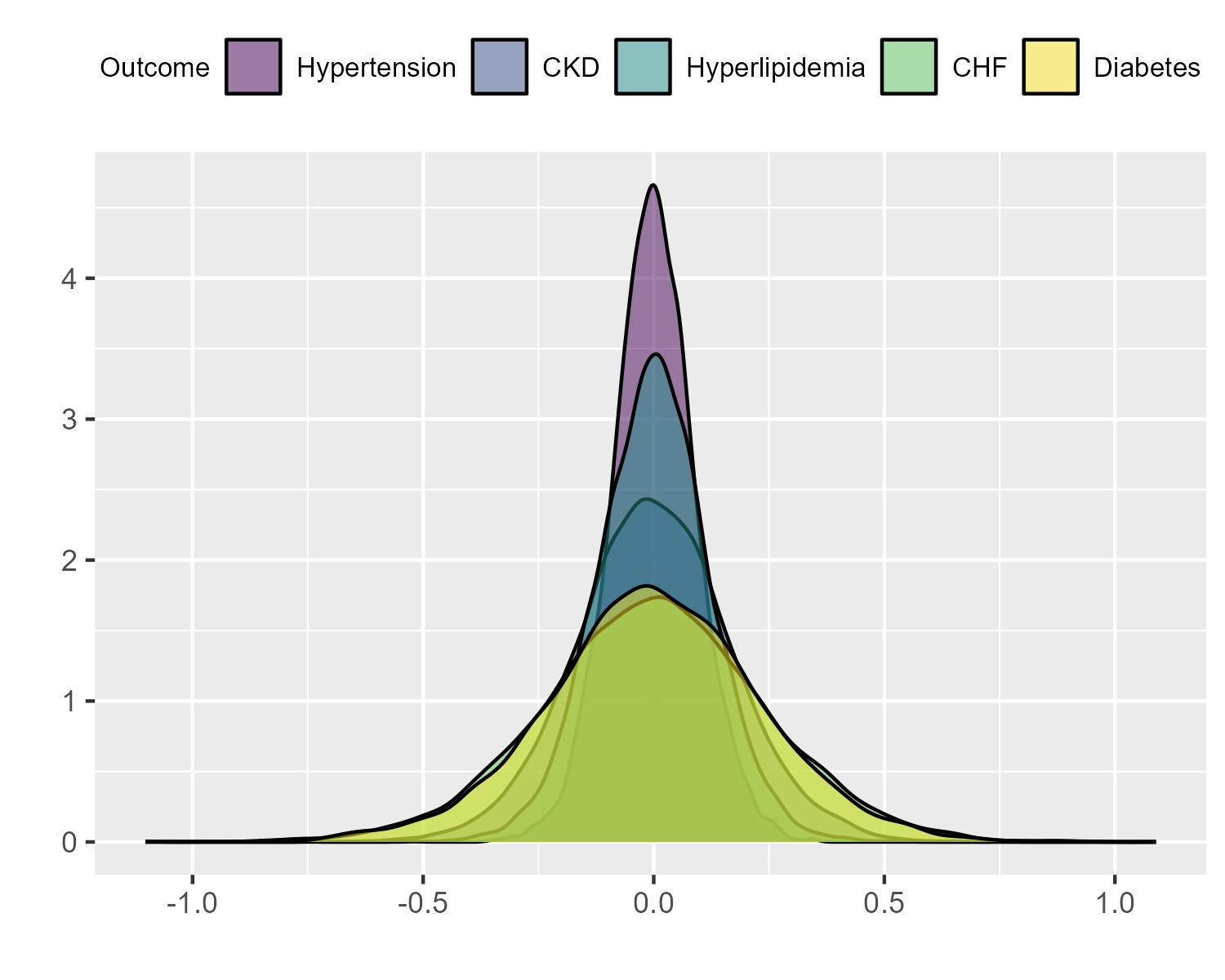}}
    \subfigure[Variograms of the residuals]{\includegraphics[width=0.49\linewidth]{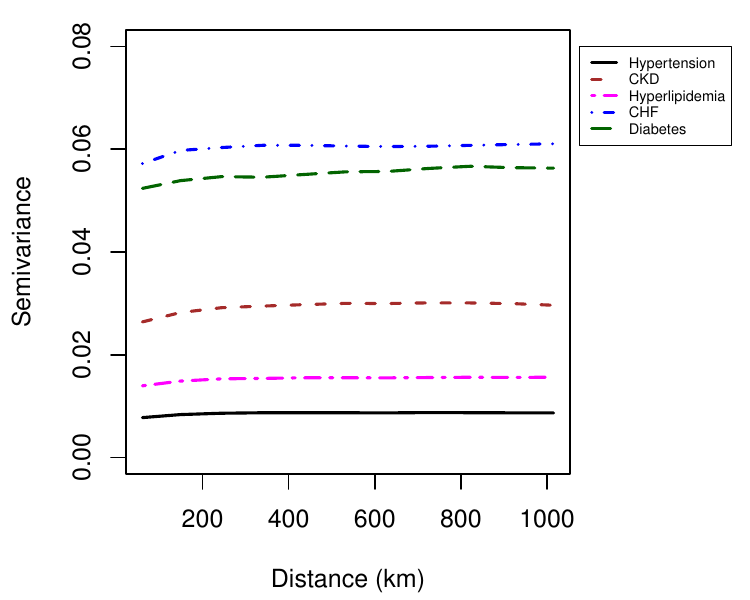}}
    \label{fig:residuals2}
\end{figure}

\begin{figure}[H]
    \centering
    \caption{{\bf Scatter plots of residuals and log population}: The blue lines are loess curves.}\includegraphics[width=0.8\linewidth]{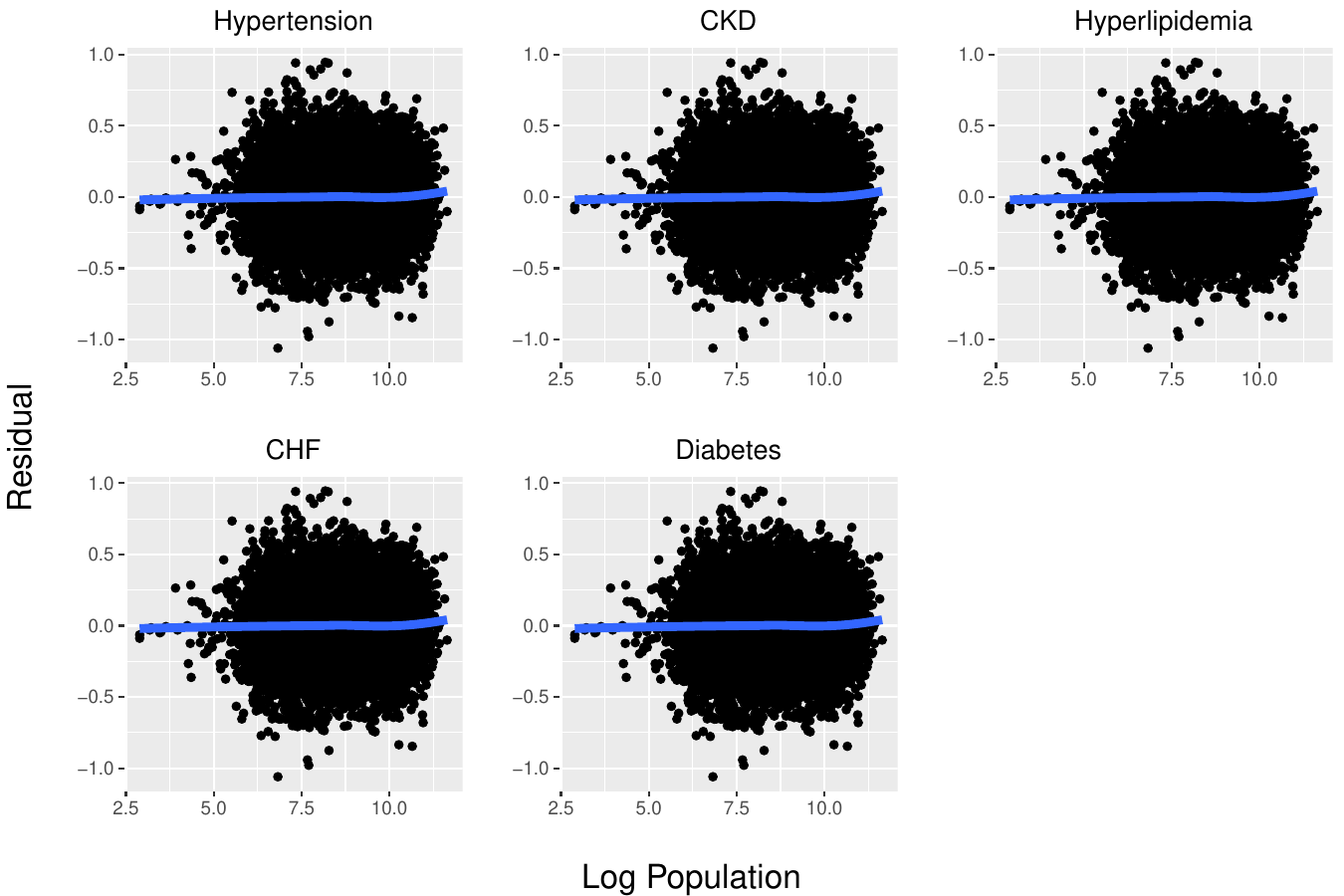}
    
    \label{fig:residual_population}
\end{figure}

    To ensure that the final model was adequate, we conducted residual analyses. Figure \ref{fig:residuals} shows the residuals in the spatial domain; none of the outcomes display any apparent spatial patterns or outliers. The density plots in Figure \ref{fig:residuals2} show the residuals are normally distributed and centered around zero, and the variograms show very low correlation across locations. Figure~\ref{fig:residual_population} plots the residuals versus the only observed continuous confounder, log population density, and shows no evidence of non-linearity.

\section{Sensitivity analysis derivations}
\label{A:sensitivity}
Let $X^*_{ie} = \left( \sqrt{1-\rho^2} \bU_{iX} + \rho \bU_{iC} \right) \bB_{eX}$, where $\bU_{iX}$ is the $i^{\text{th}}$ row of $\bU_X$ and $\bB_{eX}$ is the $e^{\text{th}}$ row of $\bB_X$. Then
\begin{eqnarray*}
    \text{Cov}\left( X^*_{ie}, \bU_{iC} \bB_{eX}^\top \right) &=& \text{Cov}\left( \rho \bU_{iC} \bB_{eX}^\top, \bU_{iC} \bB_{eX}^\top \right) 
    = \rho \text{Var} \left( \bU_{iC} \bB_{eX}^\top  \right) = \rho \bB_{eX} \bB_{eX}^\top\\
    \text{Var}\left(X^*_{ie}\right) &=& (1-\rho^2) \text{Var} \left( \bU_{iX} \bB_{eX}^\top \right) + \rho^2 \text{Var} \left( \bU_{iC} \bB_{eX}^\top \right) = \bB_{eX} \bB_{eX}^\top.
\end{eqnarray*}
Thus, $\text{Cor}\left(X^*_{ie}, \bU_{iC} \bB_{eX}^\top \right) = \rho.$   Using similar definitions for $\Theta_{ir}^*$, $U_{i\Theta}$ and $B_{r\Theta}$, 
\begin{equation*}
    \text{Cov}\left( \Theta^*_{ir}, X_{ie}^* \right) = \text{Cov}\left( \rho \bU_{iC} \bB_{r\Theta}^\top, \rho\bU_{iC} \bB_{eX}^\top \right) 
= \rho^2 \bB_{r\Theta} \bB_{eX}^\top.
\end{equation*}
Combining these results gives the expression for the bias, $\alpha_{ier}.$

\end{document}